\documentclass[prx, twocolumn, longbibliography]{revtex4-1}
\usepackage{graphicx}
\usepackage{amsmath}
\usepackage{amssymb}
\usepackage{color}
\usepackage{appendix}
\usepackage{braket}
\usepackage{bbm}
\usepackage{subfigure}
\usepackage{mathrsfs}

\usepackage[colorlinks,urlcolor=blue,citecolor=blue,linkcolor=blue]{hyperref}
\usepackage{lipsum}

\begin{document}
\title{Anatomy of open-boundary bulk in multiband non-Hermitian systems}
\author{Yongxu Fu}
\email{yongxufu@pku.edu.cn}
\affiliation{International Center for Quantum Materials, School of Physics, Peking University, Beijing, 100871, China}
\author{Yi Zhang}
\email{frankzhangyi@pku.edu.cn}
\affiliation{International Center for Quantum Materials, School of Physics, Peking University, Beijing, 100871, China}

\begin{abstract}
Although the non-Bloch band theory is a milestone in elaborating bulk energy bands of non-Hermitian systems under the open-boundary condition~(OBC), vital issues related to multivalued functions of non-Hermitian energy bands remain unsolved. In this paper, we anatomize the bulk properties of one-dimensional multiband non-Hermitian systems under OBC. We put forward the energy-band branches~(EBBs) to settle the multivalued functions of non-Hermitian energy bands, which become gapped or gapless corresponding to disconnected or connected EBBs in the complex energy plane, where the branch points and branch cuts play a crucial role. We clarify the precise significance of the non-Hermitian skin effect, which illustrates the asymptotic behavior of EBB eigenstates~(bulk eigenstates) in the deep bulk and compensates previous non-Bloch band theory. We also obtain a general expression of open-boundary Green's functions based on such EBBs and generalized Brillouin zones, useful for studies on non-Hermitian dynamical evolution. 
\end{abstract}

\maketitle

\section{Introduction}
\label{section1}

The latest developments of the fundamental theories of non-Hermitian systems~\cite{ashida2020,bergholtzrev2021}, including the energy band theory~\cite{shen2018,kunst2018,yao2018,yao201802,gong2018,kawabataprx,song2019,yokomizo2019,longhi2019,lee2019an,zhang2020,slager2020,origin2020,yang2020}, the recast of bulk-boundary correspondence~\cite{kunst2018,yao2018,yao201802,kawabata2021,zirnstein02021,zirnstein2021,yang2020}, the exceptional points of non-Hermitian systems~\cite{kawabata2019,okugawa2019,bergholtz201901,bergholtz201902,yang2019,li2019,rui2019,zhang2020ep,xue2020dirac,yokomizo2020,jones2020,knot2021,yang2021,crippa2021,ghorashi202101,ghorashi202102,fu2022}, and the non-Hermitian higher-order topological phases~\cite{kawabata2019second,edvardsson2019,ezawa2019,okugawa2019ho,lee2019ho,zhang2019ho,luo2019h0,yutaro2020,kawabatahigher,okugawa2020,fu2021,yu2021ho,wang2021ho,palacios2021,st2022}, have received much research attention in condensed matter physics. Recently, research on dynamical evolution phenomena~\cite{longhi2022,longhihealing,xue2021simple,edgeburst2022,guo2021,exact2022}, many-body properties~\cite{leemanybody2020,mu2020,kawabata2022,zhang2022,fatenonher2022,cluster2022}, non-Bloch band theory in both one and higher dimensions~\cite{zhang2022uni,wu2022,yokomizo2022,hu2022}, etc., has further broadened our scope and avenue on non-Hermitian systems. However, there remain several key issues in the foundation of non-Bloch band theory~\cite{yao2018,yokomizo2019} when it comes to one-dimensional~(1D) multiband non-Hermitian systems under the open-boundary condition~(OBC), where energy bands take the form of multivalued functions~\cite{book1,book2,book3}. Even with the concept of sub-generalized Brillouin zones (sub-GBZs)~\cite{yang2020}, an unambiguous bridge between non-Hermitian energy bands and multivalued functions is still lacking due to the latter's branch points and branch cuts. In addition, it is always the default that the part of the GBZ inside (outside) the unit circle indicates the left~(right) localized energy-band branch~(EBB) eigenstates in the terminology of the non-Hermitian skin effect~(NHSE), which needs to be more precisely elaborated.

Many functions show multivalued properties with variables lying in the complex plane, such as radical functions, logarithmic functions, inverse trigonometric functions, etc. Let us take the square-root function $w(z)=\sqrt{z-a}$ as a simple example, where $z\in\mathbb{C}$ is the variable and $a$ is a complex constant. The phase increment of $z-a$ along a closed, counterclockwise loop $C$ is $2\pi$~($0$) with $a$ inside (outside) $C$, indicating the multivalued nature of $z-a$. The special point $z=a$ is a branch point in the multivalued function $w(z)$, while the other branch point is implicit $z=\infty$. In polar coordinates, $w(z)=\rho e^{i\theta/2}$, $\rho=\sqrt{|z-a|}$, and different settings of single-valued branches are determined by the different choices of $\theta$ ranges, such as $\theta\in\left[0,2\pi\right)$ and $\theta\in\left[2\pi,4\pi\right)$ corresponding to the two branches with the argument of $w(z)$ in the range $\left[0,\pi\right)$ and $\left[\pi,2\pi\right)$, respectively. The two single-valued branches of $w(z)$ are divided by the so-called branch cut, which is constructed by connecting $a$ and $\infty$ through a proper path~(such as a straight line); that is, a path crossing the branch cut brings $w(z)$ from one single-valued branch to the other in the Riemann surface and is prohibited in a single-valued branch~\cite{book1,book2,book3}. In a non-Bloch band theory, the energy bands $E(\beta)$, $\beta\in\mathbb{C}$, of non-Hermitian systems under OBC are usually multivalued functions mathematically~\cite{math1,math2,math3,math4,math5,math6}, and the branch points and branch cuts of $E(\beta)$ thus play important and inevitable roles in non-Hermitian systems.

In this paper, we aim to address these remaining issues on 1D multiband non-Hermitian systems. We focus on the properties of non-Hermitian systems in the deep bulk and assume that the non-Hermitian 1D chains are sufficiently long. In Sec.~\ref{section2}, based on single-valued branches of multivalued functions, we put forward the concept of EBBs as exact manifestations of the energy bands in non-Hermitian systems under OBC. The multivalued functions' branch points and branch cuts play crucial roles: they are responsible for the transition between gapped and gapless bands and the stability of localized edge states in multiband non-Hermitian systems. Following the GBZs, we discover the precise significance of the NHSE, which provides a description of EBB eigenstates in the deep bulk in Sec.~\ref{section3}, and obtain a general expression of open-boundary Green's functions in the presence of NHSE useful for non-Hermitian dynamical evolution in Sec.~\ref{section4}. Finally, the conclusion is given in Sec.~\ref{section5}.

\section{Energy-band branches with branch points and branch cuts}
\label{section2}
\subsection{Energy-band branches}
The tight-binding Hamiltonian of a 1D noninteracting non-Hermitian chain of length $L$ reads 
\begin{align}
    \label{generaltba}
    \hat{H}=\sum_{x}\sum_{n\in \mathscr{D}} c_{x}^{\dagger} t_{n} c_{x+n},
\end{align}
with the matrix elements representing the internal degrees of freedoms $\mu \in (1, \mathcal{M})$, and the hopping amplitude matrices $t_{n}$ range $\mathscr{D}=\left\{-\mathcal{P},-\mathcal{P}+1,\ldots,\mathcal{Q}-1,\mathcal{Q}\right\}$. The non-Hermiticity is induced by $t_{n}\neq t_{-n}^{\dagger}$ for some non-negative $n$. The energy bands $E_{\mu}(k)$ are obtainable via the Bloch Hamiltonian $H(k)$ under the periodic boundary condition~(PBC). When it comes to non-Hermitian systems under OBC, the bulk energy bands are obtainable from the non-Bloch Hamiltonian $H(\beta)=\sum_{n\in \mathscr{D}}t_{n}\beta^{n}$ with $\beta$ lying on the GBZ in the complex $\beta$ plane~\cite{yao2018,yokomizo2019,yang2020}. On the other hand, due to the complex-valued nature of eigenvalues of non-Hermitian matrices, multiple energy bands arise from the single-valued branches $E_{\mu}(\beta)$, which are the roots of the characteristics equation 
\begin{align}
    \label{characteristics}
    ch(\beta,E)\equiv \det{\left[E-H(\beta)\right]}=0,
\end{align}
constituting a multivalued function with respect to $\beta$. Each branch $E_{\mu}(\beta)$ is a single-valued function of $\beta \in \mathbb{C}$, occupying a continuous region~(open set) $\mathbb{C}_{\mu}$ in the complex $E$ plane. After ordering the solutions of the characteristic Eq.~(\ref{characteristics}) as $|\beta_{1}(E)|\leq|\beta_{2}(E)|\leq\ldots\leq|\beta_{p+q}(E)|$ with $p=\mathcal{M}\mathcal{P}$, $q=\mathcal{M}\mathcal{Q}$ in general cases, the bulk spectra under OBC are given by those $E\in\mathbb{C}$ satisfying $|\beta_{p}(E)|=|\beta_{p+1}(E)|$; the corresponding $\beta$ values outline the GBZ in the complex $\beta$ plane. In general, the bulk spectra are composed of distinct EBBs corresponding to their respective GBZs, which are dubbed sub-GBZs denoted as $GBZ_{\mu}$~\cite{yang2020}. The EBBs are exactly the sub-GBZ spectra $E_{\mu}[GBZ_{\mu}]\equiv\left\{E_{\mu}(\beta),\beta\in GBZ_{\mu}\right\}$ located in $\mathbb{C}_{\mu}$~(Appendix \ref{appendixA}). Moreover, each $GBZ_{\mu}$ is a closed curve, which encloses $p$ zeros of $ch(\beta,E)$ for $E\in\mathbb{C}_{\mu}$ and $E\notin E_{\mu}[GBZ_{\mu}]$, thus leading to the vanishing of the winding number of $ch(\beta,E)$ surrounding $GBZ_{\mu}$~(see Appendix \ref{appendixA} for details).

\begin{figure}   
    \subfigure[]{\includegraphics[width=4cm, height=4.5cm]{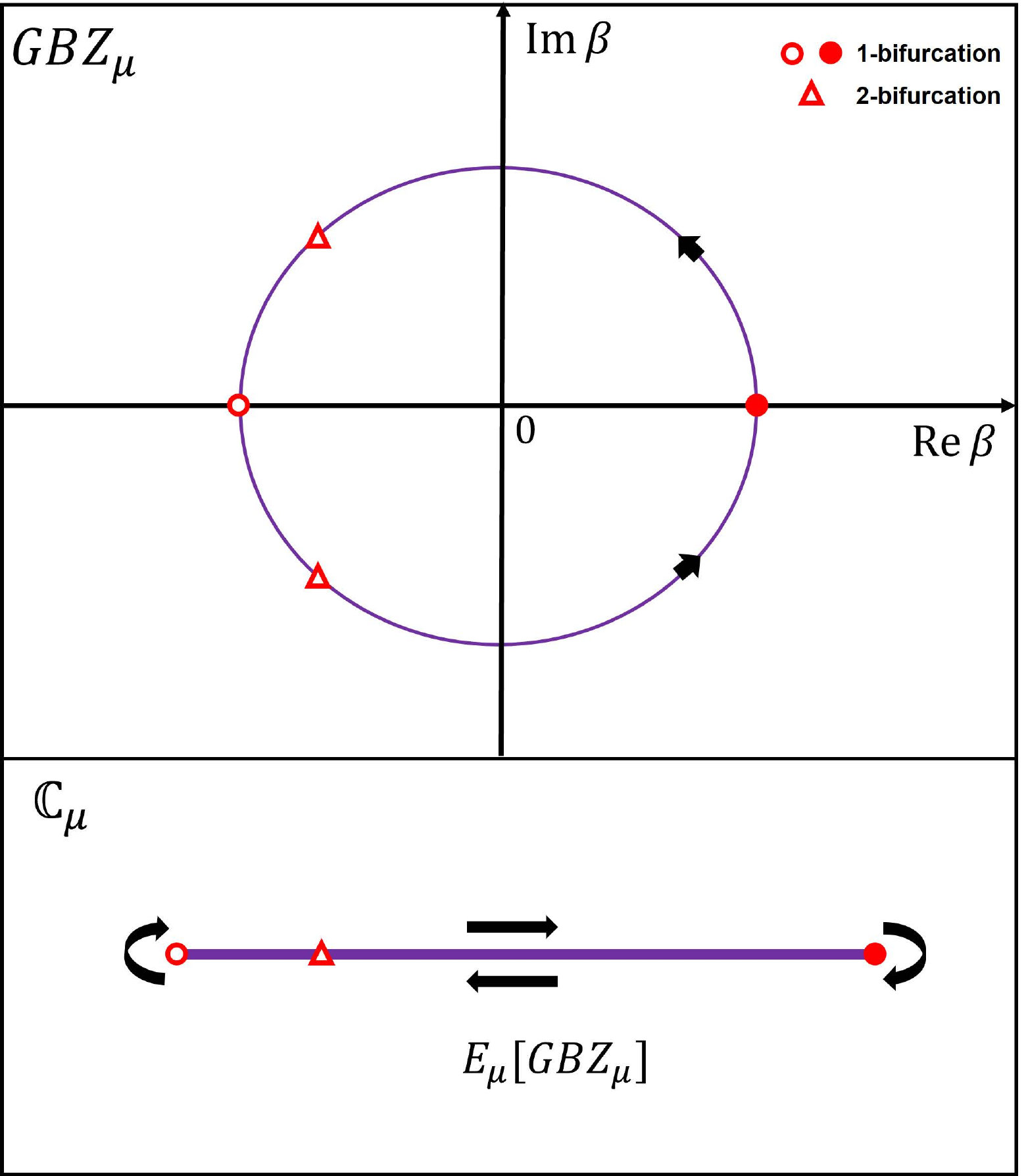}}\quad
    \subfigure[]{\includegraphics[width=4cm, height=4.5cm]{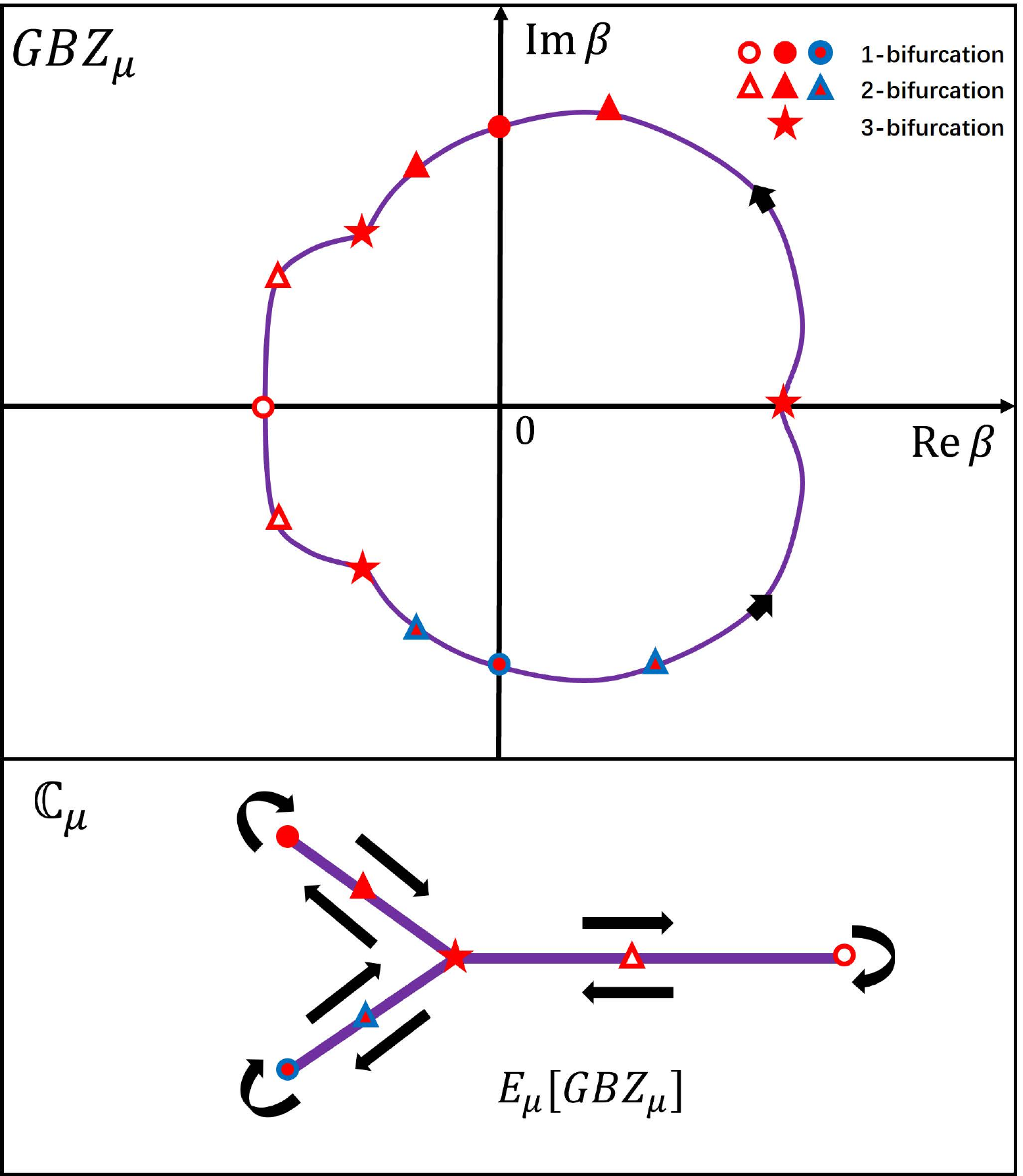}}
    \caption{We label several examples of $1$-, $2$-, and $3$-bifurcation states in the schematic illustrations of $GBZ_{\mu}$ (upper panels) and $E_{\mu}[GBZ_{\mu}]$ (lower panels): (a) a case with only $1$- and $2$-bifurcation states, and (b) a case with $1$-, $2$-, and $3$-bifurcation states. The black arrows denote the moving directions in $GBZ_{\mu}$ and $E_{\mu}[GBZ_{\mu}]$ as we circle counterclockwise around the GBZs. }
    \label{branch_conn}
\end{figure}

Consider an arbitrary point $E_{0}$ on an EBB $E_{\mu}[GBZ_{\mu}]$; the solutions of Eq.~(\ref{characteristics}) with respect to $E_{0}$ are $|\beta_{1}|\leq\ldots\leq|\beta_{p}|=|\beta_{p+1}|\leq\ldots\leq|\beta_{p+q}|$, and $\beta_{p}, \beta_{p+1}$ lie on GBZ~\cite{yao2018,yokomizo2019,yang2020}. We expect that $\beta_{p}, \beta_{p+1}$ both belong to $GBZ_{\mu}$, and to address this vital property, we observe the connectedness between two arbitrary EBBs $E_{\mu}[GBZ_{\mu}]$ and $E_{\nu}[GBZ_{\nu}]$. Intuitively, the two EBBs are either completely disconnected or overlapping at some points in the complex $E$ plane. The former corresponds to two gapped EBBs, which implies $E_{0}=E_{\mu}(\beta_{p})=E_{\mu}(\beta_{p+1})$; i.e., $\beta_{p}, \beta_{p+1}$ both lie on $GBZ_{\mu}$ only, and vice versa for $GBZ_{\nu}$. The latter is more subtle, of which we consider two situations assuming that $E_{0}$ is an overlapping point between two EBBs. First, $\beta_{p}, \beta_{p+1}$ lie on both $GBZ_{\mu}$ and $GBZ_{\nu}$ (therefore their intersections), which makes the two EBBs degenerate at the points, i.e., gapless with the possible emergence of exceptional points under OBC~\cite{fu2022}. Second, $E_{0}=E_{\mu}(\beta_{p})=E_{\nu}(\beta_{p+1})$; however, we can always avoid such a situation through suitable settings of the single-valued branches and keeping each EBB and its sub-GBZ continuous. To see this, we note that, in general, the points on the GBZ with $n$ equal norms of $\beta's$ corresponding to an eigenenergy are called $n$-bifurcation states~\cite{wu2022}. Most of the points on the GBZ with only two equal norms $|\beta_{p}|=|\beta_{p+1}|$ are 2-bifurcation states~\cite{wu2022}, while the points $\beta_{p}=\beta_{p+1}$ corresponding to the end points of EBBs are 1-bifurcation states; see illustrations in Fig.~\ref{branch_conn}. As we circle counterclockwise around $GBZ_{\mu}$ and pass through those $1$-, $2$-, and $3$-bifurcation points, the corresponding eigenenergies are visited $1$, $2$, and $3$ times, respectively, during the continuous movement along $E_{\mu}[GBZ_{\mu}]$. Without loss of generality, we can always set the single-valued branches to include the continuous EBBs $E_{\mu}[GBZ_{\mu}]$, which consequently ensures that all points associated with the $n$-bifurcation states with respect to $E\in E_{\mu}[GBZ_{\mu}]$, including the aforementioned 2-bifurcation points $\beta_{p}, \beta_{p+1}$, lie on the same $GBZ_{\mu}$.

\begin{figure}
    \subfigure[]{\includegraphics[width=4cm, height=3cm]{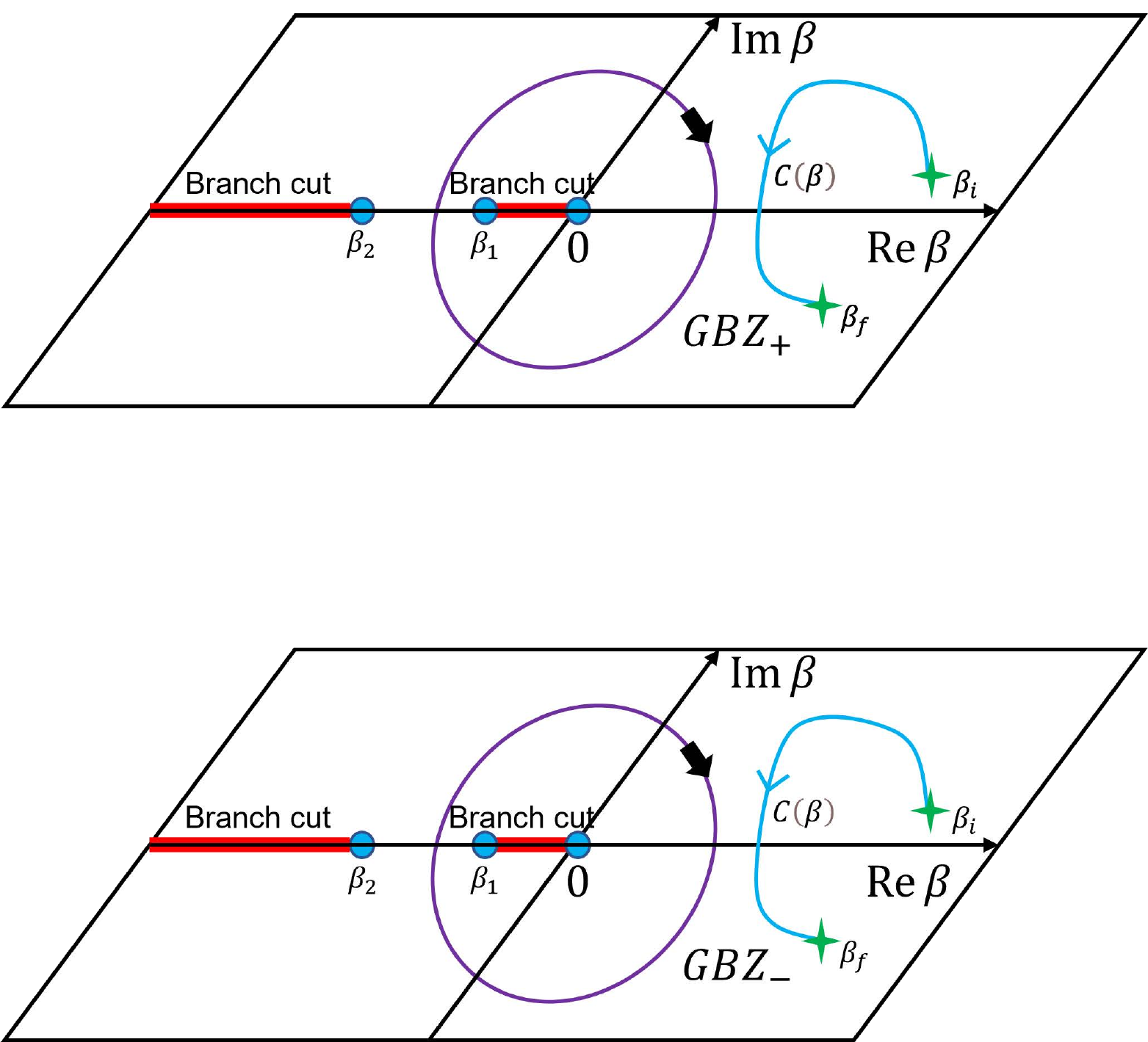}}
    \subfigure[]{\includegraphics[width=4cm, height=3cm]{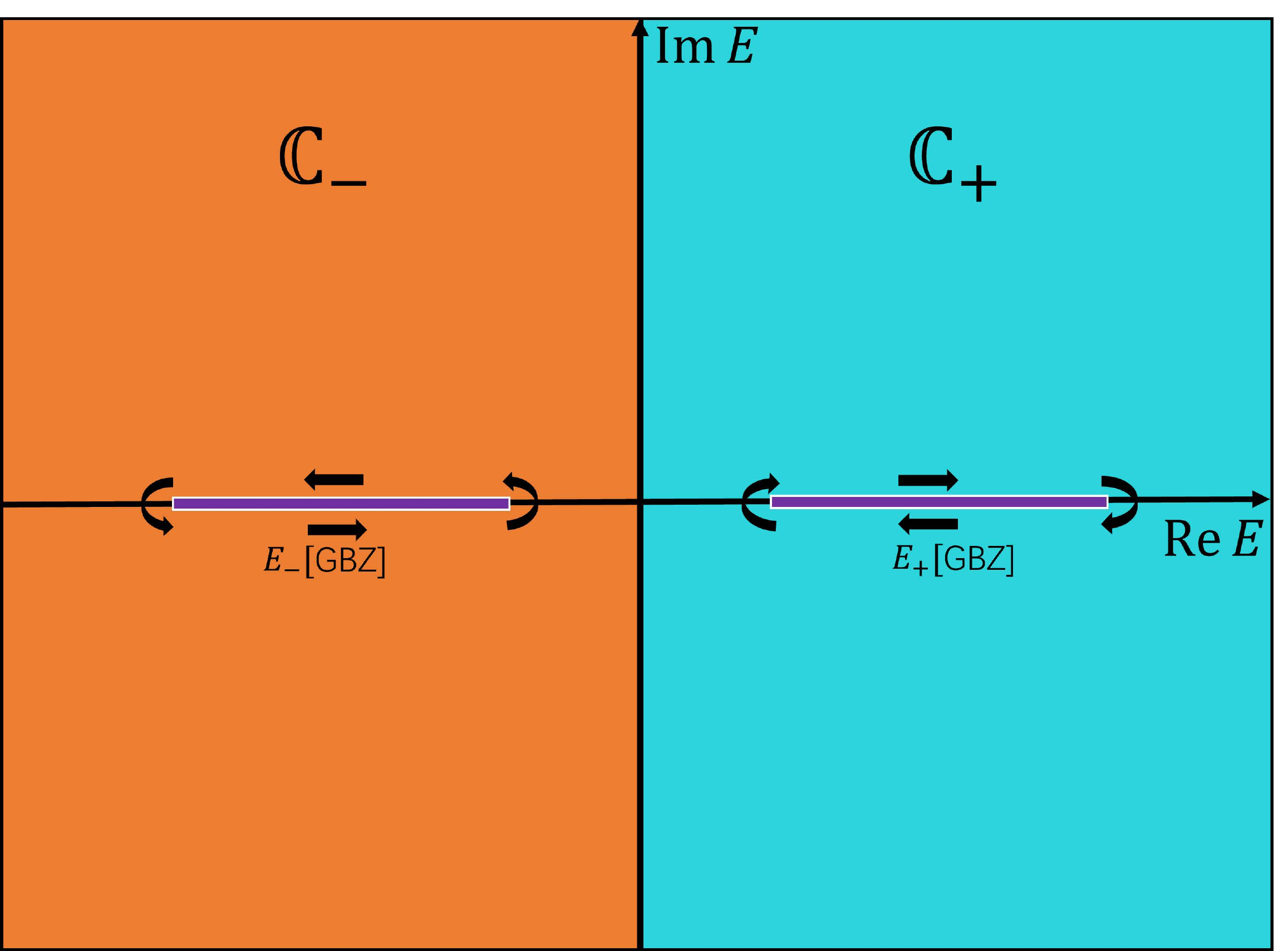}}
    \subfigure[]{\includegraphics[width=4cm, height=3cm]{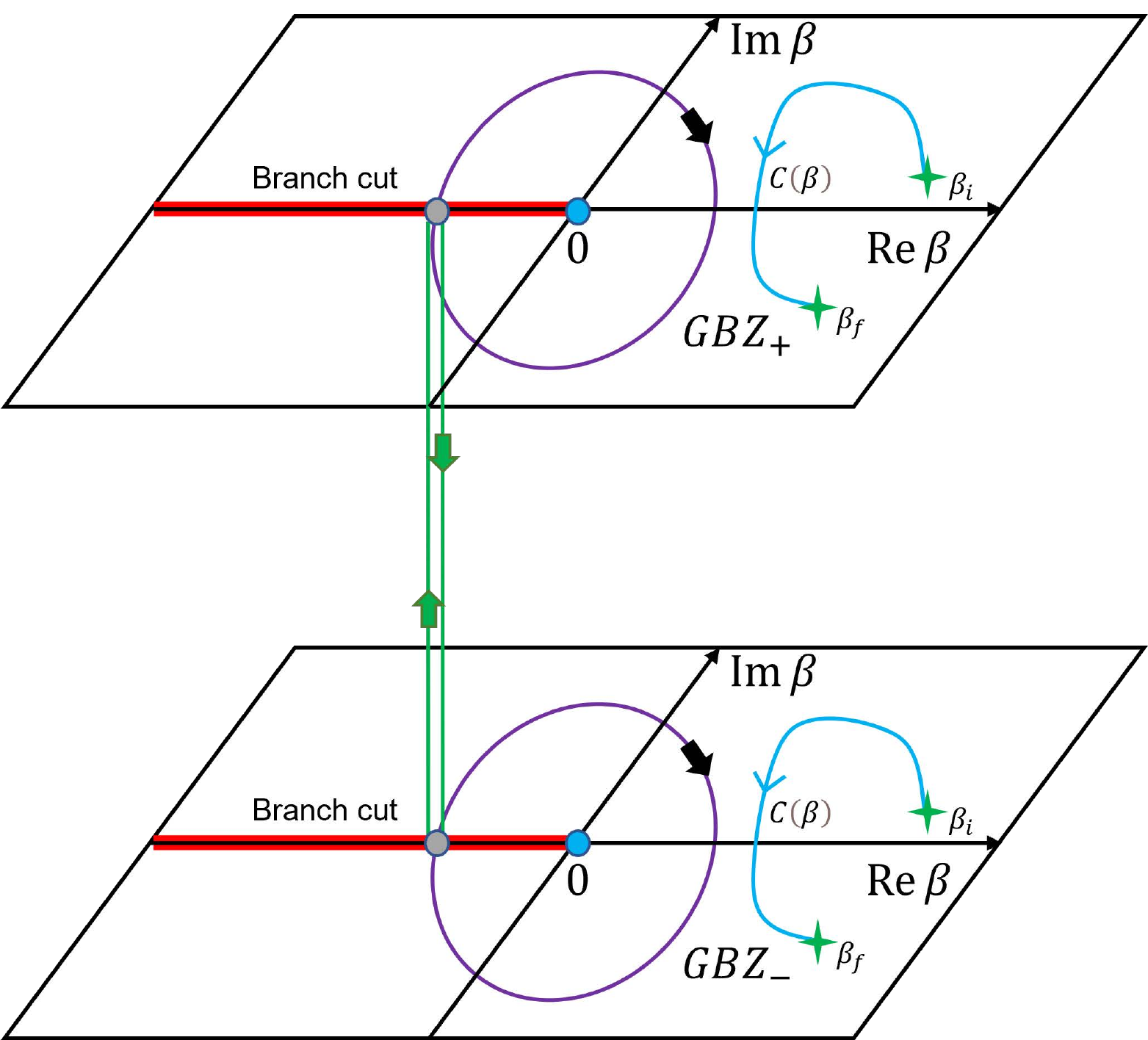}}
    \subfigure[]{\includegraphics[width=4cm, height=3cm]{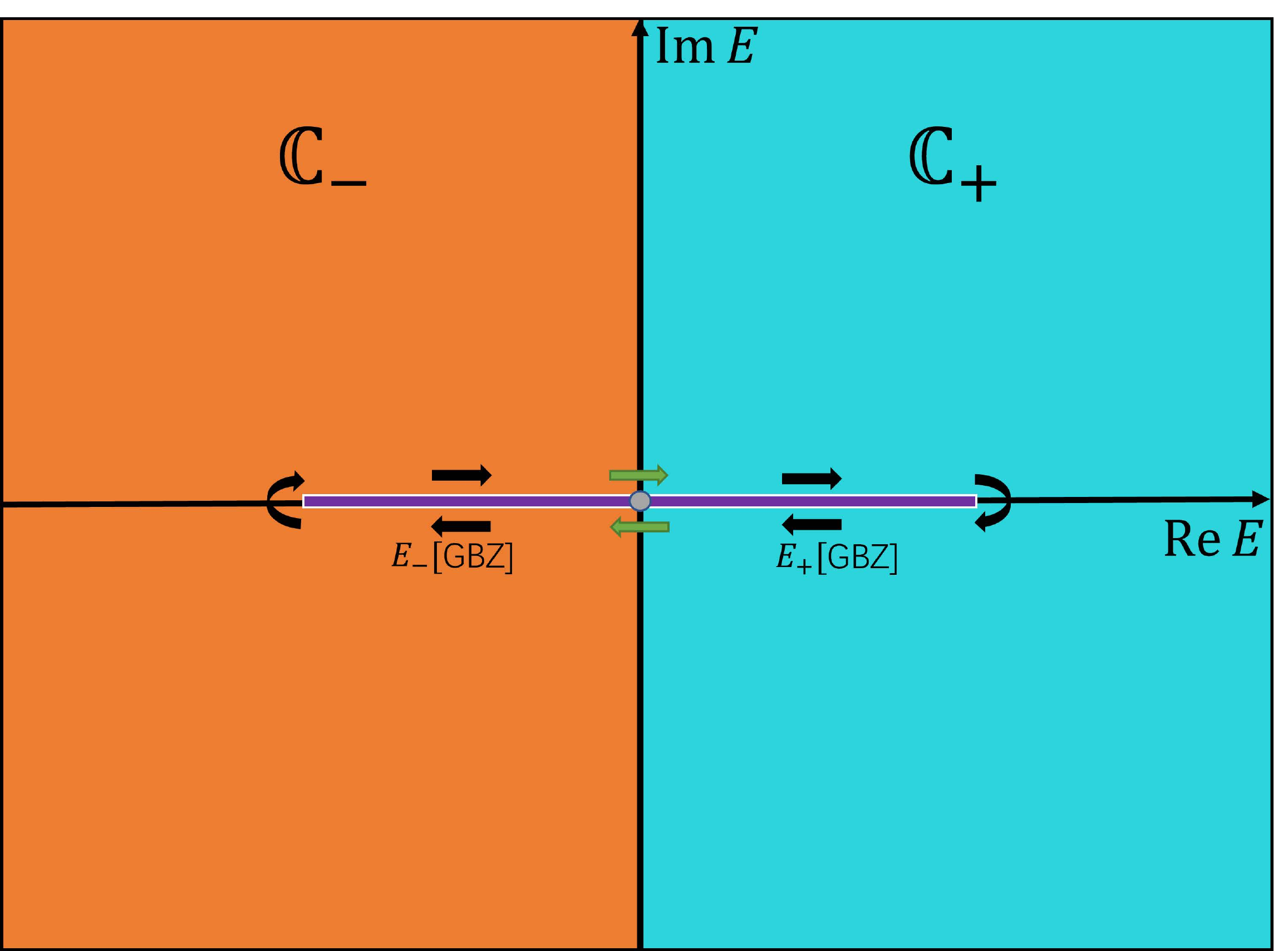}}
    \caption{Schematic illustrations of the GBZs [(a), (c)] and EBBs [(b), (d)] for the NH-SSH model in the gapped case [(a), (b)] and gapless case [(c), (d)]: (a) The $GBZ_{\pm}$ in the upper and lower panels are independent, and (b) the  EBBs $E_{\pm}[GBZ]$ are disconnected and gapped, each within its single-valued branch $\mathbb{C}_{\pm}$; (c) $GBZ_{+}$ switches to $GBZ_{-}$ after crossing the branch cut, and vice versa, and (d) the corresponding EBBs $E_{\pm}[GBZ]$ are connected and gapless. The blue dots and red lines are branch points and branch cuts, respectively. The cyan and orange regions in the complex $E$ plane are the two corresponding single-valued branches $\mathbb{C}_{\pm}$, respectively. There exists a continuous curve $C(\beta)$~(blue line) connecting two arbitrary points $\beta_{i}$, $\beta_{f}$~(green stars) in each single-valued branch.} 
    \label{nhssh_branch}
\end{figure}

\begin{figure*}  
    \subfigure[]{\includegraphics[width=4.2cm, height=3.5cm]{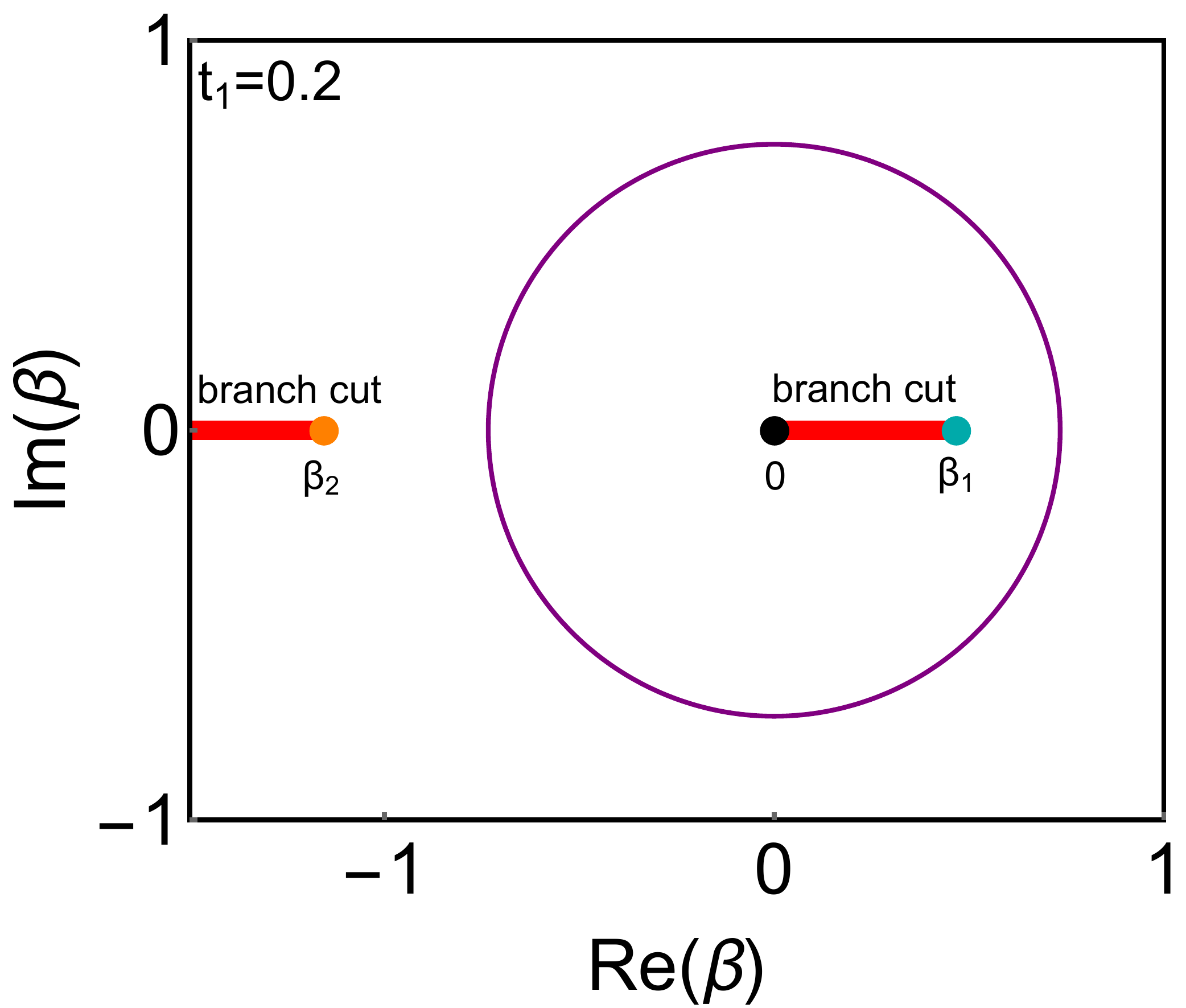}}
    \subfigure[]{\includegraphics[width=4.2cm, height=3.5cm]{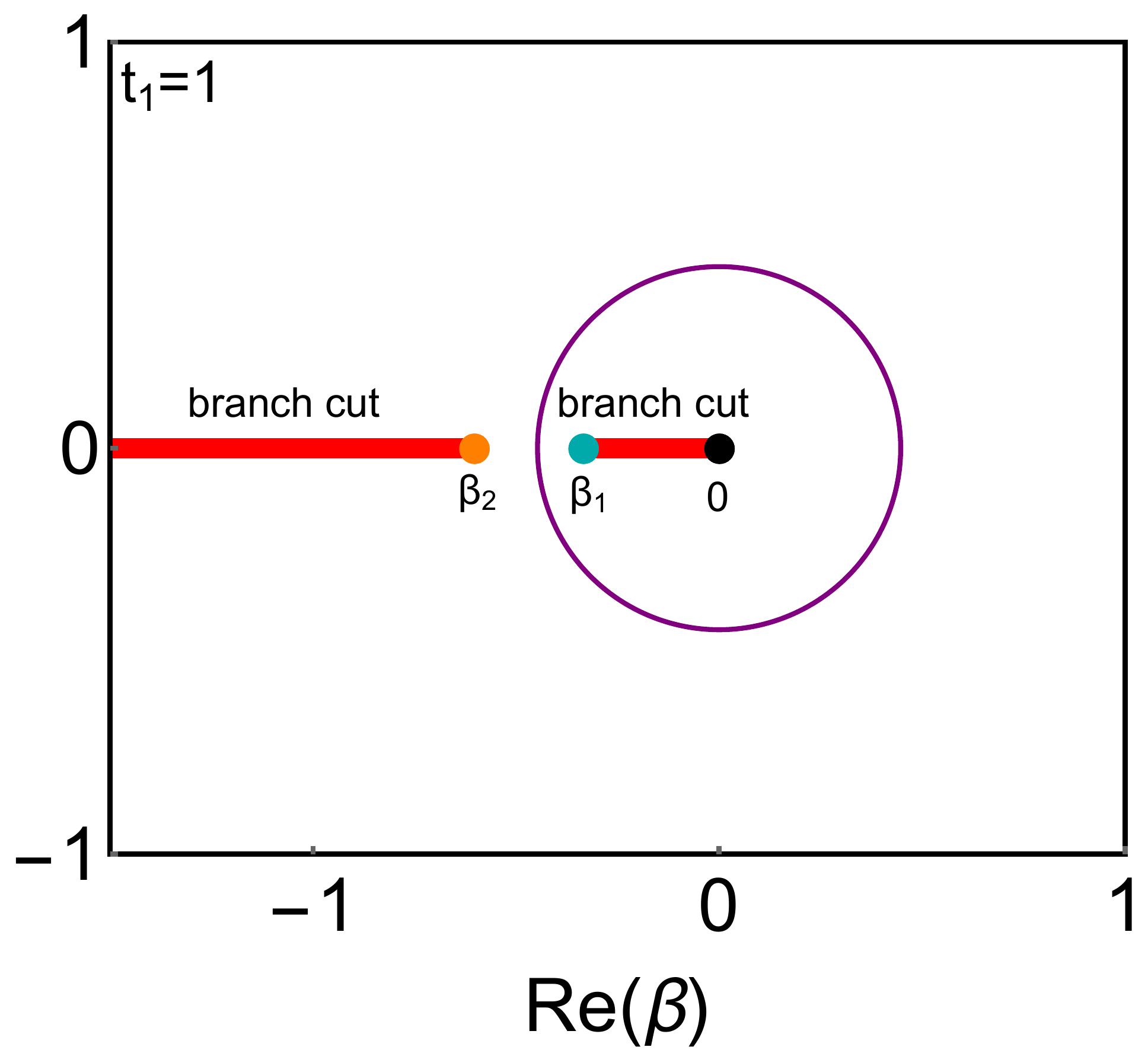}}
    \subfigure[]{\includegraphics[width=4.2cm, height=3.5cm]{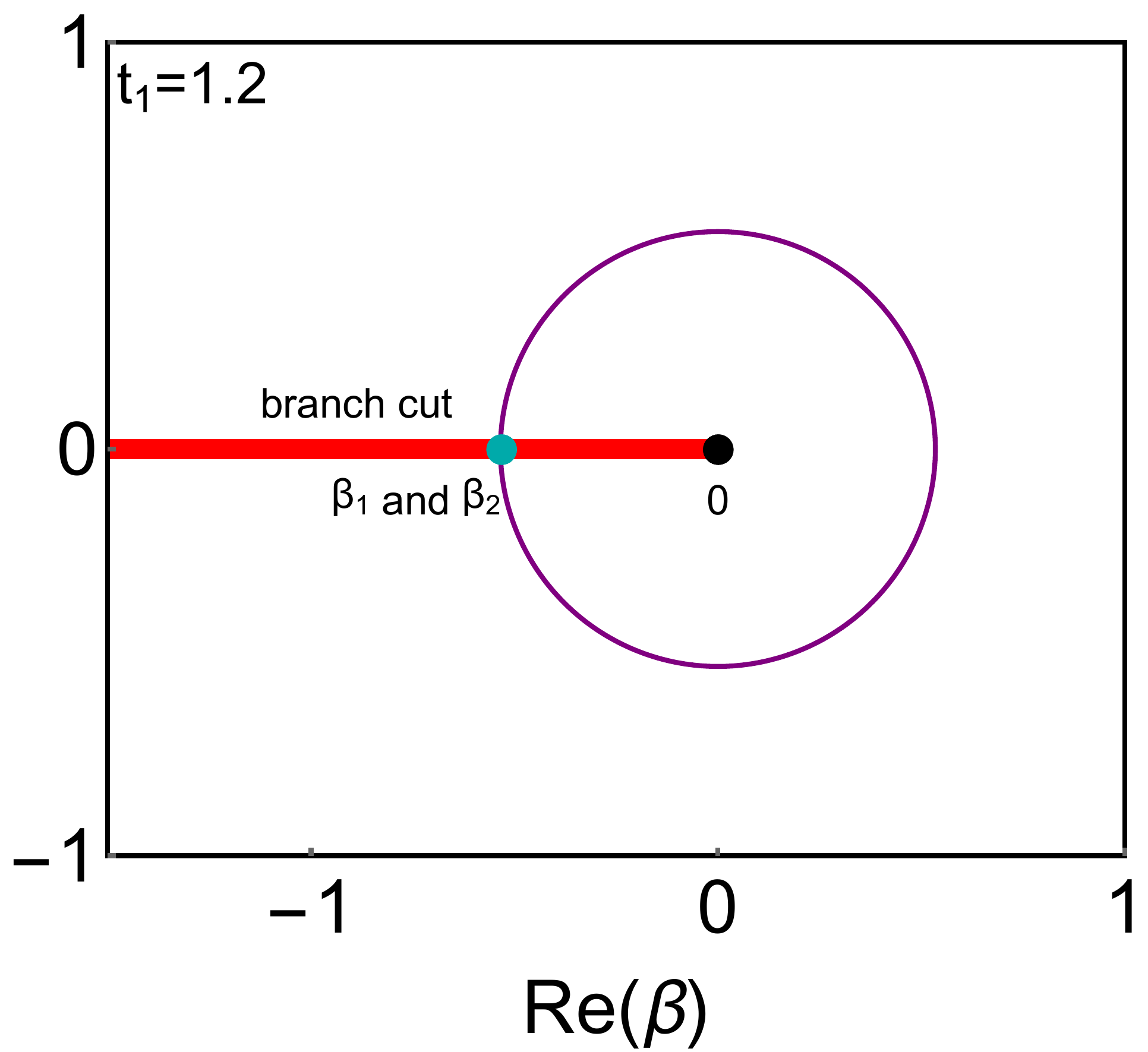}}
    \subfigure[]{\includegraphics[width=4.2cm, height=3.5cm]{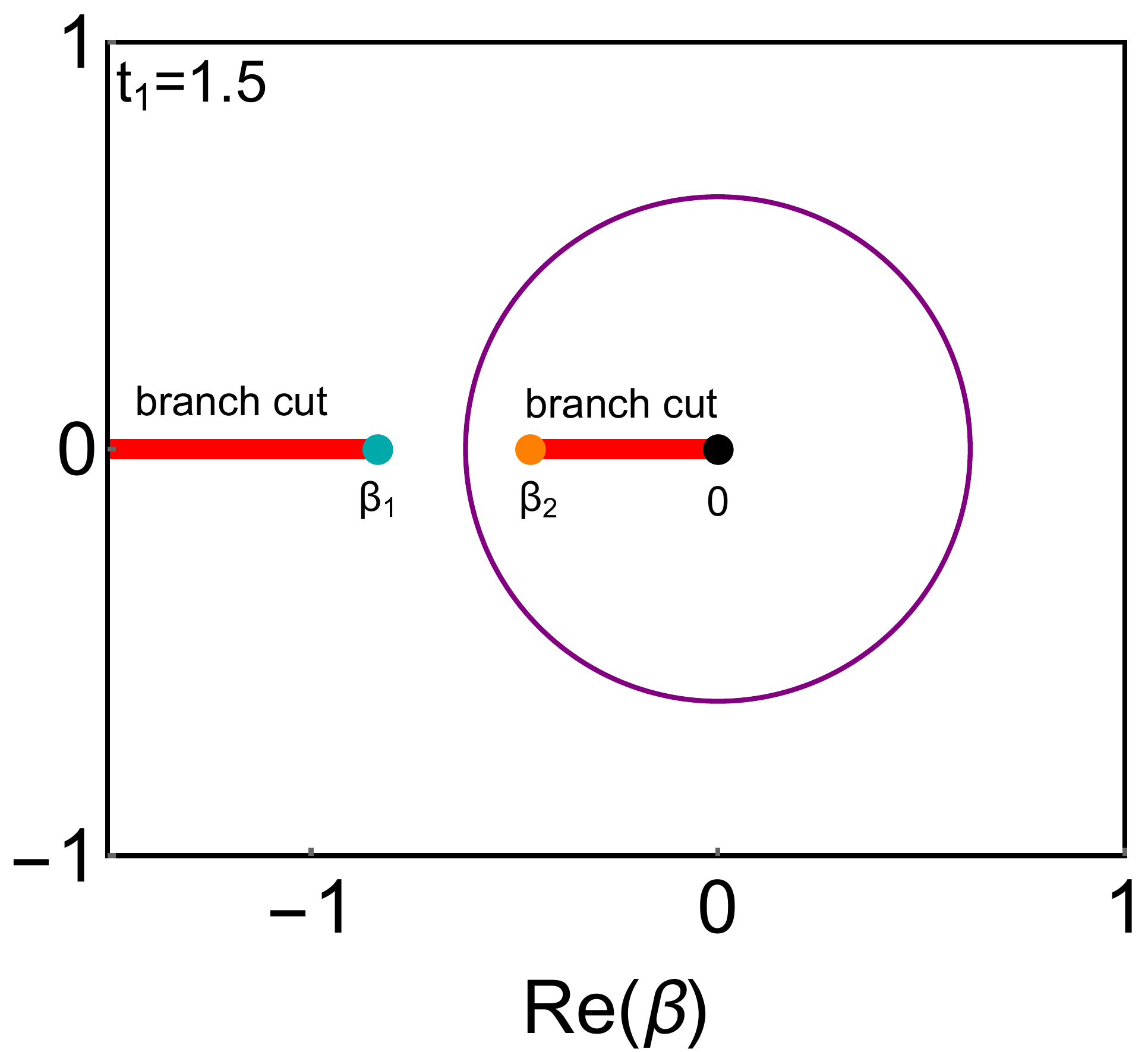}}
    \subfigure[]{\includegraphics[width=4.7cm, height=3.5cm]{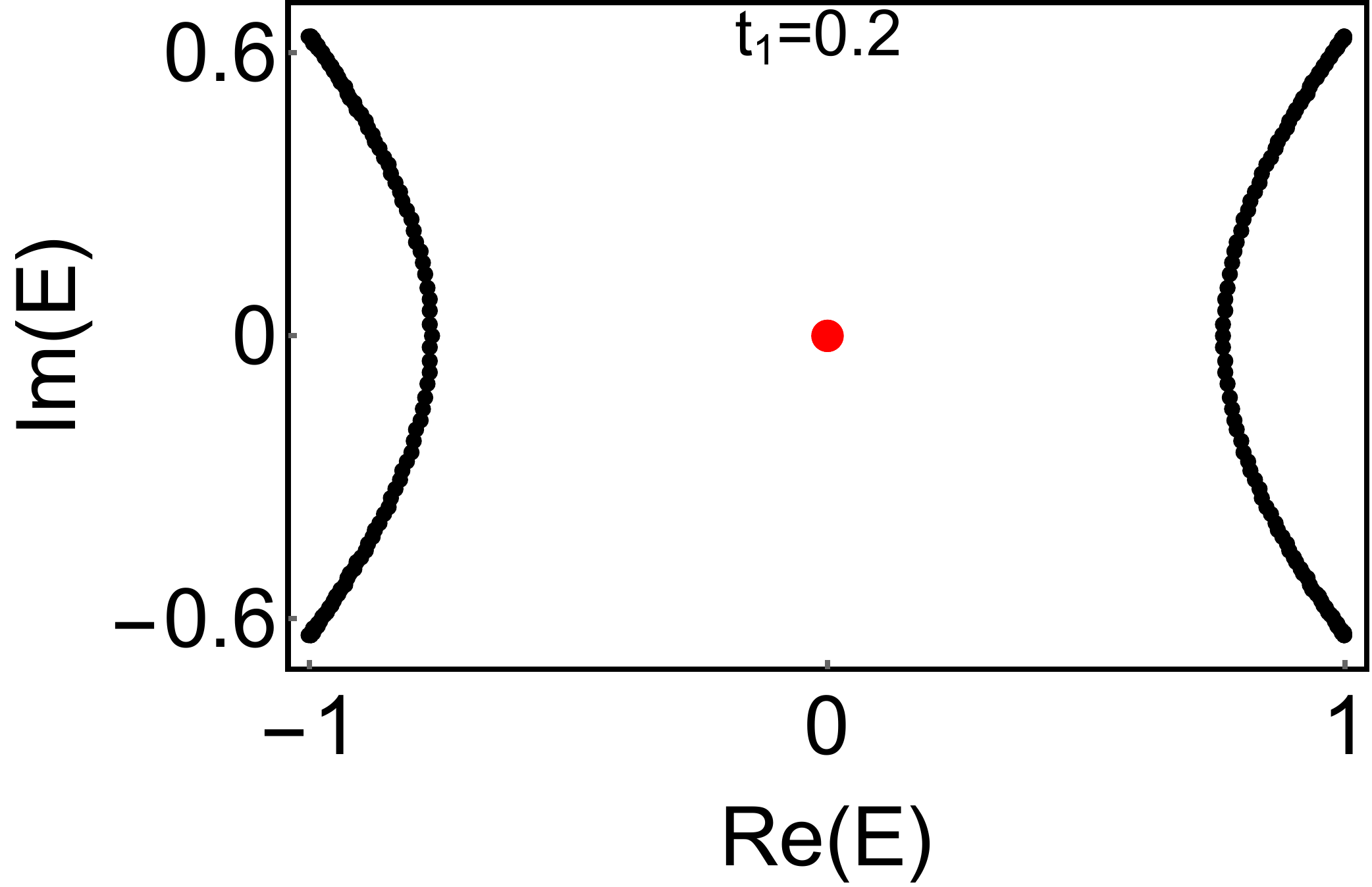}}
    \subfigure[]{\includegraphics[width=4.3cm, height=3.5cm]{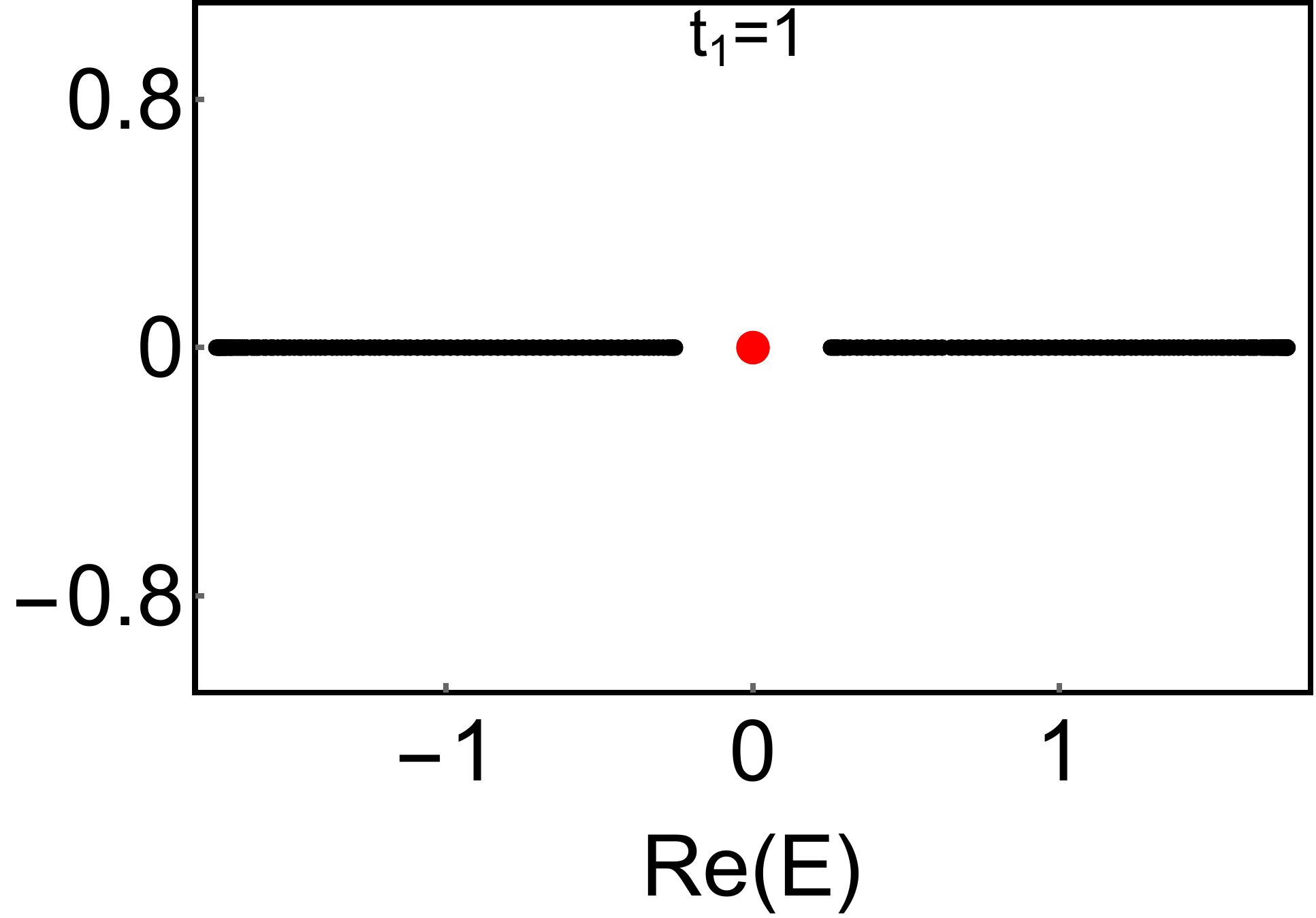}}
    \subfigure[]{\includegraphics[width=4.3cm, height=3.5cm]{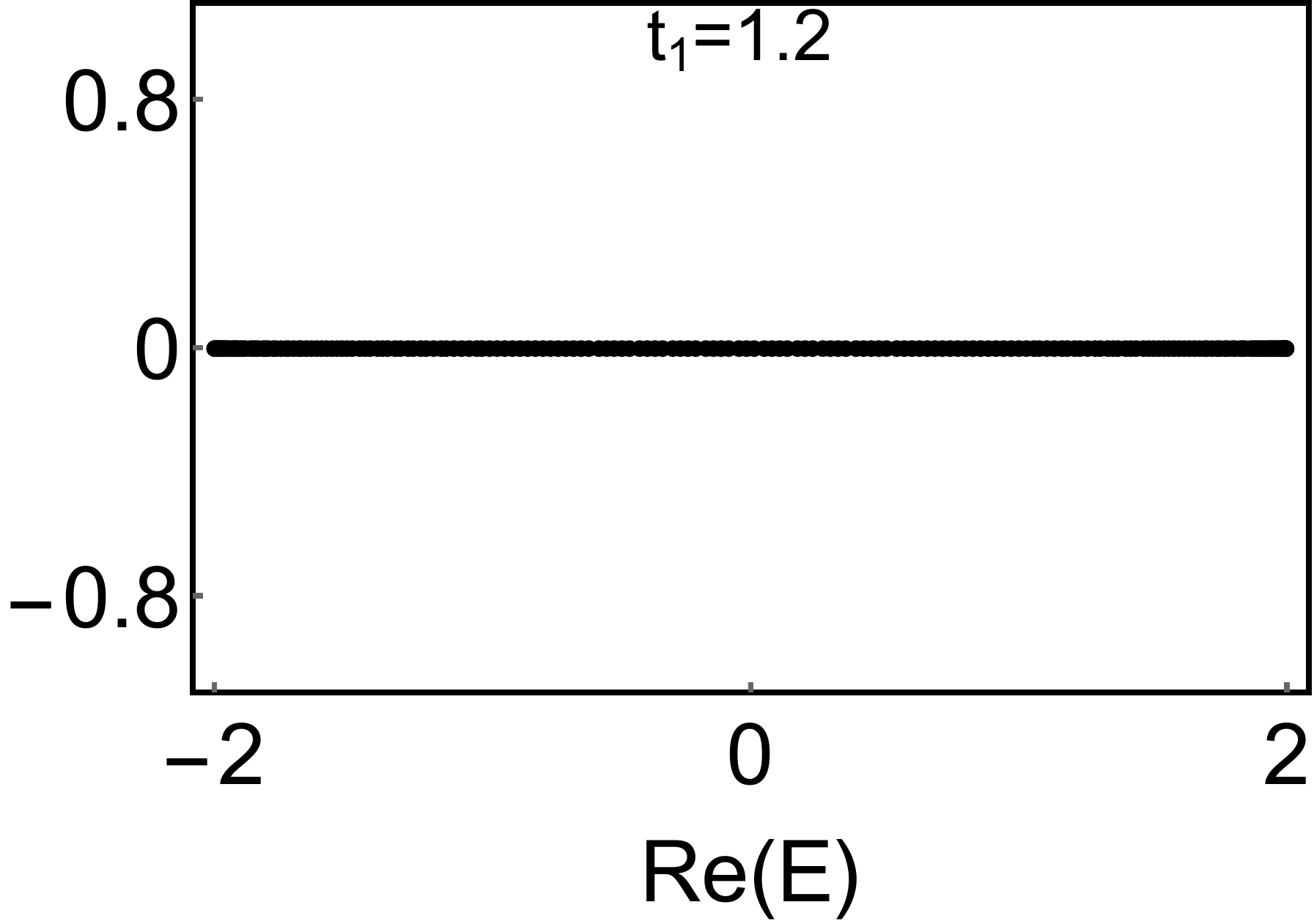}}
    \subfigure[]{\includegraphics[width=4.3cm, height=3.5cm]{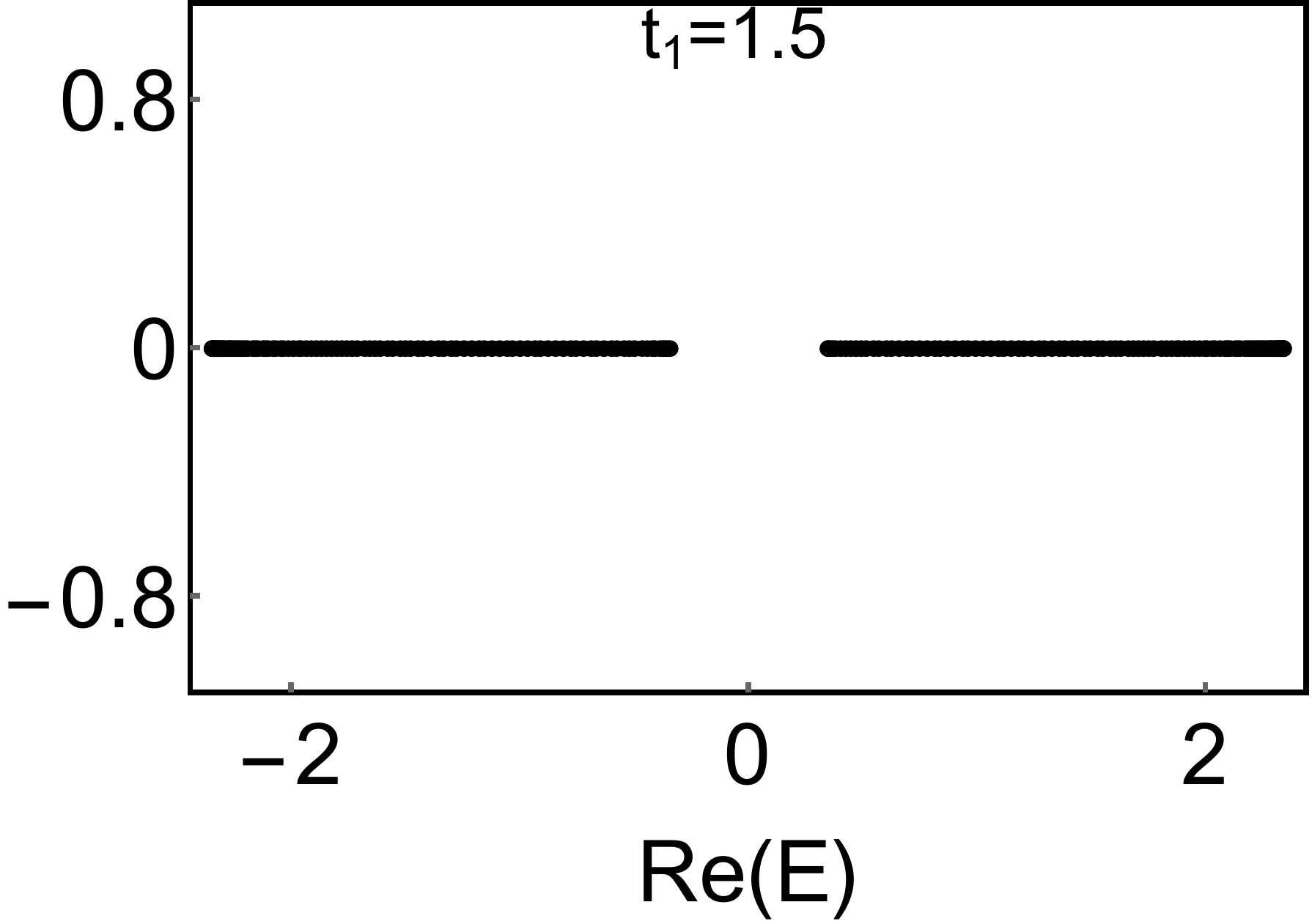}}
    \caption{Results on the GBZs [purple circles in (a)-(d)] and EBBs of the NH-SSH model~[Eq.~(\ref{nhsshnonbloch})], as well as the branch points ($0$, $\beta_{1}$, and $\beta_{2}$ as black, cyan, and orange dots) and branch cuts (red lines), demonstrate the occurrence of distinctive phases and the in-between phase transition. We set $t_{2}=1$, $\gamma=4/3$ and vary the value of $t_1$. The black lines and red dots in (e)-(h) are the bulk spectra and edge states, respectively.}
    \label{nhssh_num}
\end{figure*}

\subsection{Roles of branch points and branch cuts}
The branch points and branch cuts are crucial in setting single-valued branches of a multivalued function, thus depicting the connectedness~(gap and gaplessness) of EBBs. For example, we consider the well-known non-Hermitian Su-Schrieffer-Heeger~(NH-SSH) model~\cite{yao2018}, whose non-Bloch Hamiltonian is given by
\begin{align}
    \label{nhsshnonbloch}
    H(\beta)= \left(\begin{matrix}
       0&t_{1}+\gamma/2+t_{2}\beta^{-1}\\
        t_{1}-\gamma/2+t_{2}\beta&0
    \end{matrix}\right).
\end{align}
Let us take its parity-time~(PT) symmetric EBBs as an example: the two EBBs correspond to the single-valued branches of the multivalued square-root function $E(\beta)=Q(\beta)^{1/2}$, $Q(\beta)=(t_{1}+\gamma/2+t_{2}\beta^{-1})(t_{1}-\gamma/2+t_{2}\beta)$ with four branch points: $0$, $\beta_{1}=-(t_{1}-\gamma/2)/t_{2}$, $\beta_{2}=-t_{2}/(t_{1}+\gamma/2)$, and $\infty$~[shown as the blue dots in Fig.~\ref{nhssh_branch}(a)]. Here, we set the single-valued branches $E_{+}(\beta)$ and $E_{-}(\beta)$ as the right and left half-planes $\mathbb{C}_{\pm}$~[cyan and orange regions in Fig.~\ref{nhssh_branch}(a)] separated by the imaginary axis, and introduce branch cuts by connecting $0$ and $\beta_{1}$, $\beta_{2}$, and $\infty$, in each single-valued branch $E_{\pm}(\beta)$, respectively~[upper and lower panels in Fig.~\ref{nhssh_branch}(a)]. In each branch, two arbitrary points $\beta_{i}$ and $\beta_{f}$~[green stars in Fig.~\ref{nhssh_branch}(a)] can be connected by a continuous curve $C(\beta)$~[blue curves in Fig.~\ref{nhssh_branch}(a)] without crossing the branch cuts.

In the NH-SSH model, the $GBZ_{\pm}$ corresponding to the two EBBs are identical~[purple loops in Fig.~\ref{nhssh_branch}(a)]. Since the number of branch points inside each $GBZ_{\pm}$ is two in Fig.~\ref{nhssh_branch}(a), we can always avoid the crossing of the branch cuts by the $GBZ_{\pm}$, thus keeping $GBZ_{\pm}$ independent from each other and the EBBs $E_{\pm}[GBZ]$  disconnected and gapped~[purple lines in Fig.~\ref{nhssh_branch}(b)], lying within their respective single-valued branches $\mathbb{C}_{\pm}$. In comparison, when there exists only one branch point inside $GBZ_{\pm}$, $GBZ_{\pm}$ must cross the branch cuts~[Fig.~\ref{nhssh_branch}(c)]. Consequently, as we circle counterclockwise around $GBZ_{+}$, it switches to $GBZ_{-}$ at the branch cut, and vice versa. Simultaneously, the corresponding EBB also switches to its partner within the pair, resulting in gapless PT-symmetric EBBs~[Fig.~\ref{nhssh_branch}(d)]. 

Further, we numerically verify these schematic properties of the NH-SSH model and summarize key results in Fig.~\ref{nhssh_num}. Without loss of generality, we assume $t_{1}>0$. When $t_{1}<(t_{2}^{2}+\gamma^{2}/4)^{1/2}$~[Figs.~\ref{nhssh_num}(a)(b)], the branch points $0$ and $\beta_{1}$ ($\beta_{2}$ and $\infty$) are inside (outside) the GBZ, and the resulting EBBs are gapped with two degenerate edge states at zero energy~[Figs.~\ref{nhssh_num}(e)(f)]. Noteworthily, $0$ and $\beta_{1}$ coalesce at $t_{1}=\gamma/2$, an infernal point which is the critical point between PT-symmetry preserving and spontaneous breaking, and the theory of GBZ and EBB is invalid~\cite{fu2022}. When $t_{1}>(t_{2}^{2}+\gamma^{2}/4)^{1/2}$ [Fig.~\ref{nhssh_num}(d)], the branch points $0$ and $\beta_{2}$ ($\beta_{1}$ and $\infty$) are inside (outside) the GBZ, and the EBBs are also gapped yet without the edge states~[Fig.~\ref{nhssh_num}(h)], corresponding to a topologically trivial phase~\cite{yao2018}. At the point of the topological phase transition $t_{1}=(t_{2}^{2}+\gamma^{2}/4)^{1/2}$~\cite{yao2018}, the branch points $\beta_{1}$ and $\beta_{2}$ coalesce and annihilate~[Fig.~\ref{nhssh_num}(c)] \footnote{The resulting point is no longer a branch point since circling it gives rise to $Q(\beta)$ a $4\pi$ phase.}. There remains a single branch point inside the GBZ, which inevitably crosses the branch cut, leading to gapless EBBs~[Fig.~\ref{nhssh_num}(g)].

As demonstrated above, branch points and branch cuts play crucial roles in our theory of EBBs. In general, a sub-GBZ enclosing an odd number of branch points in the complex $\beta$ plane will inevitably cross the branch cuts irrespective of the choices of single-valued branches; circling such a sub-GBZ switches an EBB to another, leading to the emergence of connected EBBs~(gapless bands) under OBC. On the other hand, the presence of disconnected EBBs~(gapped bands) under OBC requires that all sub-GBZs enclose an even number of branch points and no branch point on the sub-GBZs. The transition between gapped and gapless bands must accompany the change in the number of branch points inside the GBZs; simultaneously, degenerate points between the two EBBs appear on the GBZ and at the branch cuts. These are one of the main conclusions of the paper. 

\begin{figure*}
    \subfigure[]{\includegraphics[width=4cm, height=4cm]{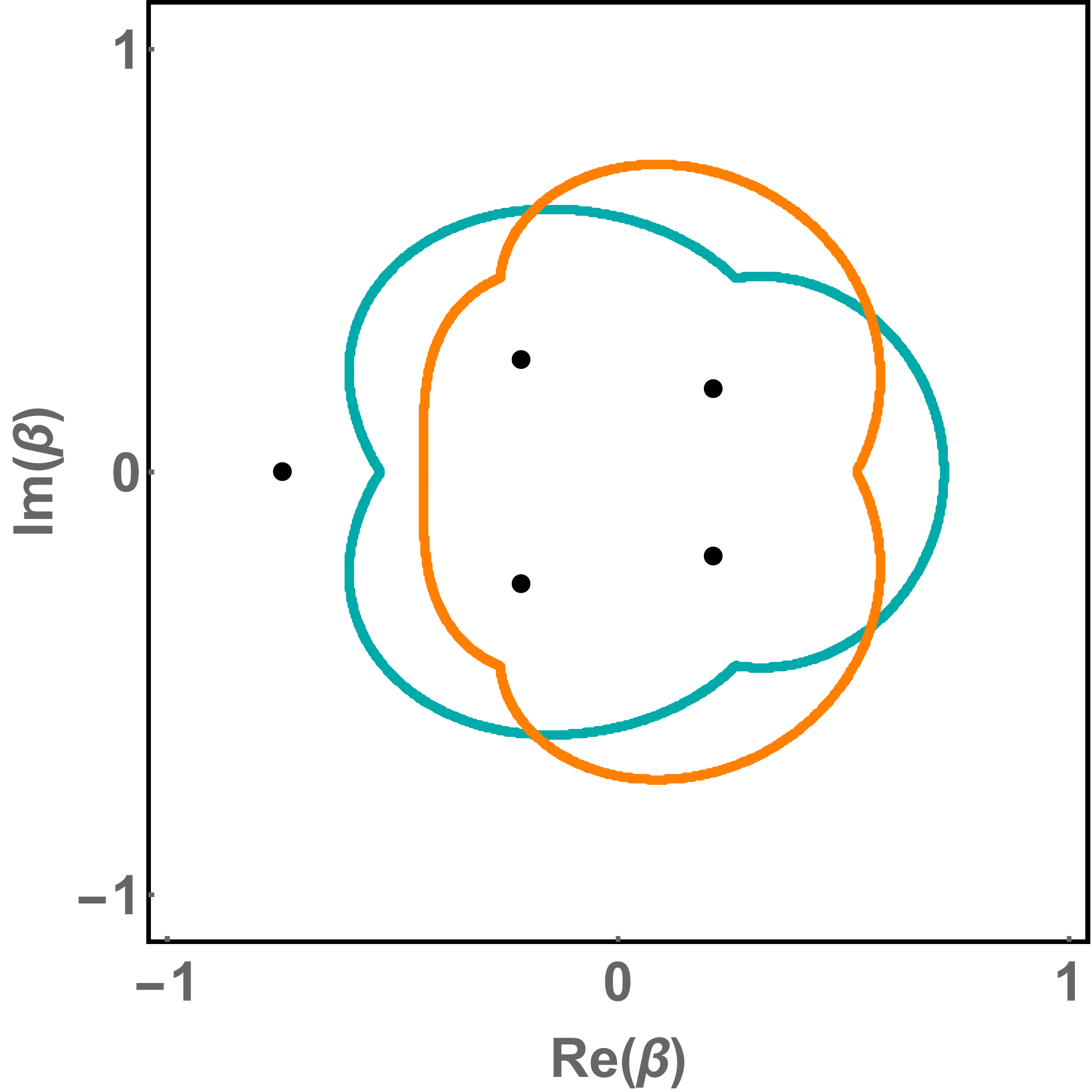}}
    \subfigure[]{\includegraphics[width=4cm, height=4cm]{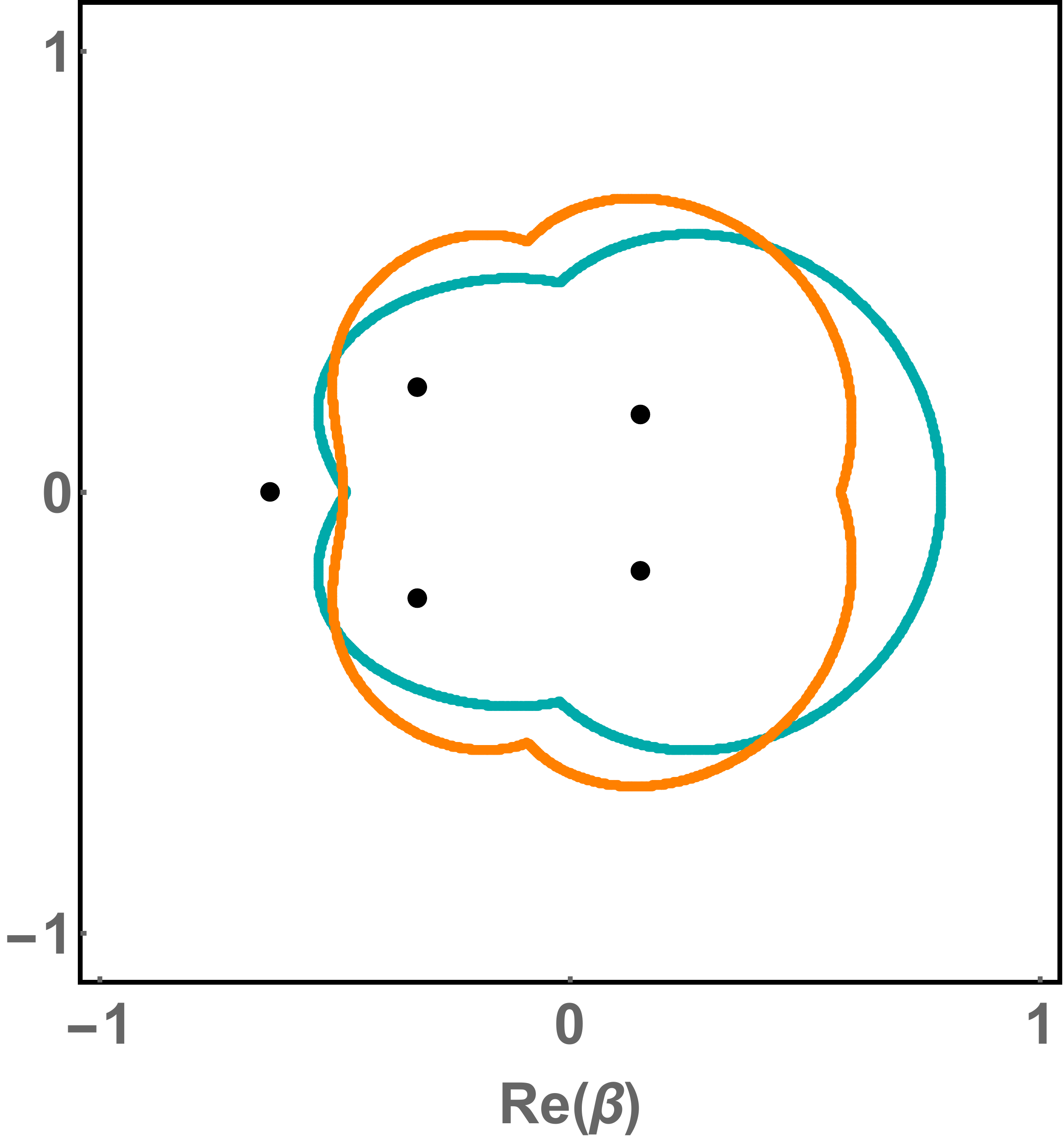}}
    \subfigure[]{\includegraphics[width=4cm, height=4cm]{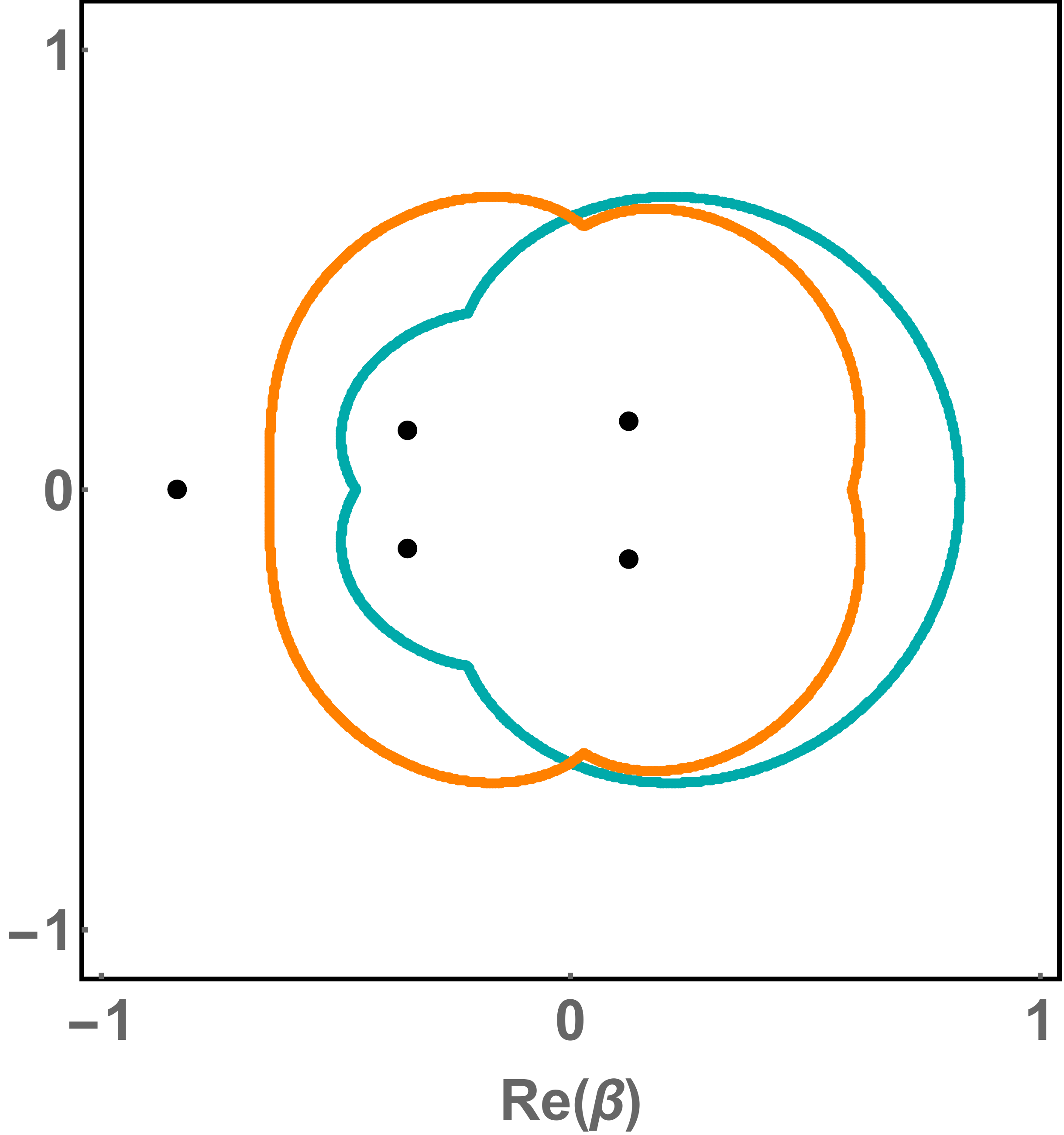}}\\
    \subfigure[]{\includegraphics[width=4.3cm, height=3.5cm]{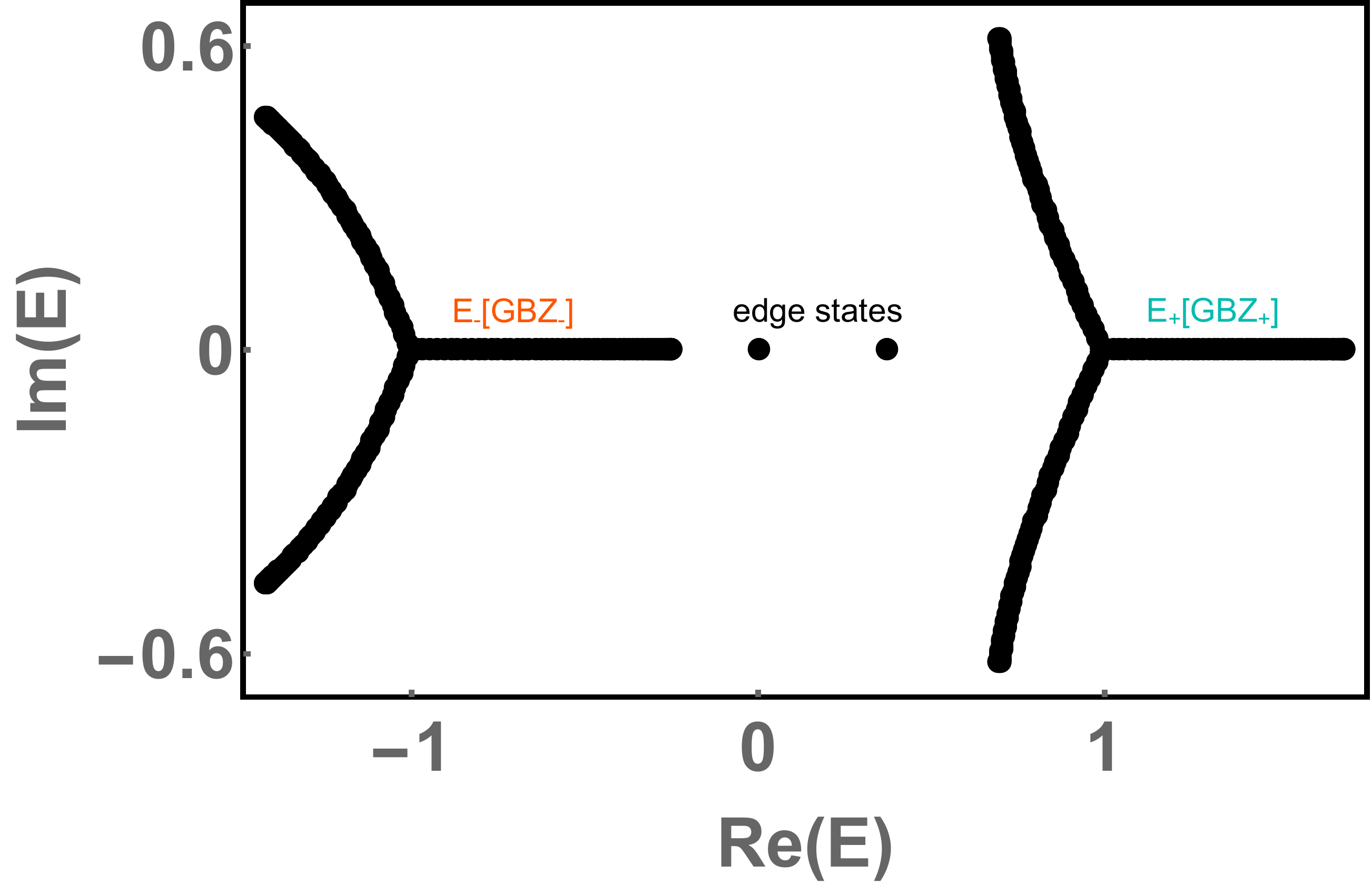}}
    \subfigure[]{\includegraphics[width=4.3cm, height=3.5cm]{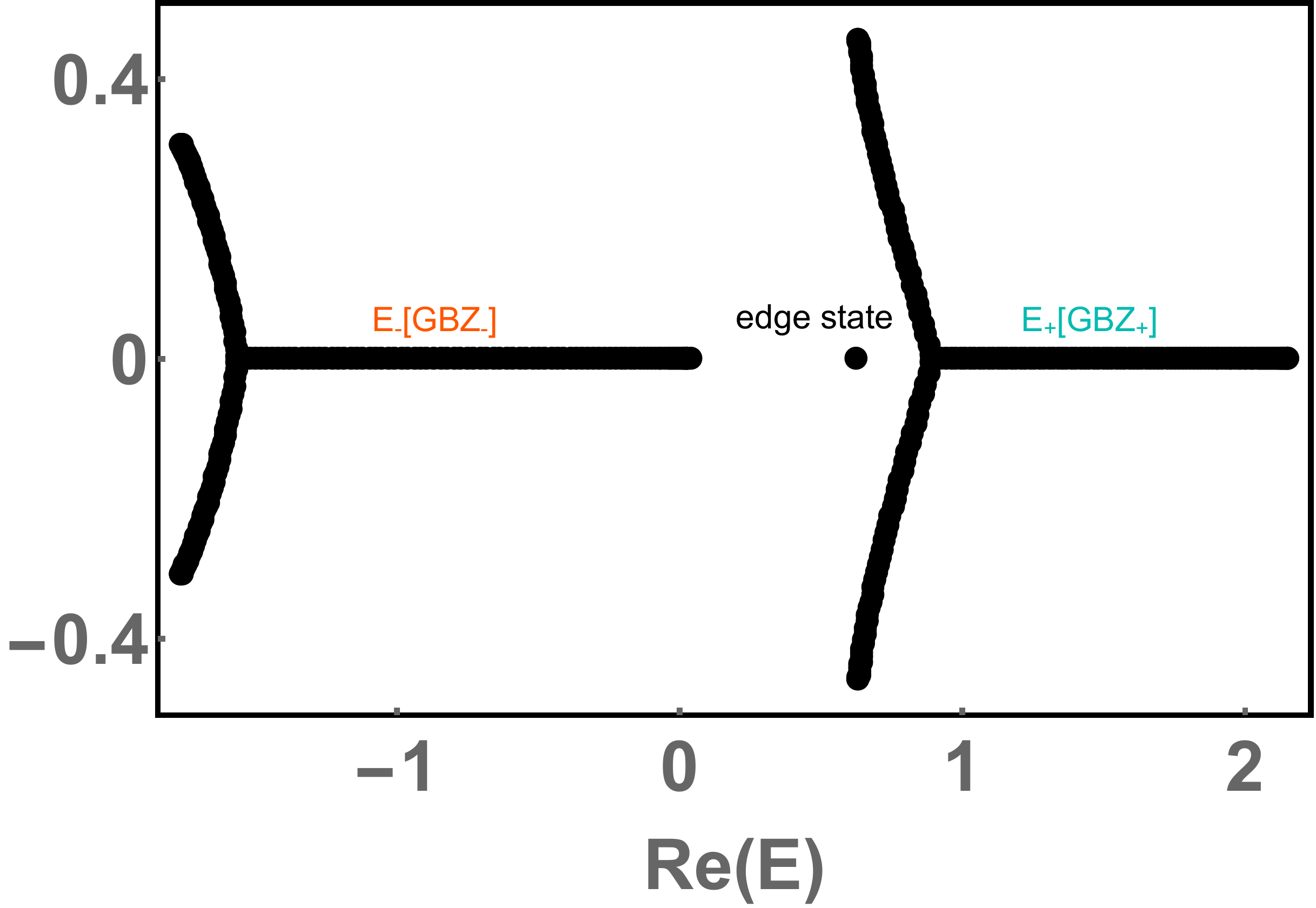}}
    \subfigure[]{\includegraphics[width=4.3cm, height=3.5cm]{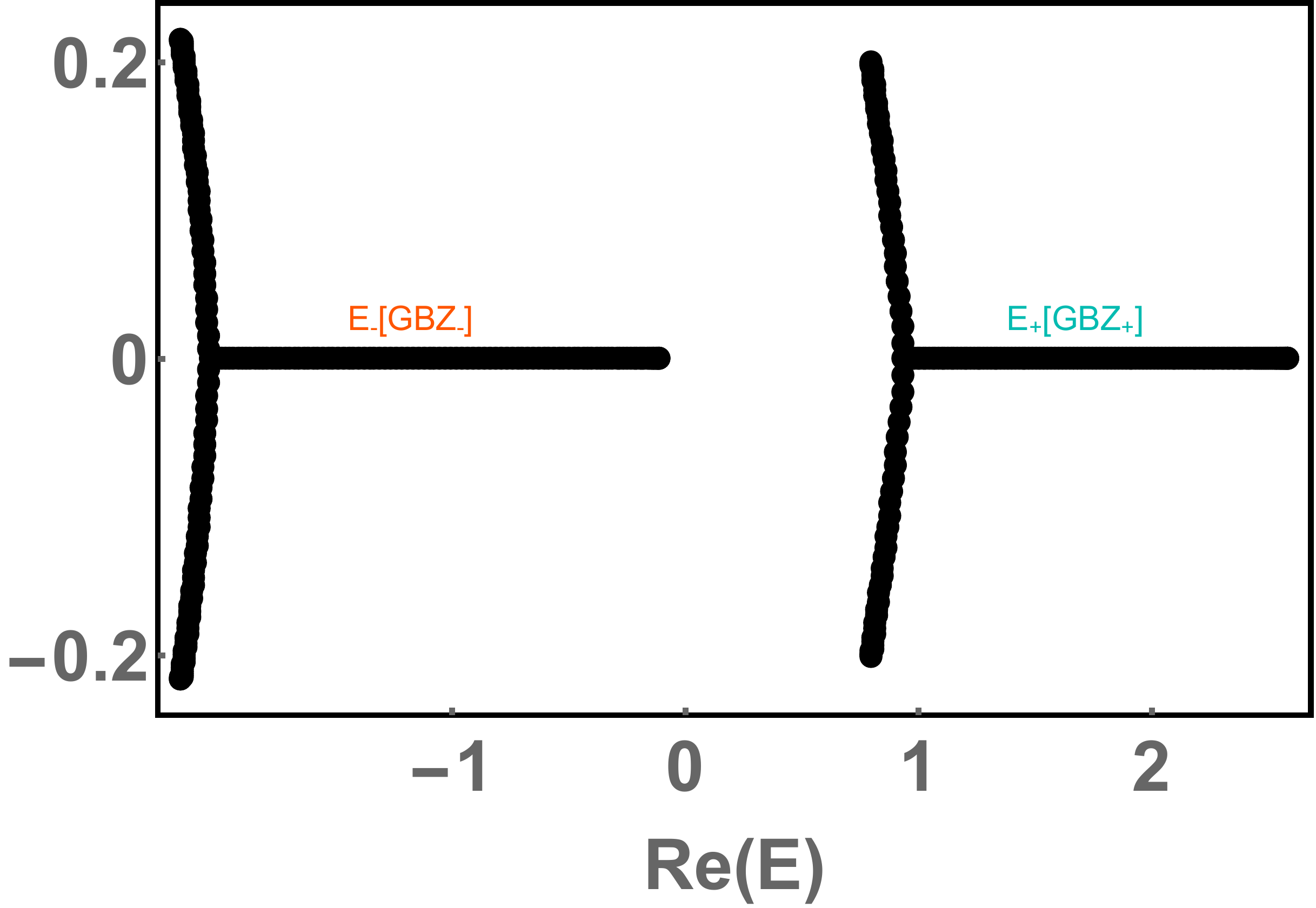}}
    \caption{(a)-(c): The branch points~(black dots) and $GBZ_{\pm}$~(cyan and orange loops, respectively) of the two-band model in Eq.~(\ref{fwmodelnonbloch}) with selected $t_{1}=0.7, 1.1, 1.5$, respectively, show that there are always four branch points inside $GBZ_{\pm}$. There are also three branch points outside the displayed region. (d)-(f): The energy spectra under OBC corresponding to (a)-(c), respectively, show unstable edge states that emerge from and disappear into the bulk without topological protection. $\gamma=2/3$, $t_{2}=1$, and $t_{3}=1/5$.}
    \label{fig4}
\end{figure*}

Such analysis generalizes straightforwardly. Next, we consider a non-Hermitian two-band model with the following non-Bloch Hamiltonian~(see Ref.~\cite{fu2022} and Appendix \ref{appendixA} for details),
\begin{align}
    \label{fwmodelnonbloch}
    H_{FW}(\beta)= \left(\begin{matrix}
        t_{3}\beta^{2} & t_{1}+\gamma+t_{2}\beta^{-1}\\
        t_{1}-\gamma+t_{2}\beta & t_{3}\beta^{-2}
    \end{matrix}\right),
\end{align}
which possesses two bands with distinct sub-GBZs. The multivalued function concerning EBBs can be expressed as
\begin{widetext}
\begin{align}
    \label{fwebbs}
    E(\beta)=\frac{1}{2}\left[t_{3}\left(\beta^{2}+\beta^{-2}\right)
    +\sqrt{\left(t_{3}(\beta^{2}+\beta^{-2})\right)^{2}-4\left(t_{3}^{2}-(t_{1}+\gamma+t_{2}\beta^{-1})(t_{1}-\gamma+t_{2}\beta)\right)}\right],
\end{align} 
\end{widetext}
which induces two single-valued branches $\mathbb{C}_{\pm}$ following the square-root function and eight branch points. With the variation of $t_{1}$~($t_{1}>0$ without loss of generality) and other fixed parameters, there are always four branch points inside both sub-GBZs $GBZ_{\pm}$ [Figs.~\ref{fig4}(a)-(c)], and the other four branch points outside $GBZ_{\pm}$. Thus, we can always arrange the branch cuts to avoid crossing with the sub-GBZs $GBZ_{\pm}$, indicating the existence of a robust trivial phase with two gapped EBBs. There are also emergent edge states in the gap between the two EBBs with varying parameters~[Figs.~\ref{fig4}(d)(e)]. As $t_{1}$ increases from $0$, these two edge states merge into the bulk sequentially [Figs.~\ref{fig4}(d)(e)], and eventually leave a fully gapped energy spectrum without an edge state~[Fig.~\ref{fig4}(f)]. Since the EBBs remain gapped during this process without a (topological) phase transition, we determine that these edge states should not be the consequence of a topological phase or guaranteed to remain stable, in contrast to the stable zero-energy edge states of the NH-SSH model. In other words, if these edge states were topologically nontrivial, their emergence and disappearance must have been accompanied by an EBB gap closure, a change in the number of branch points, and a topological phase transition. 

\begin{figure*}
    \subfigure[]{\includegraphics[width=5cm, height=5cm]{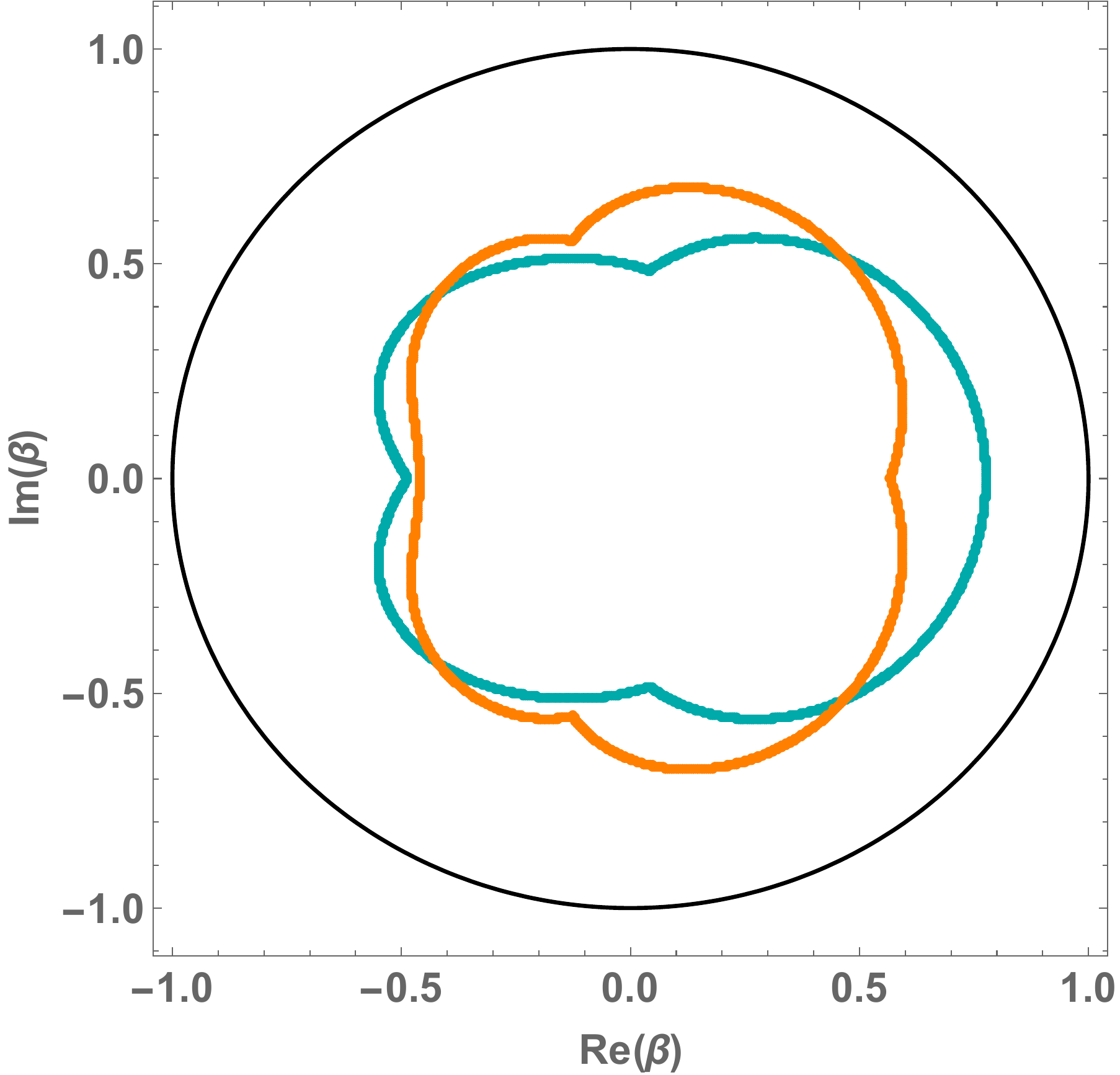}}\qquad
    \subfigure[]{\includegraphics[width=5.5cm, height=5cm]{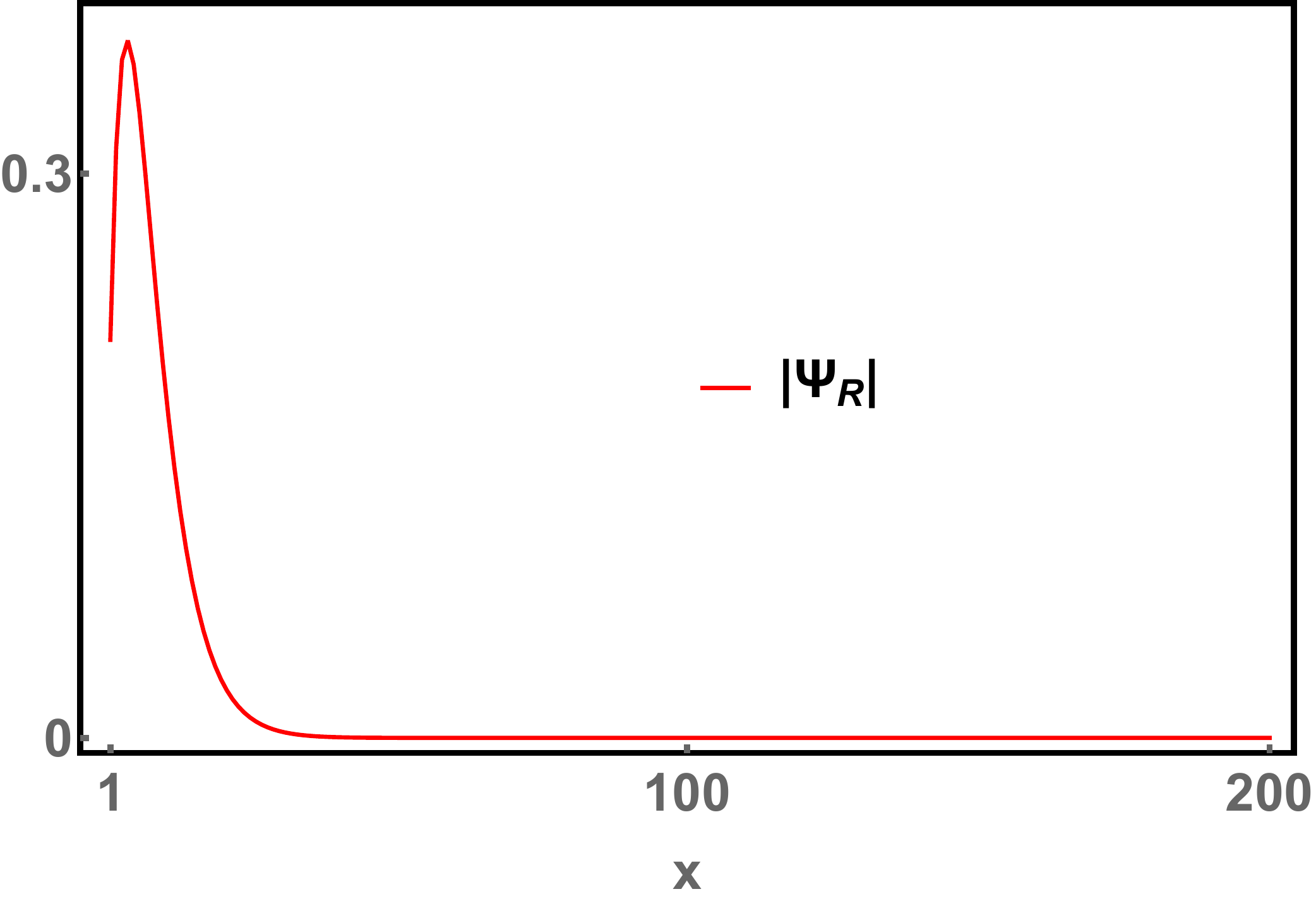}}\\
    \subfigure[]{\includegraphics[width=3.3cm, height=3.3cm]{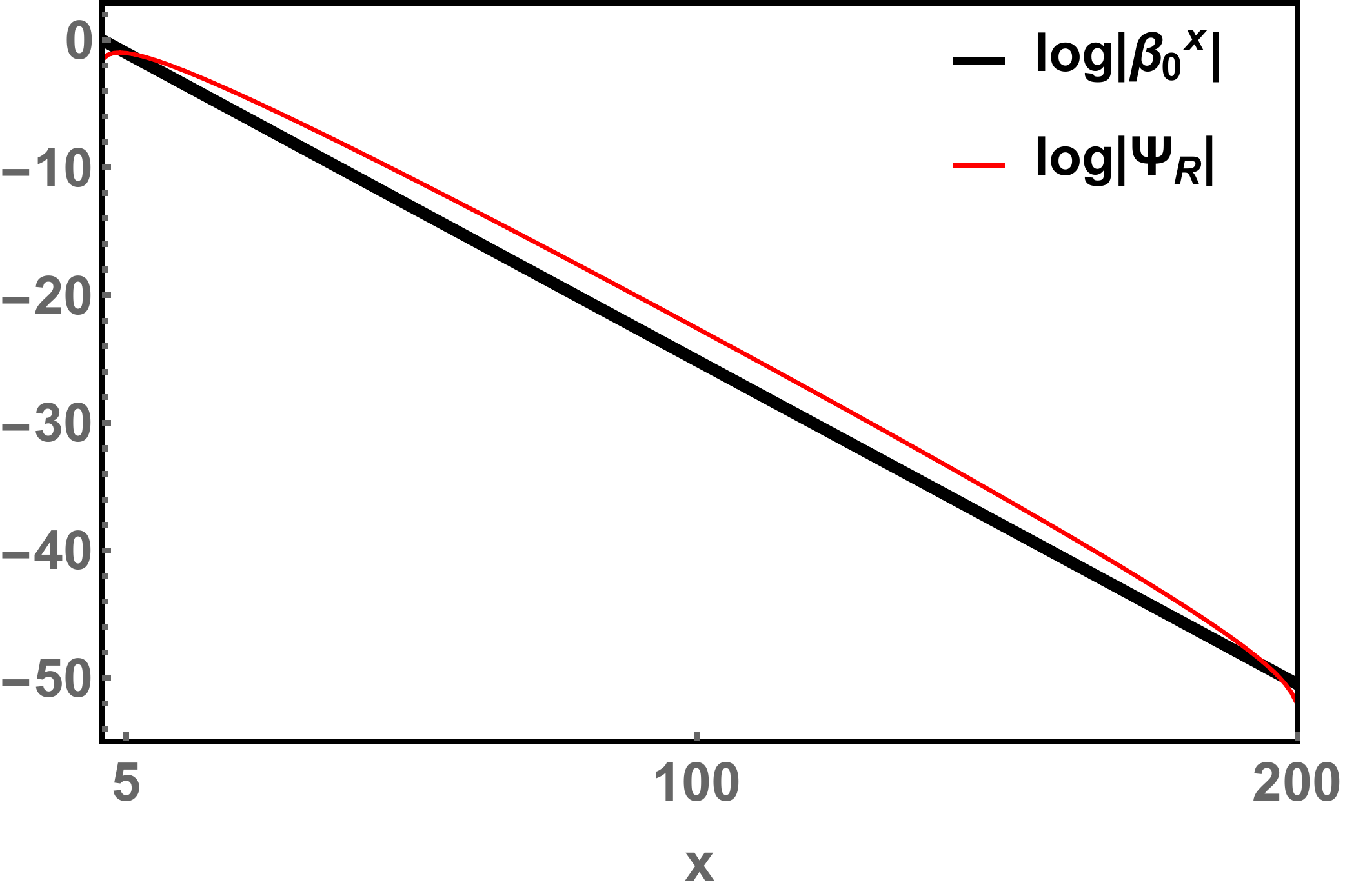}}
    \subfigure[]{\includegraphics[width=3.3cm, height=3.3cm]{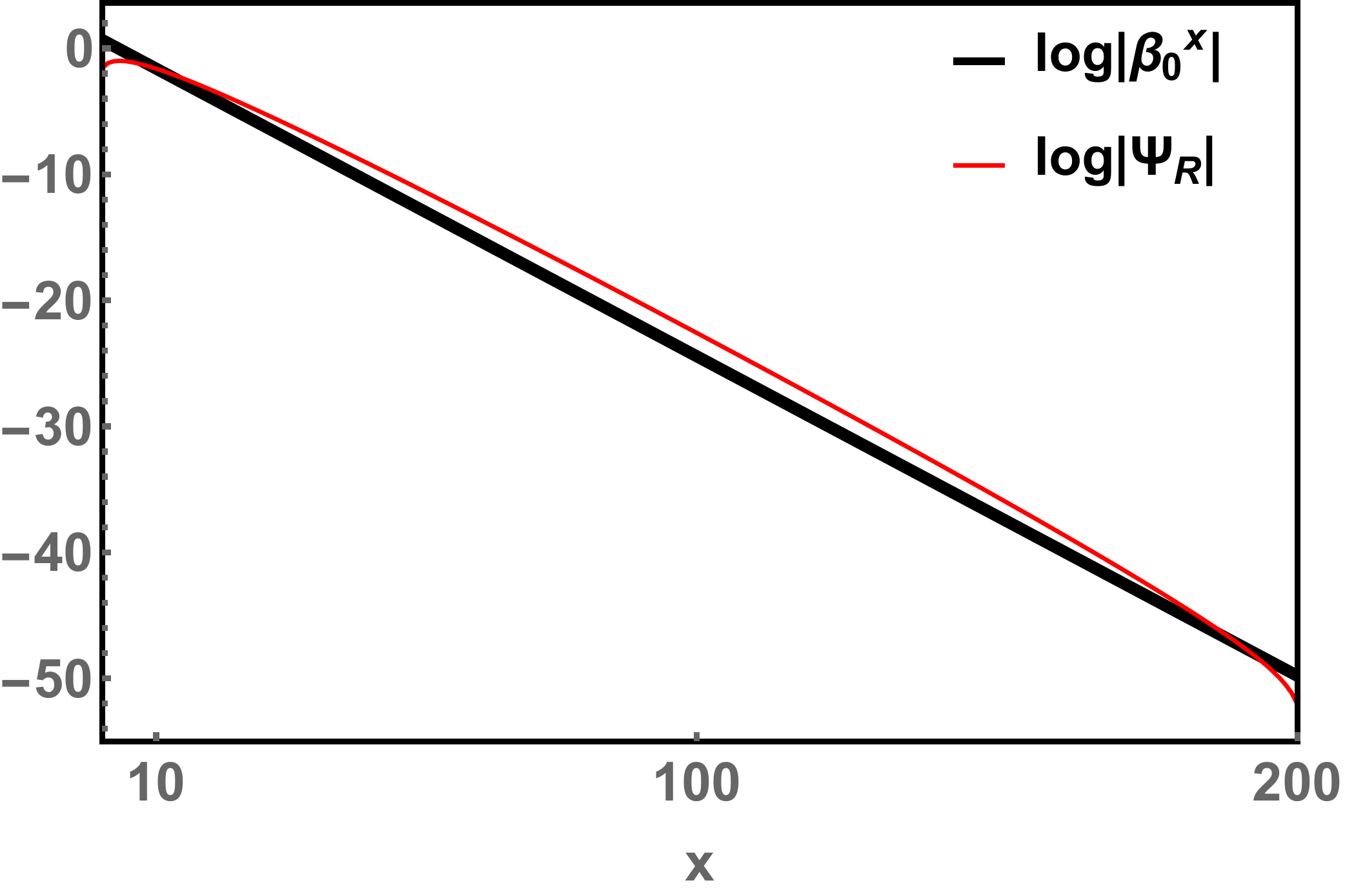}}
    \subfigure[]{\includegraphics[width=3.3cm, height=3.3cm]{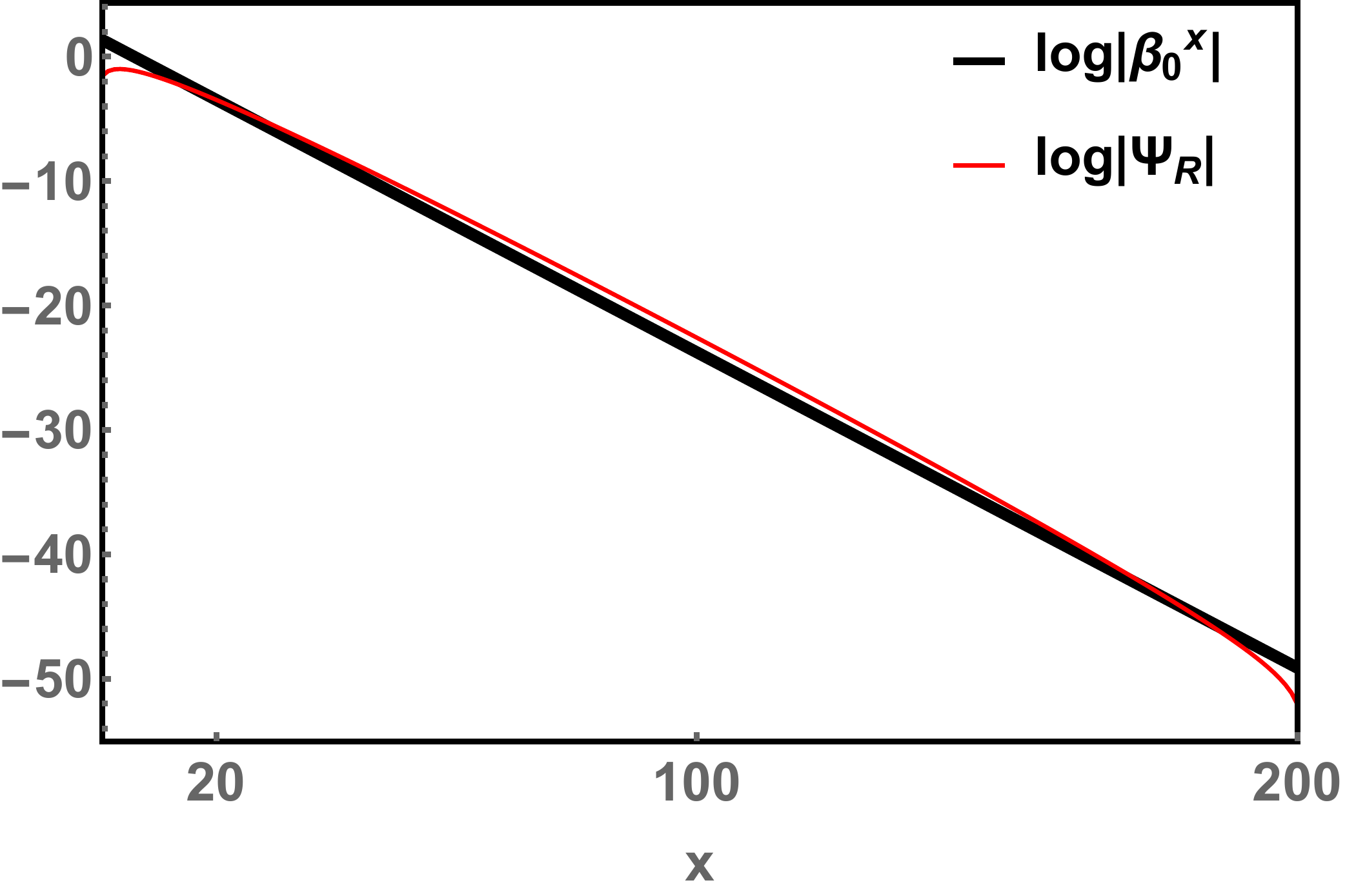}}
    \subfigure[]{\includegraphics[width=3.3cm, height=3.3cm]{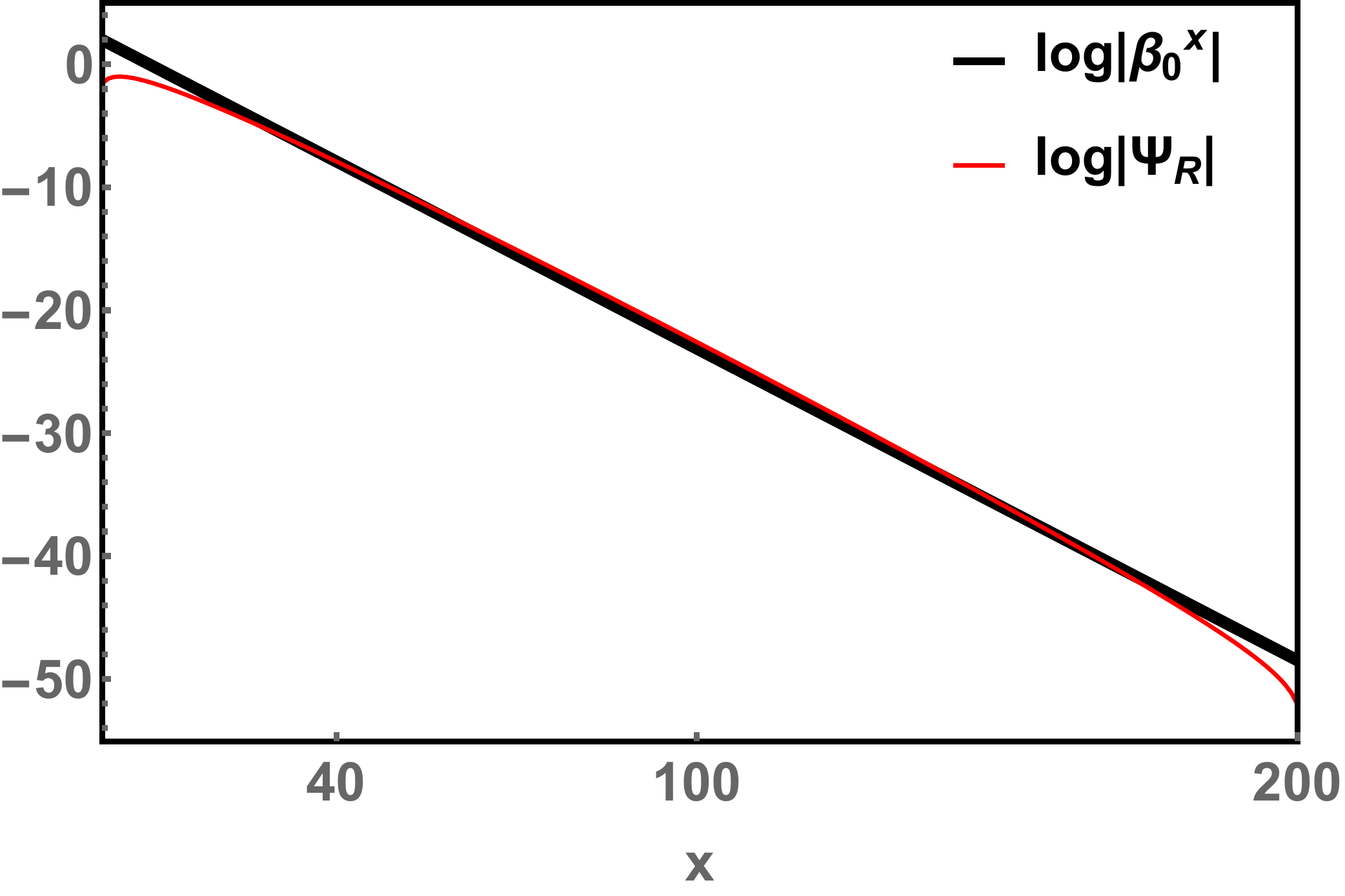}}
    \subfigure[]{\includegraphics[width=3.3cm, height=3.3cm]{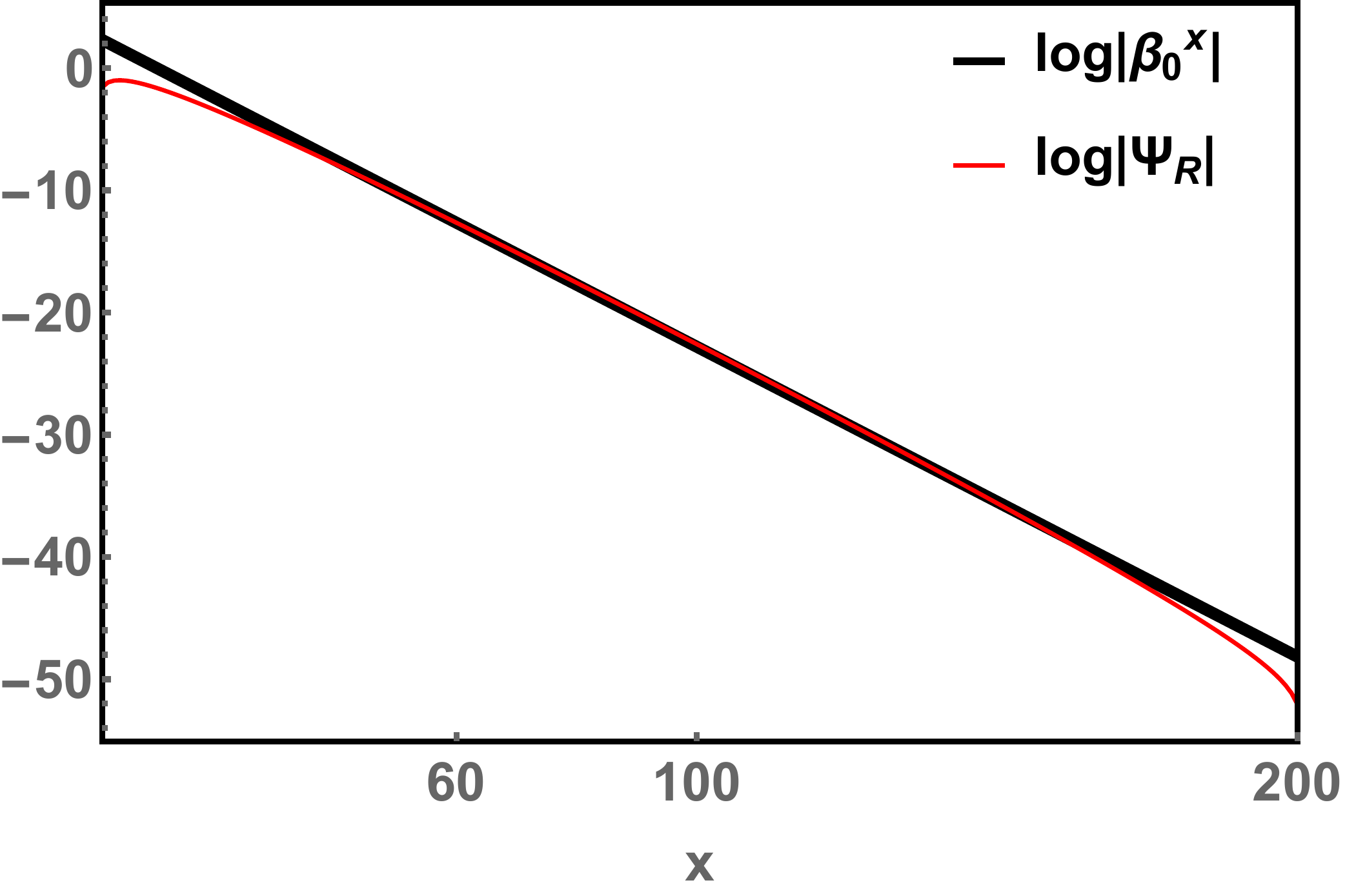}}
    \caption{We illustrate the precise significance of NHSE with the two-band model in Eq.~(\ref{fwmodelnonbloch}): (a) $GBZ_{\pm}$~(cyan and orange loops) and the unit circle~(black), (b) the weight distribution of the right eigenstate $\Psi_{R}(x)$ throughout the system with eigenenergy $E_{0}=2.0393$, and (c)-(g) the logarithm of the right eigenstate $\Psi_{R}(x)$~(red lines) and $\beta_{0}^{x}$~(black lines) from close to the left boundary to deep bulk, fitted through $|\Psi_{R}(x_{0})|=|\beta_{0}^{x_{0}}|$ with $x_{0}=5,10,20,40,60$, respectively. $t_{1}=1$, $\gamma=2/3$, $t_{2}=1$, and $t_{3}=1/5$.}
    \label{fig5}
\end{figure*}

\section{Precise significance of non-Hermitian skin effect}
\label{section3}

An essential phenomenon of 1D non-Hermitian systems is the NHSE, which supports the existence of localized bulk eigenstates~\cite{yao2018,yokomizo2019}. Intuitively, the part of the GBZ inside~(outside) the unit circle in the complex $\beta$ plane indicates the presence of left~(right) localized eigenstates, providing an alternative definition of the NHSE from the perspective of the non-Bloch band theory. Such terminology, however, lacks rigor, especially for scenarios where the EBB eigenstates may possess more than one $\beta$ solution of the characteristic equation in Eq.~(\ref{characteristics}). 

To complement the previous pictures, we clarify the precise significance of NHSE as the asymptotic behavior of EBB eigenstates in the deep bulk $\mathcal{P}, \mathcal{Q}\ll x\ll L-\mathcal{Q}+1$ under OBC in the thermodynamics limit $L \rightarrow \infty$. More specifically, the asymptotic behavior of an EBB eigenstate concerning $E_{\mu}(\beta_{0})$ is depicted by $\beta_{0}\in GBZ_{\mu}$, with its wave function approaching $\Psi_{R}(x)\sim\beta_{0}^{x}$ and $\Psi_{L}^{*}(x)\sim\beta_{0}^{-x}$~(see Appendix~\ref{appendixB} for details) in the deep bulk for a right and a left EBB eigenstate, respectively. 

We illustrate such NHSE's precise significance in Fig.~\ref{fig5} for the two-band model in Eq.~(\ref{fwmodelnonbloch}). With parameters $t_{1}=1$, $\gamma=2/3$, $t_{2}=1$, $t_{3}=1/5$, both $GBZ_{\pm}$ are inside the unit circle~[Fig.~\ref{fig5}(a)], leading to left-localized asymptotic behaviors in the deep bulk for all right EBB-eigenstates. For any particular right EBB eigenstate $\Psi_{R}(x)$, its weight distribution in the deep bulk compares consistently with the expected asymptotic behavior $\beta_{0}^{x}$, where $\beta_{0}$ is the solution of characteristic equation $\det{\left[E_{0}-H_{FW}(\beta)\right]}=0$ with the second smallest norm~[Fig.~\ref{fig5}(b)-(g)]. The deviations between $\log|\Psi_{R}(x)|$ and $\log|\beta_{0}^{x}|$ near the left and right boundaries arise from the contributions of the right eigenstates with characteristic equation solutions away from GBZs~[Fig.~\ref{fig5}(c)-(g)]. These eigenstates fade away, tending to the deep bulk, and are challenging to track analytically. In addition, stable or not, isolated edge states may present at the boundaries, together with the EBBs under OBC.

\section{Open-boundary Green's functions}
\label{section4}
Dynamical evolution, usually encoded in single-particle Green's functions, is indispensable for a comprehensive study of noninteracting non-Hermitian systems. Due to the breakdown of the Bloch band theory under OBC, open-boundary Green's functions for non-Hermitian systems are no longer accessible through their usual expressions in the Brillouin zone~(BZ) and require scrutiny in the GBZ for proper generalization. Motivated by the pioneer study of GBZ-based Green's functions for single-band non-Hermitian systems~\cite{xue2021simple,exact2022}, we derive a general expression of open-boundary Green's functions for multiband non-Hermitian systems. 

Instead of starting from the biorthogonal eigenstates of non-Hermitian tight-binding Hamiltonians, we construct a set of minimally biorthogonal basis~(MBB), a natural non-Hermitian generalization of the Bloch orthogonal basis under PBC~(Appendix \ref{appendixC}),
\begin{align}
    \label{mbb}
    \ket{\beta}_{R}&=\frac{1}{\sqrt{L}}\sum_{x=1}^{L}\beta^{x}\ket{x},\nonumber\\
    _{L}\bra{\beta}&=\frac{1}{\sqrt{L}}\sum_{x=1}^{L}\beta^{-x}\bra{x},
\end{align}
where $\beta=\mathscr{R}e^{i\theta}$ with the a real, positive modulus $\mathscr{R}$ and a phase $\theta=\frac{2\pi}{L}m$, $m=0,1,2,\ldots,L-1$. $\ket{x}$ contains internal degrees of freedom. The MBB follows the biorthogonality and completeness conditions in the thermodynamics limit~(Appendix \ref{appendixC}),
\begin{align}
    \label{biorthogonality}
    &_{L}\braket{\beta|\beta'}_{R}=\delta_{\beta\beta'},\nonumber\\
    &\frac{L}{2\pi}\int_{0}^{2\pi}d\theta\ket{\beta}_{RL}\bra{\beta}=\mathbf{1}.
\end{align}
After some algebra, we obtain the single-particle retarded Green's function $G(x,y;t)=-i\bra{x}e^{-i\hat{H}t}\ket{y}$~($t>0$) under OBC,
\begin{align}
    \label{generalgreen}
    G(x,y;t)=-i\oint_{|\beta|=\mathscr{R}}\frac{d\beta}{2\pi i\beta}\beta^{x-y}e^{-iH(\beta)t},
\end{align}
where we have eliminated the contributions that vanish in the deep bulk~(Appendix \ref{appendixD}). Transforming into the frequency space, we arrive at~(Appendix \ref{appendixE})
\begin{align}
    \label{frequencygreen}
    G(x,y;\omega)=\oint_{|\beta|=\mathscr{R}}\frac{d\beta}{2\pi i\beta}\frac{\beta^{x-y}}{\omega-H(\beta)}.
\end{align}

When it comes to the case with sub-GBZs and EBBs, we can obtain that the physical integral contour $|\beta|=\mathscr{R}$ is equivalent to $GBZ_{\mu}$ for any given $\omega\in\mathbb{C}_{\mu}$ and $\omega\not\in E_{\mu}[GBZ_{\mu}]$, denoted as $\omega_{\mu}$~(Appendix \ref{appendixF}). Thus, the open-boundary Green's function is given by~(Appendix \ref{appendixF})
\begin{align}
    \label{greenfunc}
    G(x,y;\omega_{\mu})=\oint_{GBZ_{\mu}}\frac{d\beta}{2\pi i\beta}\frac{\beta^{x-y}}{\omega_{\mu}-H(\beta)}.
\end{align}
The GBZ-based Green's function in Refs.~\cite{xue2021simple,exact2022} is a reduction of Eq.~(\ref{greenfunc}) into single-particle non-Hermitian systems.
Noteworthily, all sub-GBZs with respect to EBBs $E_{\mu}(\beta)$ become degenerate for Hermitian systems or non-Hermitian systems without NHSE. Consequently, Eq.~(\ref{greenfunc}) reduces to Green's functions' conventional form under PBC. 

\section{Conclusion}
\label{section5}
In this paper, we have anatomized the bulk properties of 1D multiband non-Hermitian systems under OBC and addressed crucial issues complementing the non-Bloch band theory. We have introduced the concept of EBBs to settle the multivalued functions of energy bands arising from the complex-valued nature of multiband non-Hermitian systems, and endowed the gapped and gapless energy bands with more rigorous terminology of disconnected and connected EBBs in the complex energy plane. We have also considered the roles of branch points and branch cuts, which depict the transition between gapped and gapless bands. Moreover, we have clarified the precise significance of NHSE, which predicts the asymptotic behavior of EBB eigenstates in the deep bulk. Based on the EBBs and sub-GBZs, we have derived a general form of open-boundary Green's functions in the deep bulk. We leave the connection between open-boundary multiband non-Hermitian systems and various symmetries for future studies. 

\section*{Acknowledgements}

We acknowledge helpful discussions with Hao-Yan Chen, Zihao Dong, and Haoshu Li. We also acknowledge support from the National Key R\&D Program of China (No.2022YFA1403700) and the National Natural Science Foundation of China (No.12174008 \& No.92270102).

\appendix
\begin{widetext}
\section{Energy-band branches and sub-generalized Brillouin zones of multiband non-Hermitian systems}
\label{appendixA}
The non-Bloch band theory with GBZs is a well-established theory to expound bulk bands of 1D non-Hermitian systems under OBC~\cite{yao2018,yokomizo2019}. Subsequently, the terminology of GBZs is generalized to the emergence of sub-generalized Brillouin zones~(sub-GBZs) for general multiband systems~\cite{yang2020}. The bulk spectra are gestated in the non-Bloch Hamiltonian $H(\beta)=\sum_{n=-\mathcal{P}}^{\mathcal{Q}}t_{n}\beta^{n},\beta\in\mathbb{C}$ through solving the characteristic equation $ch(\beta,E)\equiv\det{[E-H(\beta)]}=\sum_{j=-p}^{q}a_{j}(E)\beta^{j}=\prod_{\mu=1}^{\mathcal{M}}\left[E-E_{\mu}(\beta)\right]=0$, where $E_{\mu}(\beta)$ is an energy band of $H(\beta)$, or more strictly, a single-valued branch of a multivalued radical function. We number the solutions of the characteristic equation as $|\beta_{1}(E)|\leq|\beta_{2}(E)|\leq\ldots\leq|\beta_{p+q}(E)|$. The bulk spectra are given by $E\in\mathbb{C}$ satisfying $|\beta_{p}(E)|=|\beta_{p+1}(E)|$, and these $\beta$ values outline the GBZs in the complex $\beta$ plane. Furthermore, the branches $E_{\mu}(\beta)$ correspond to different GBZs, thus resulting in the sub-GBZ concerning each branch, denoted as $GBZ_{\mu}$. According to the theory of multivalued functions, as $\beta$ runs through the complex plane, each branch $E_{\mu}(\beta)$ occupies a continuous region~(open set) $\mathbb{C}_{\mu}$ of the complex plane, being a single-valued function of $\beta$. The set of these regions is a covering of the whole complex plane. Therefore, the sub-GBZ spectrum~(bulk energy band) $E_{\mu}[GBZ_{\mu}]\equiv\left\{E_{\mu}(\beta),\beta\in GBZ_{\mu}\right\}$ corresponding to $GBZ_{\mu}$ is located in $\mathbb{C}_{\mu}$, which we dub the energy-band branch~(EBB) in the main text. 

In general, the number of EBBs and sub-GBZs must be equal, which contains the cases that two or more EBBs correspond to one sub-GBZ. For example, we only obtain one GBZ in the well-known non-Hermitian Su-Schrieffer-Heeger~(NH-SSH) model~\cite{yao2018,yokomizo2019}, a circle with radius unequal to $1$, to which two branches $E_{\pm}(\beta)$ correspond. Noteworthily, the single-valued branches concerning the two EBBs are single-valued functions of $\beta$ located on the two complex half-planes $\mathbb{C}_{\pm}$ divided by the imaginary axis, respectively. More specially, for a single-band model, there exists one branch $H(\beta)$, i.e., the whole complex plane, which is a single-valued function of $\beta$. It has been shown that the GBZ corresponding to the single-band $H(\beta)$ must be a closed curve and encloses $p$ zeros of $E-H(\beta)=\sum_{j=-p}^{q}a_{j}(E)\beta^{j}$~\cite{zhang2020}, where $E$ is not the point on GBZ spectrum $H[GBZ]$. According to the argument principle, the winding number of $E-H(\beta)$ around GBZ vanishes due to the equal number of zeros and poles. 

Making slight modifications, we identify $H(\beta)$ in single-band models with each single-valued branch $E_{\mu}(\beta)$ in multiband models, respectively. We immediately obtain that each sub-GBZ $GBZ_{\mu}$ is a closed curve, and $GBZ_{\mu}$ encloses $p$ zeros of $ch(\beta,E)$ for $E\in\mathbb{C}_{\mu}$ and $E\notin E_{\mu}[GBZ_{\mu}]$, thus leading to the vanishing of the winding number of $ch(\beta,E)$ surrounding $GBZ_{\mu}$. We illustrate a typical non-Hermitian two-band model with two distinct sub-GBZs, which is a generalization of the NH-SSH model by adding the next nearest neighbor hopping matrix~\cite{fu2022}. The Hamiltonian of this model in real space reads~(Fig.~\ref{suppfigs0})
\begin{align}
    \label{fwmodel}
    \hat{H}_{FW}=\sum_{x}\left(c^{\dagger}_{x}Mc_{x}+c^{\dagger}_{x}T_{1}c_{x+1}+c^{\dagger}_{x+1}T_{-1}c_{x}+c^{\dagger}_{x}T_{2}c_{x+2}+c^{\dagger}_{x+2}T_{-2}c_{x}\right),
\end{align}
where 
\begin{align}
    M=\left(\begin{matrix}
        0&t_{1}+\gamma\\
        t_{1}-\gamma&0
    \end{matrix}\right),
    T_{1}=\left(\begin{matrix}
        0&0\\
        t_{2}&0
    \end{matrix}\right),
    T_{-1}=\left(\begin{matrix}
        0&t_{2}\\
       0&0
    \end{matrix}\right),
    T_{2}=\left(\begin{matrix}
        t_{3}&0\\
        0&0
    \end{matrix}\right),
    T_{-2}=\left(\begin{matrix}
        0&0\\
       0&t_{3}
    \end{matrix}\right).
    \nonumber
\end{align}
The non-Bloch Hamiltonian is 
\begin{align}
    \label{fwnonbloch}
    H_{FW}(\beta)= \left(\begin{matrix}
        t_{3}\beta^{2}&t_{1}+\gamma+t_{2}\beta^{-1}\\
        t_{1}-\gamma+t_{2}\beta&t_{3}\beta^{-2}
    \end{matrix}\right),
\end{align}
and the multivalued function concerning EBBs can be expressed as
\begin{align}
    \label{fwenergybands}
    E(\beta)=\frac{1}{2}\left[t_{3}\left(\beta^{2}+\beta^{-2}\right)+\sqrt{\left(t_{3}(\beta^{2}+\beta^{-2})\right)^{2}-4\left(t_{3}^{2}-(t_{1}+\gamma+t_{2}\beta^{-1})(t_{1}-\gamma+t_{2}\beta)\right)}\right],
\end{align}
which induces two single-valued branches $\mathbb{C}_{\pm}$ according to the square-root function. We take parameters as $t_{1}=1,\gamma=\frac{2}{3},t_{2}=1,t_{3}=\frac{1}{5}$, and $\mathbb{C}_{\pm}$ are numerically the two half complex planes divided by the imaginary axis. We plot the sub-GBZs $GBZ_{\pm}$~(cyan and orange loops, respectively) corresponding to $E_{\pm}[GBZ_{\pm}]$ and zeros~(black dots) of $ch[\beta,E]$ for selected $E\in\mathbb{C}_{\pm}$ in Fig.~\ref{suppfigs1}. The sub-GBZ loops indeed enclose $p=2$ zeros with the selected $E\in\mathbb{C}_{\pm}$ and $E\not\in E_{\pm}[GBZ_{\pm}]$, respectively, and at least two zeros lie on $GBZ_{\pm}$ when $E\in E_{\pm}[GBZ_{\pm}]$. Besides, $GBZ_{\pm}$ may not enclose $2$ zeros of $ch[\beta,E]$ for selected $E\in\mathbb{C}_{\mp}$, respectively~(Fig.~\ref{suppfigs2}).

\begin{figure}
    \includegraphics[width=16cm, height=2.7cm]{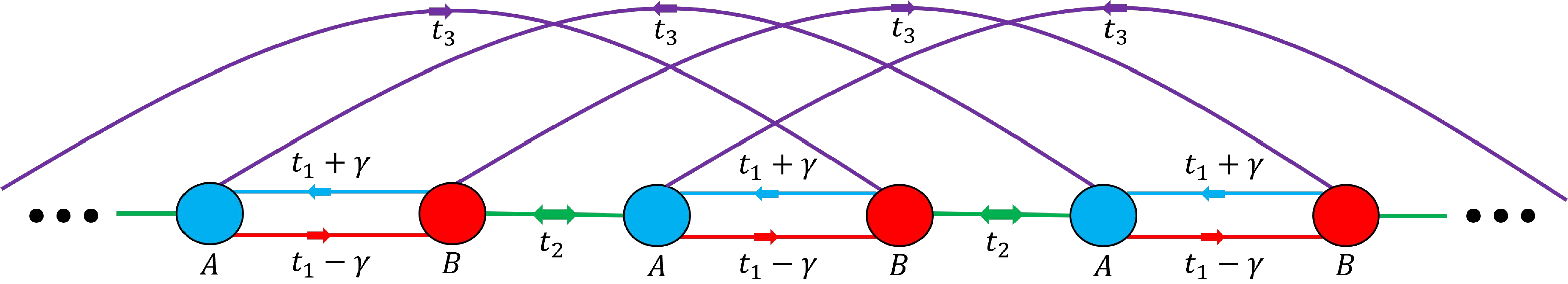}
    \caption{The two-band model Eq.~(\ref{fwmodel}). The blue and red solid circles denote sublattices A and B, respectively.} 
    \label{suppfigs0}
\end{figure}

\begin{figure}
    \subfigure[]{\includegraphics[width=3.5cm, height=3.5cm]{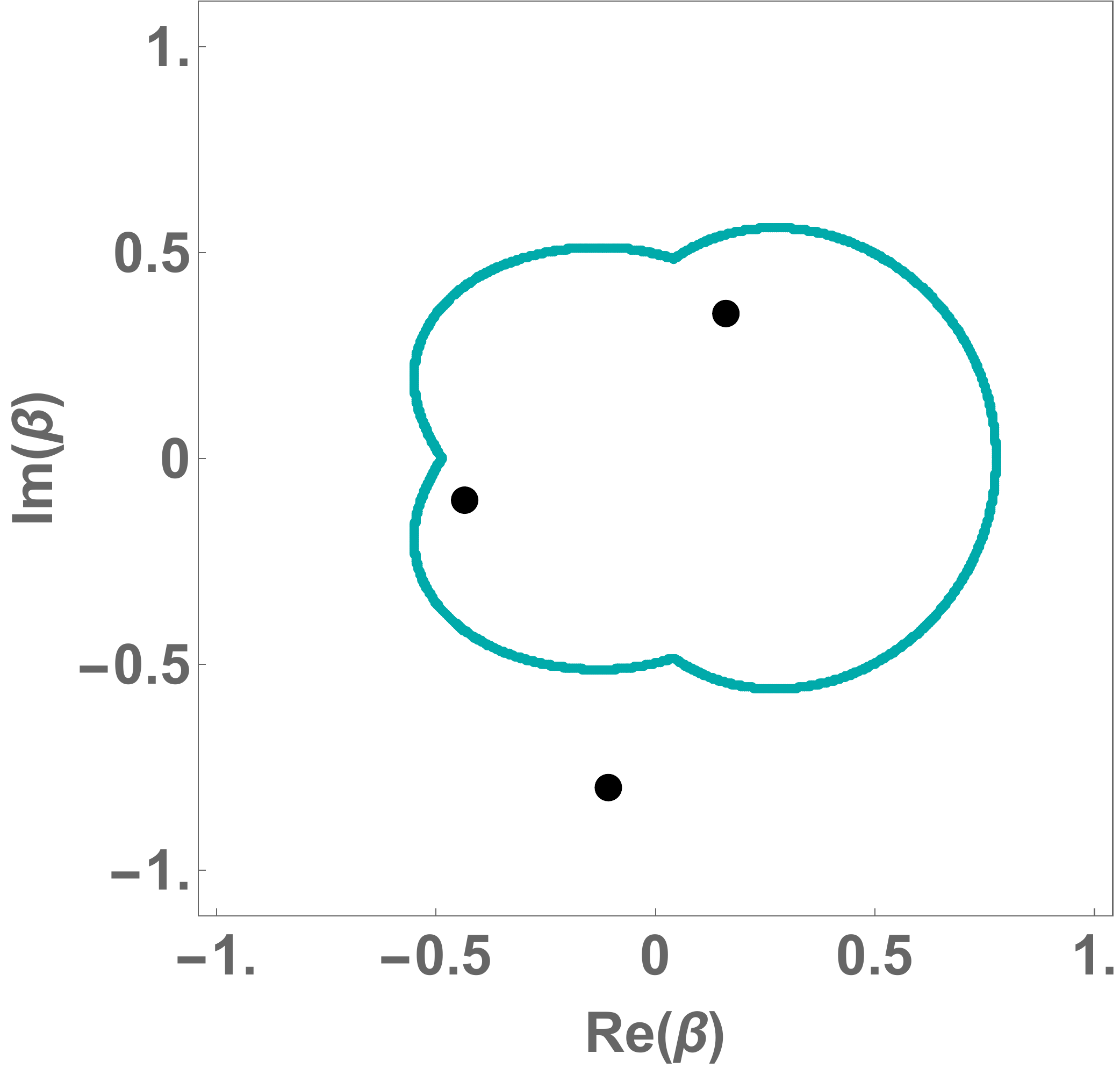}}
    \subfigure[]{\includegraphics[width=3.5cm, height=3.5cm]{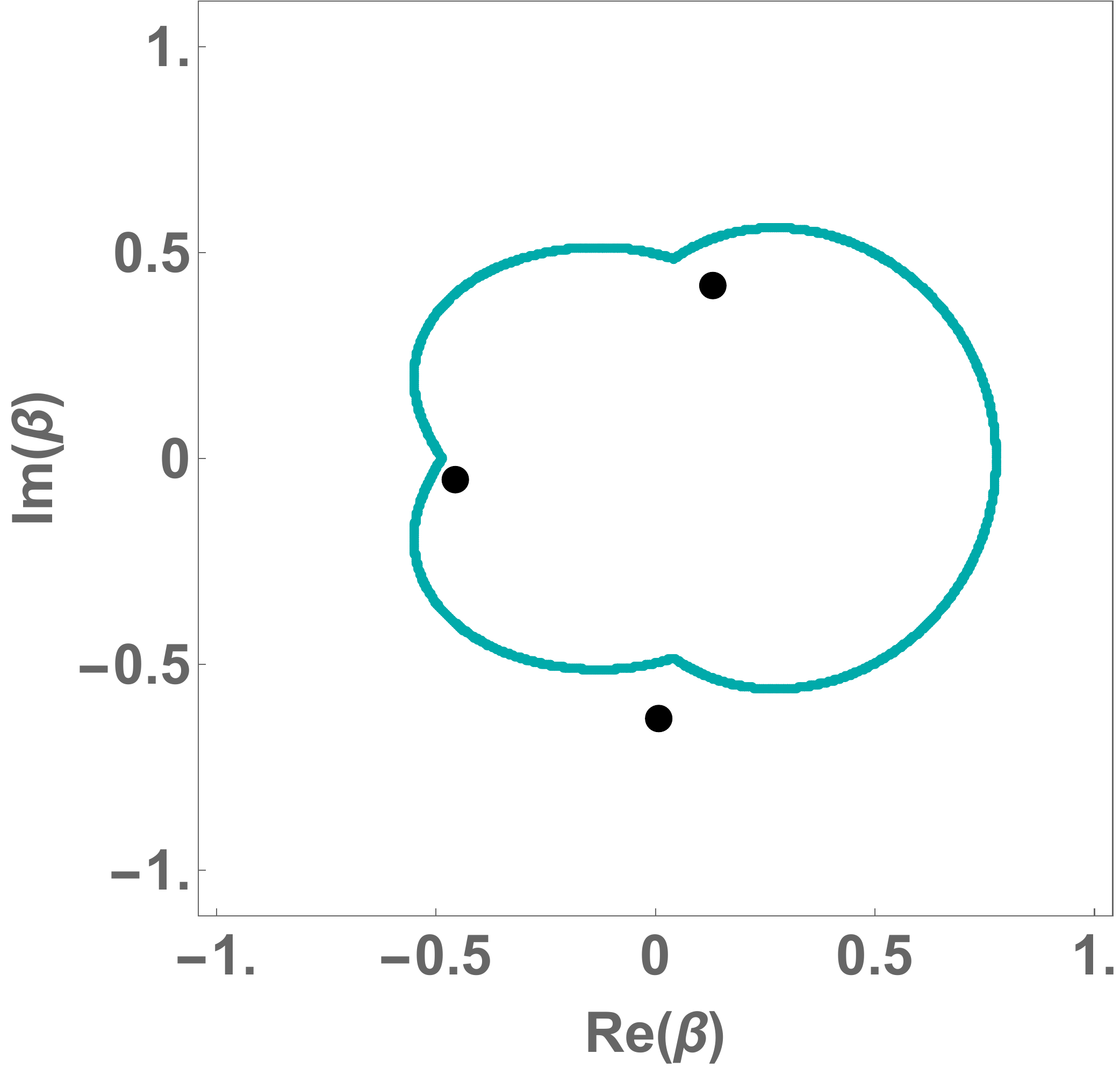}}
    \subfigure[]{\includegraphics[width=3.5cm, height=3.5cm]{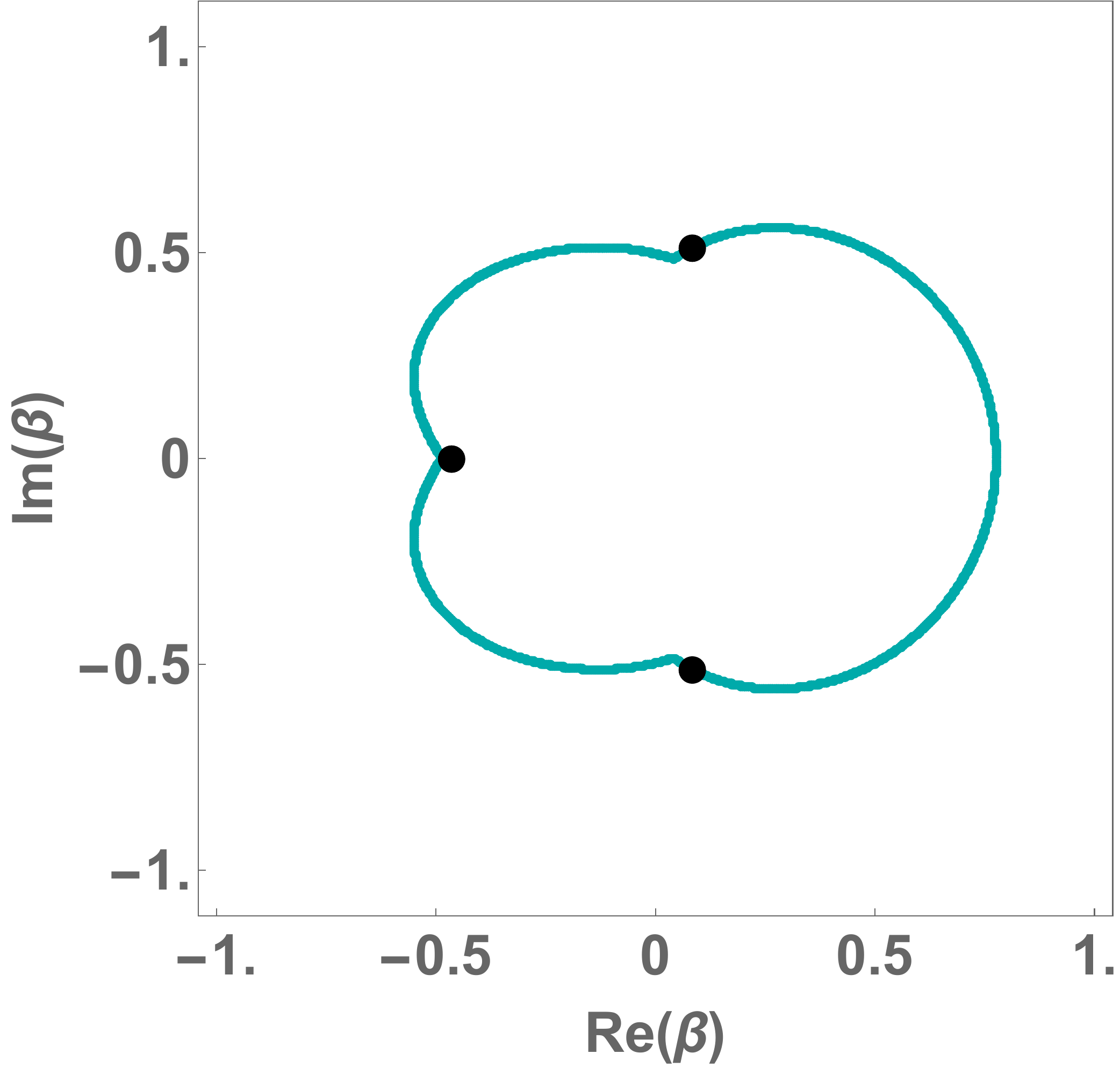}}
    \subfigure[]{\includegraphics[width=3.5cm, height=3.5cm]{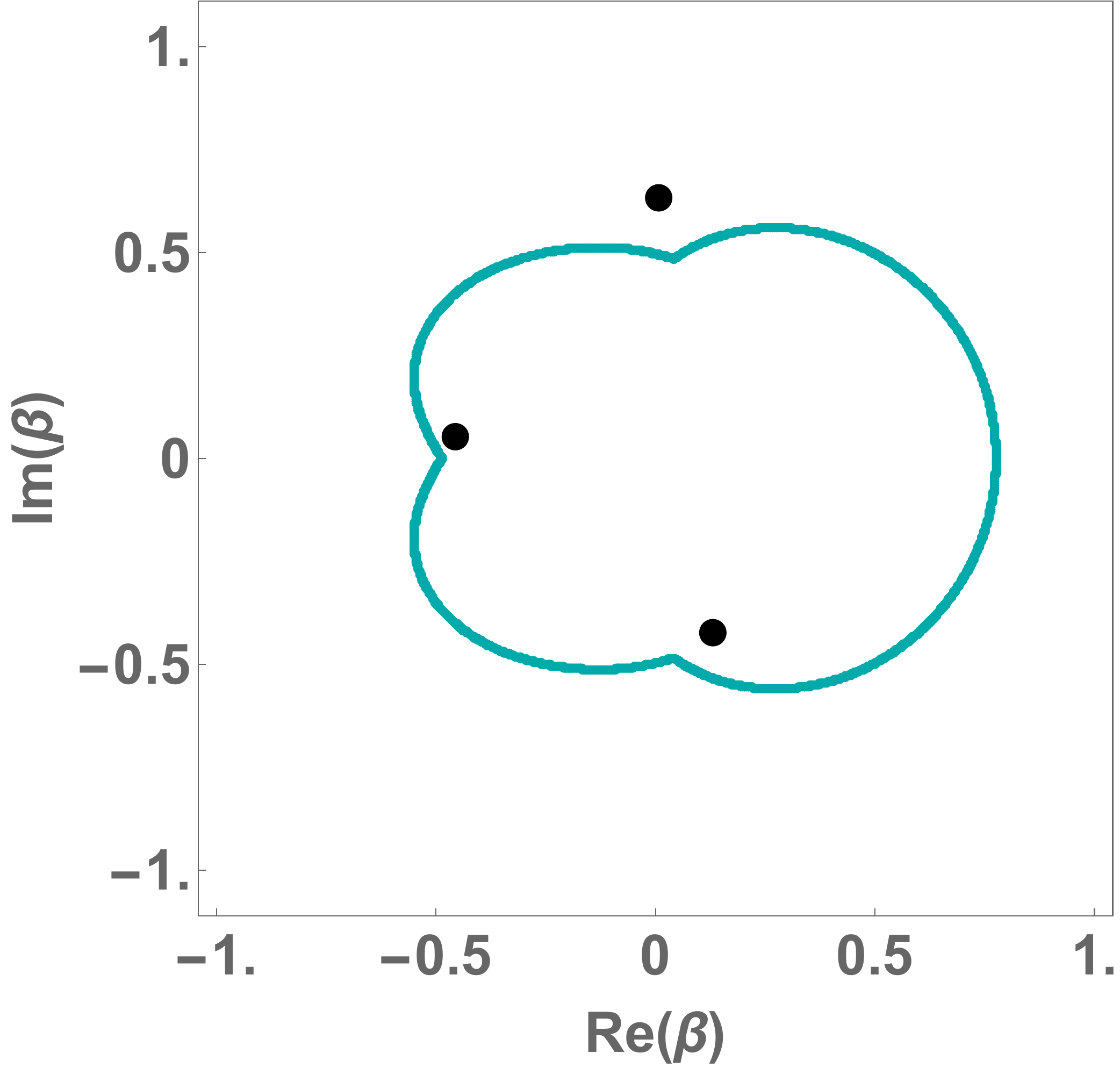}}
    \subfigure[]{\includegraphics[width=3.5cm, height=3.5cm]{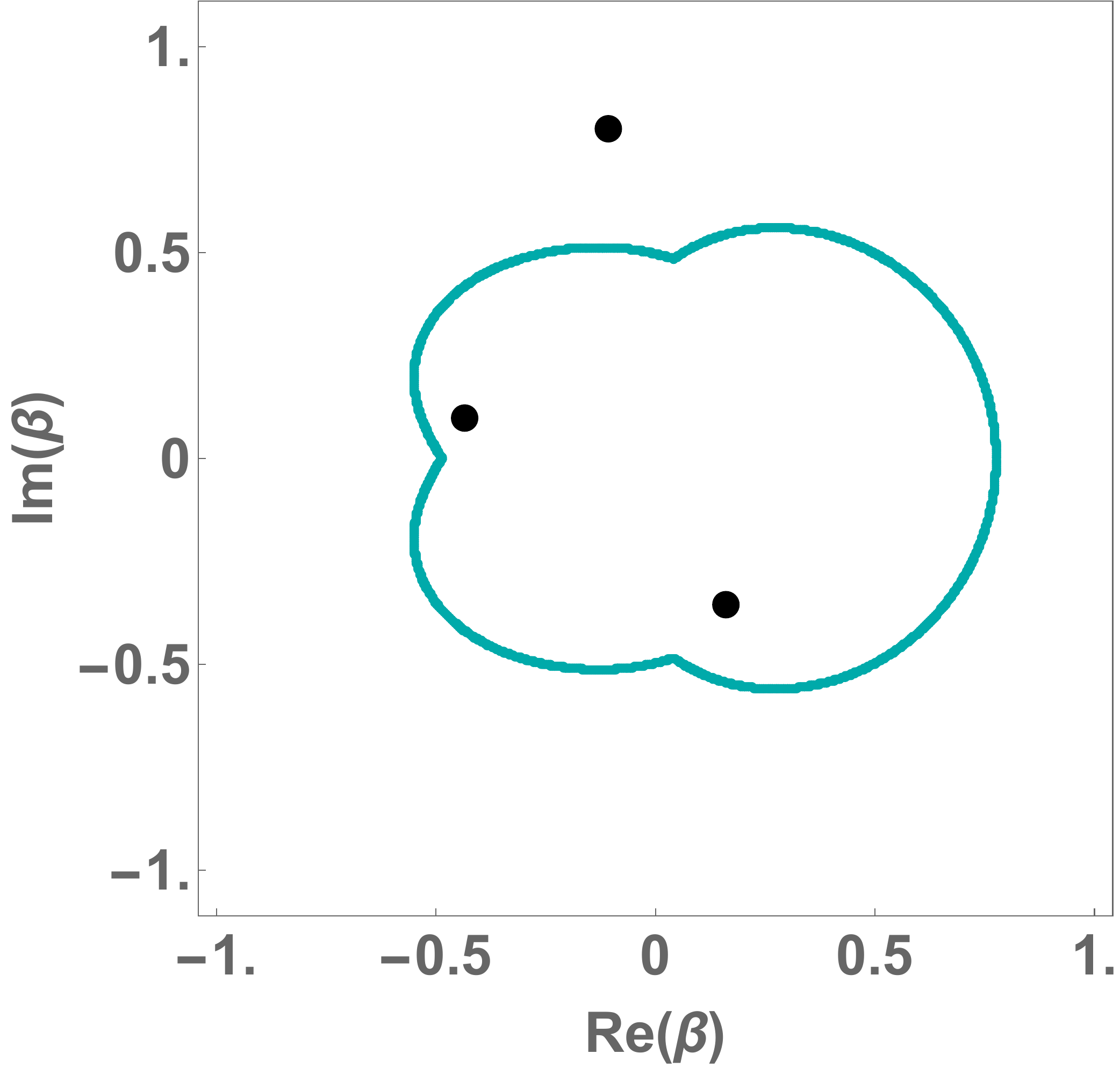}}
    \subfigure[]{\includegraphics[width=3.5cm, height=3.5cm]{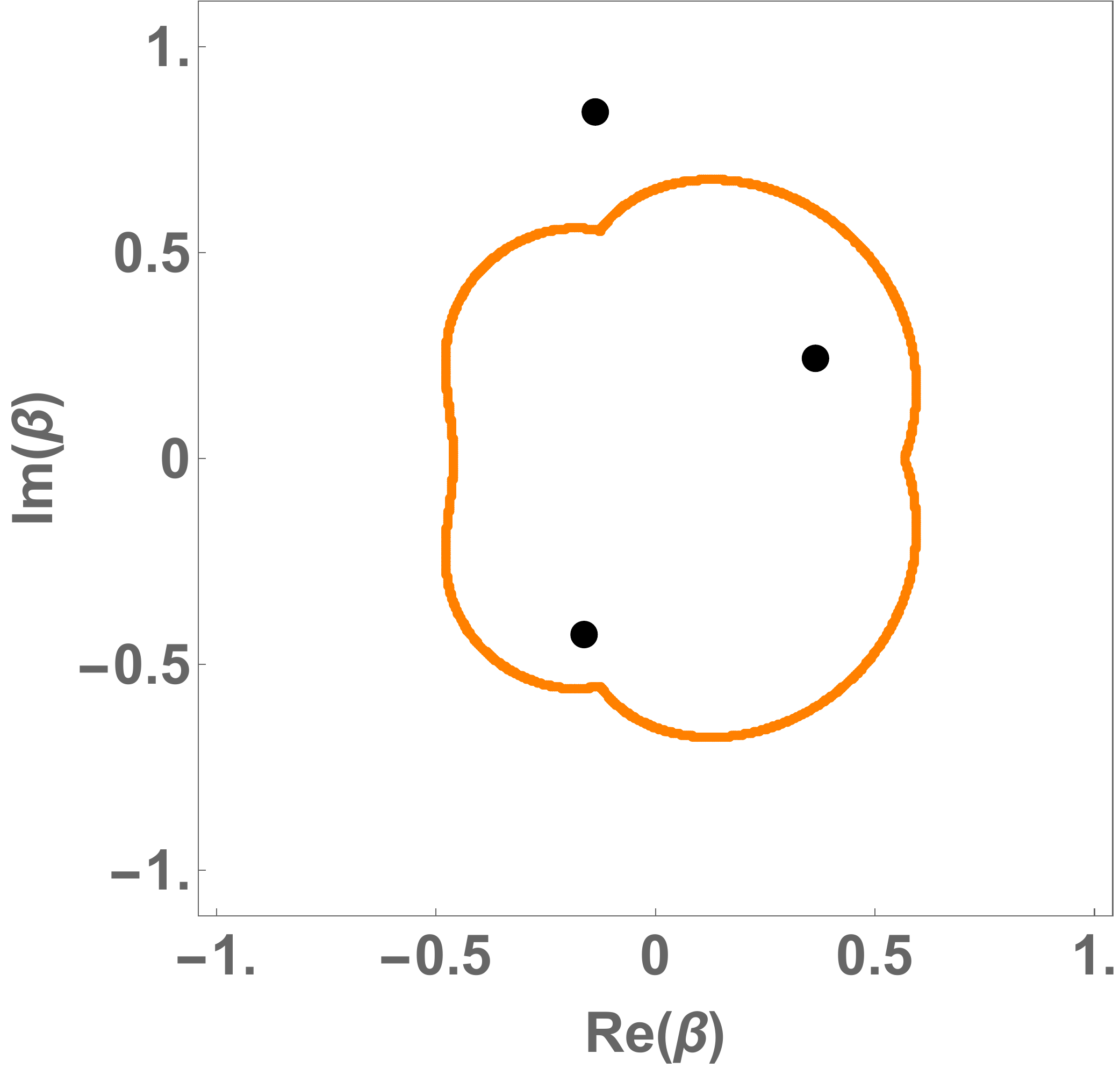}}
    \subfigure[]{\includegraphics[width=3.5cm, height=3.5cm]{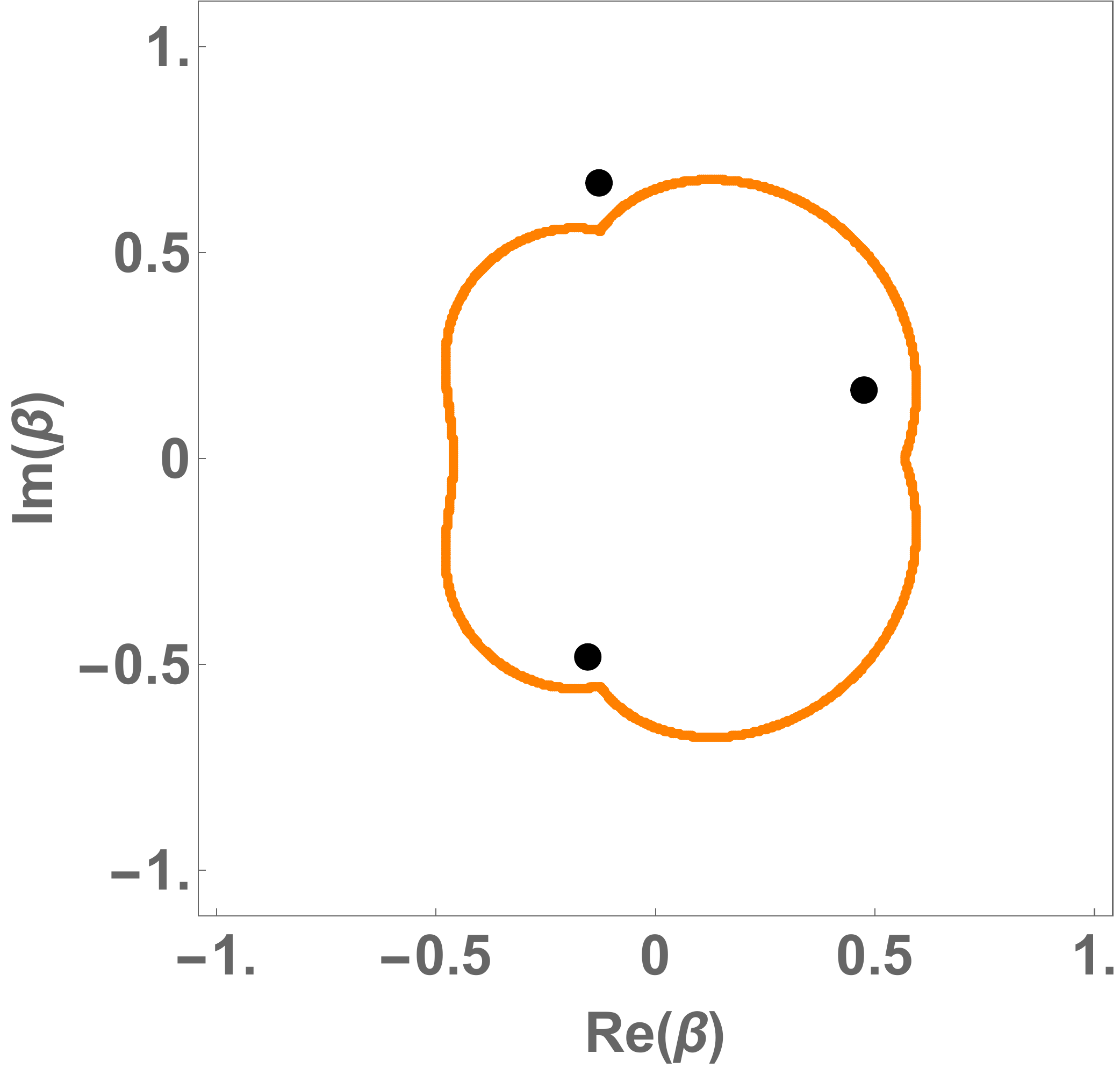}}
    \subfigure[]{\includegraphics[width=3.5cm, height=3.5cm]{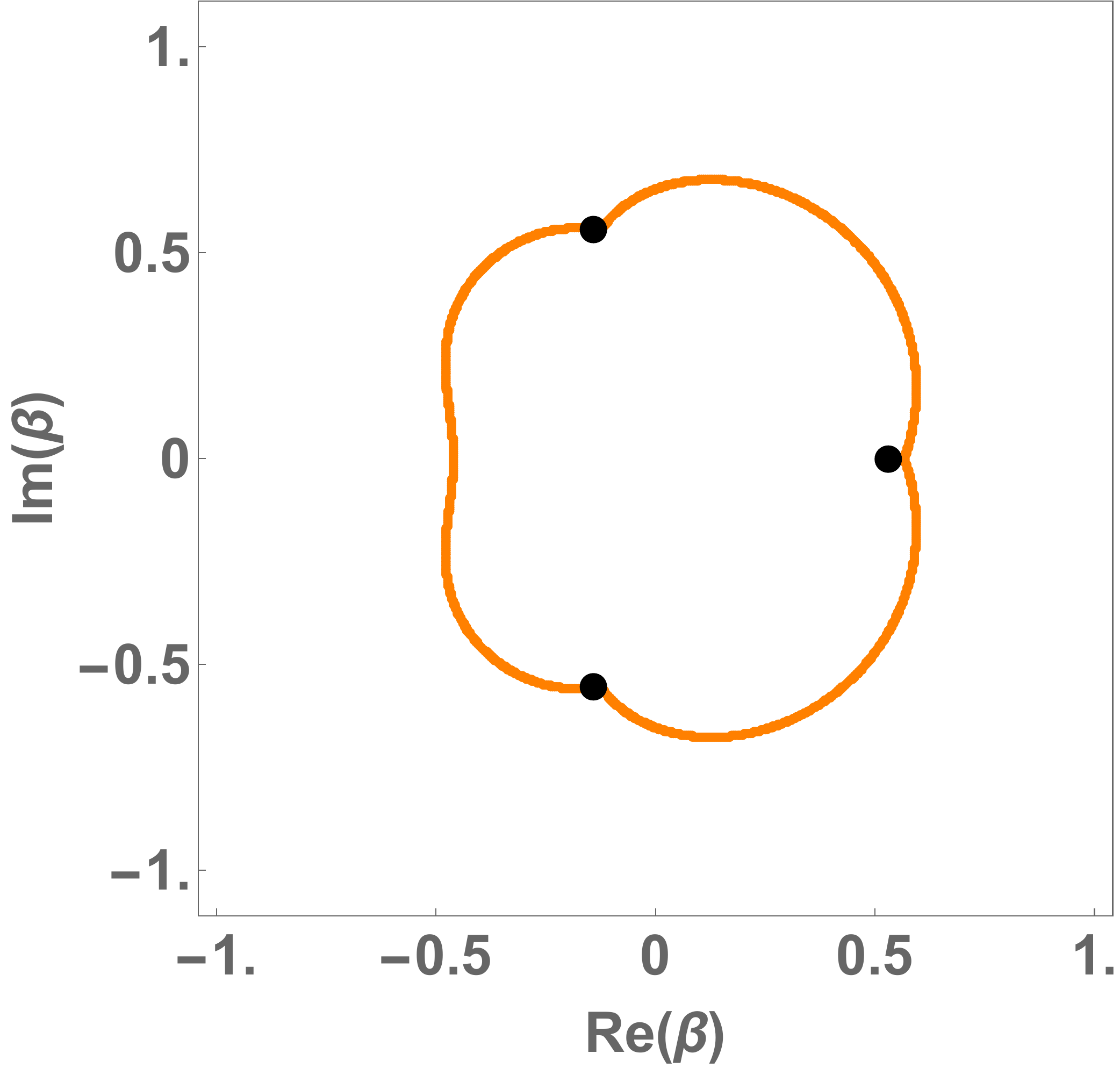}}
    \subfigure[]{\includegraphics[width=3.5cm, height=3.5cm]{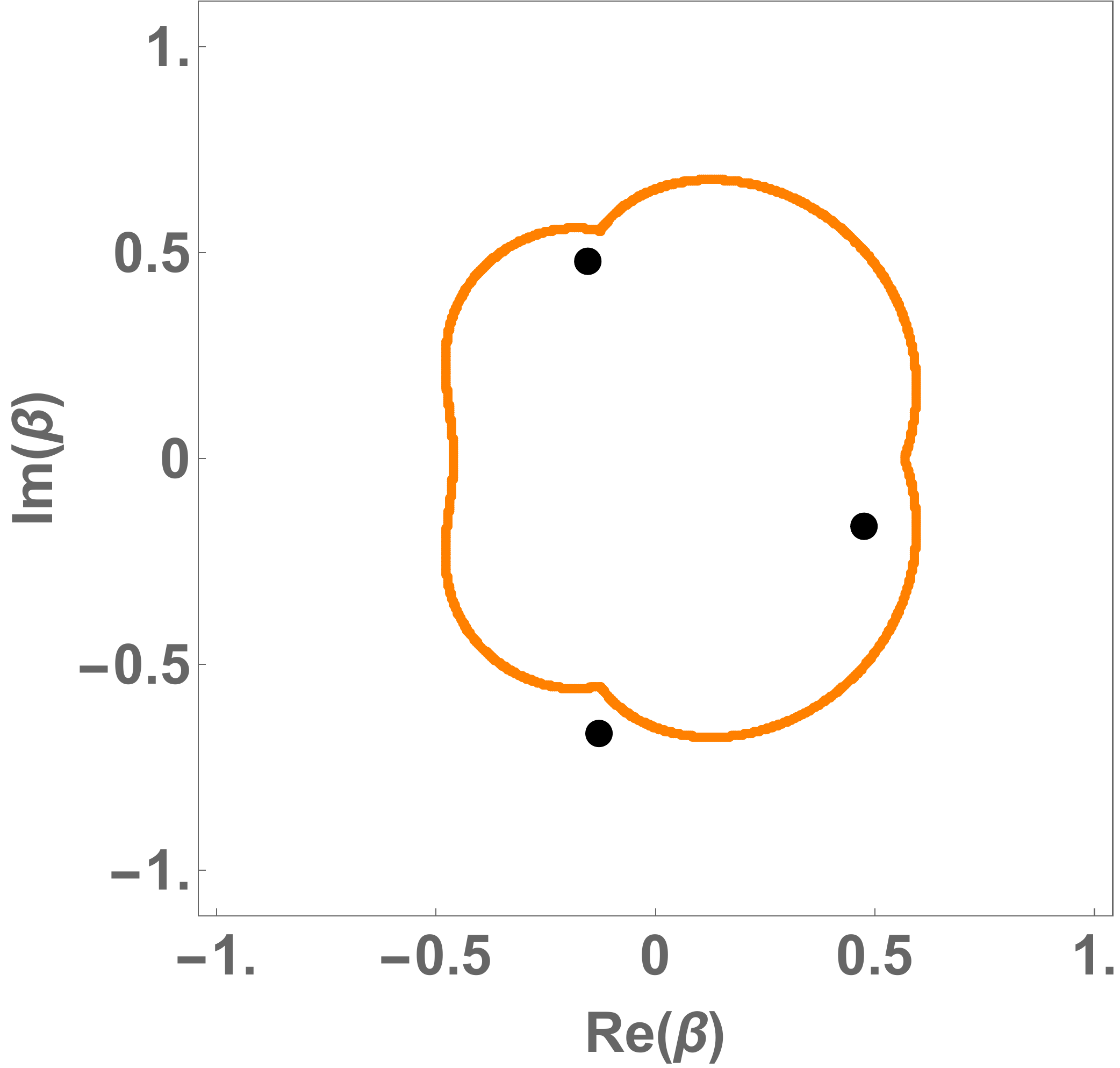}}
    \subfigure[]{\includegraphics[width=3.5cm, height=3.5cm]{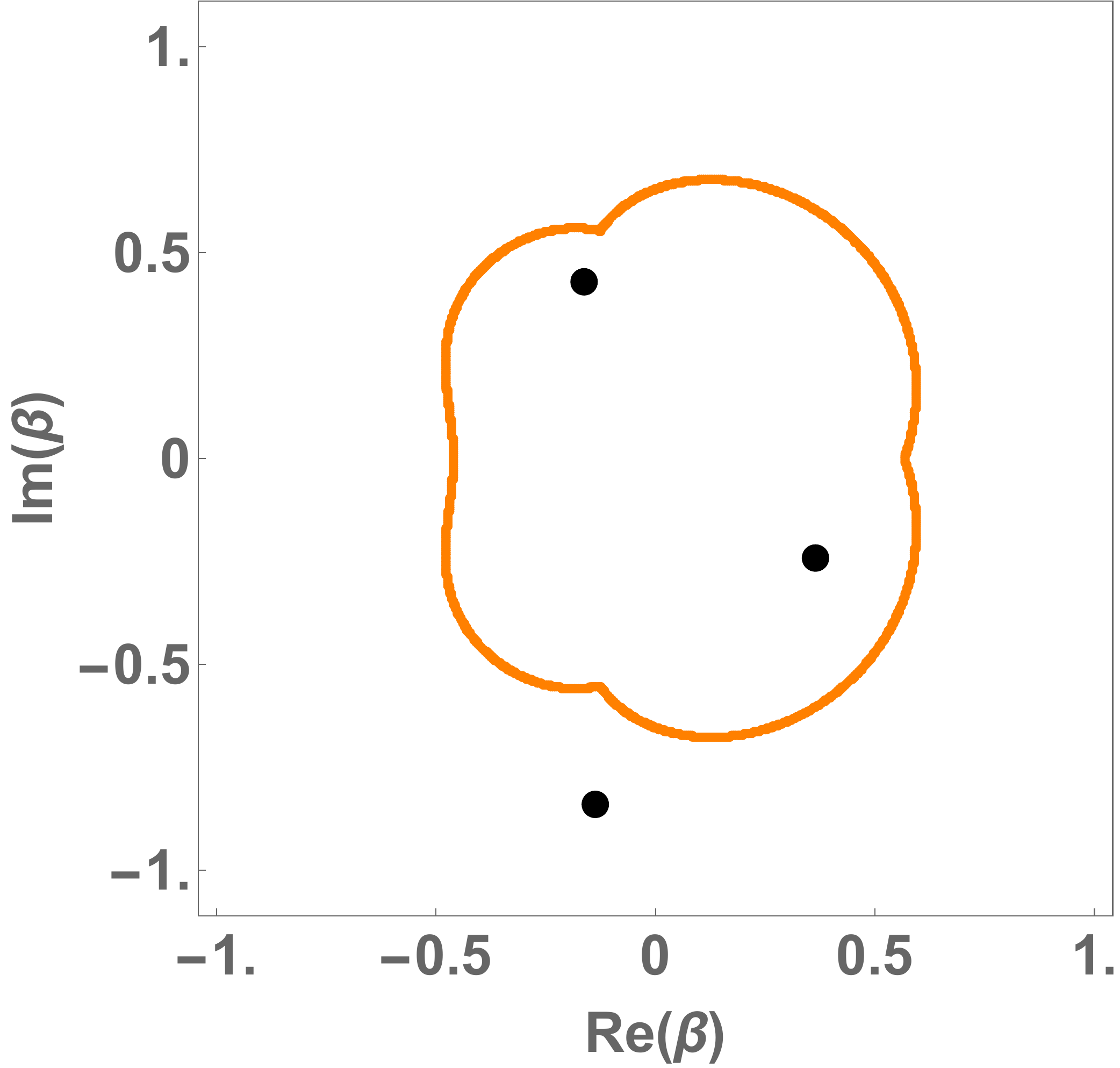}}
    \caption{The sub-GBZs $GBZ_{\pm}$~(cyan and orange loops, respectively) of the two-band model Eq.~(\ref{fwmodel}) and zeros~(black dots) of $ch[\beta,E]$ for selected $E\in\mathbb{C}_{\pm}$. (a)-(e): $E=1-0.4i,1-0.2i,1,1+0.2i,1+0.4i$ for $E\in\mathbb{C}_{+}$; (f)-(j): $E=-1.4-0.4i,-1.4-0.2i,-1.4,-1.4+0.2i,-1.4+0.4i$ for $E\in\mathbb{C}_{-}$. Three zeros corresponding to $E=1$ and $E=-1.4$ with $E\in E_{\pm}[GBZ_{\pm}]$ are located on $GBZ_{\pm}$, respectively. The fourth root is not shown for being outside this region. The parameters are $t_{1}=1,\gamma=\frac{2}{3},t_{2}=1,t_{3}=\frac{1}{5}$.}
    \label{suppfigs1}
\end{figure}

\begin{figure}
    \subfigure[]{\includegraphics[width=4.5cm, height=4.5cm]{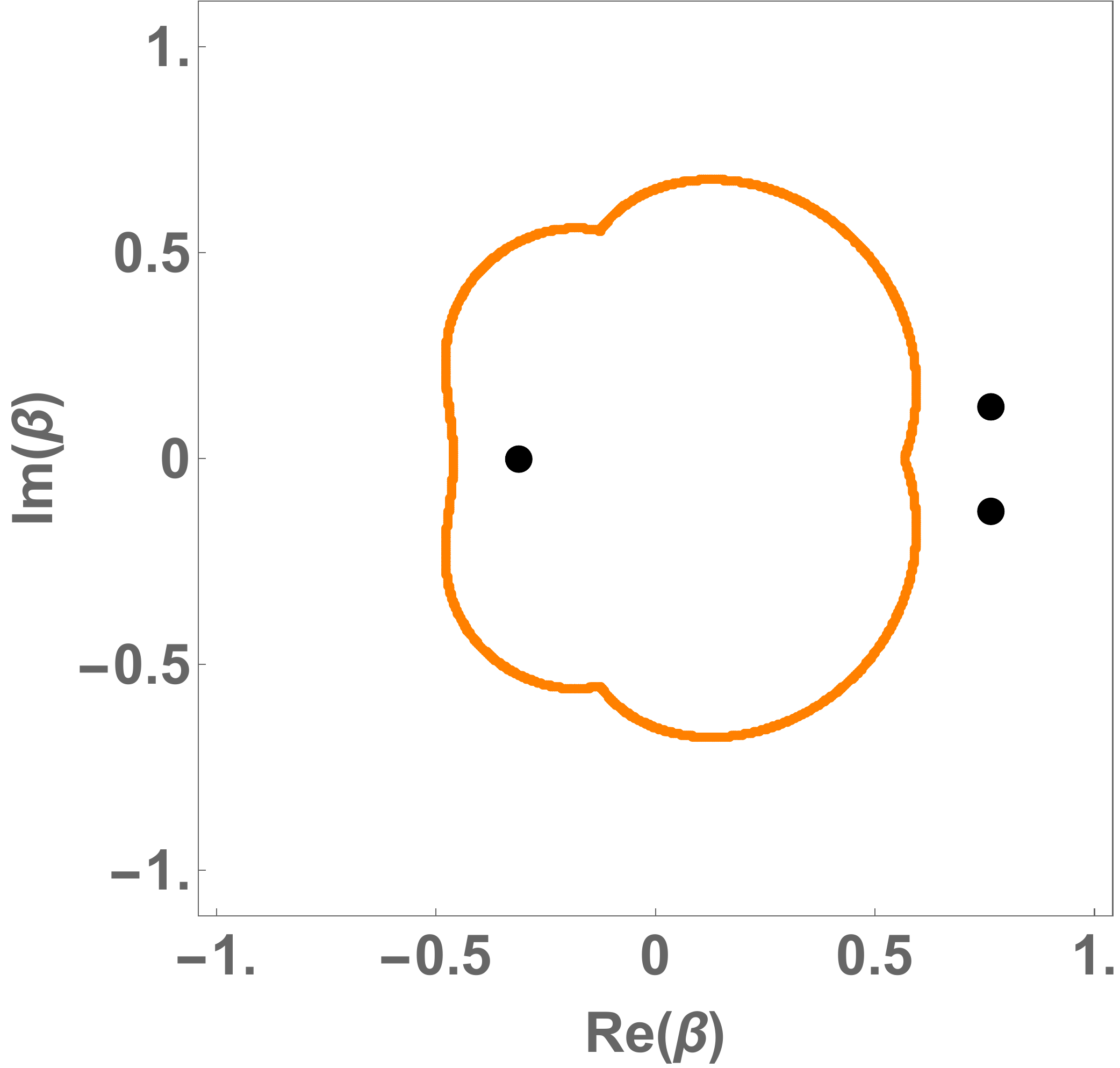}}\qquad\qquad
    \subfigure[]{\includegraphics[width=4.5cm, height=4.5cm]{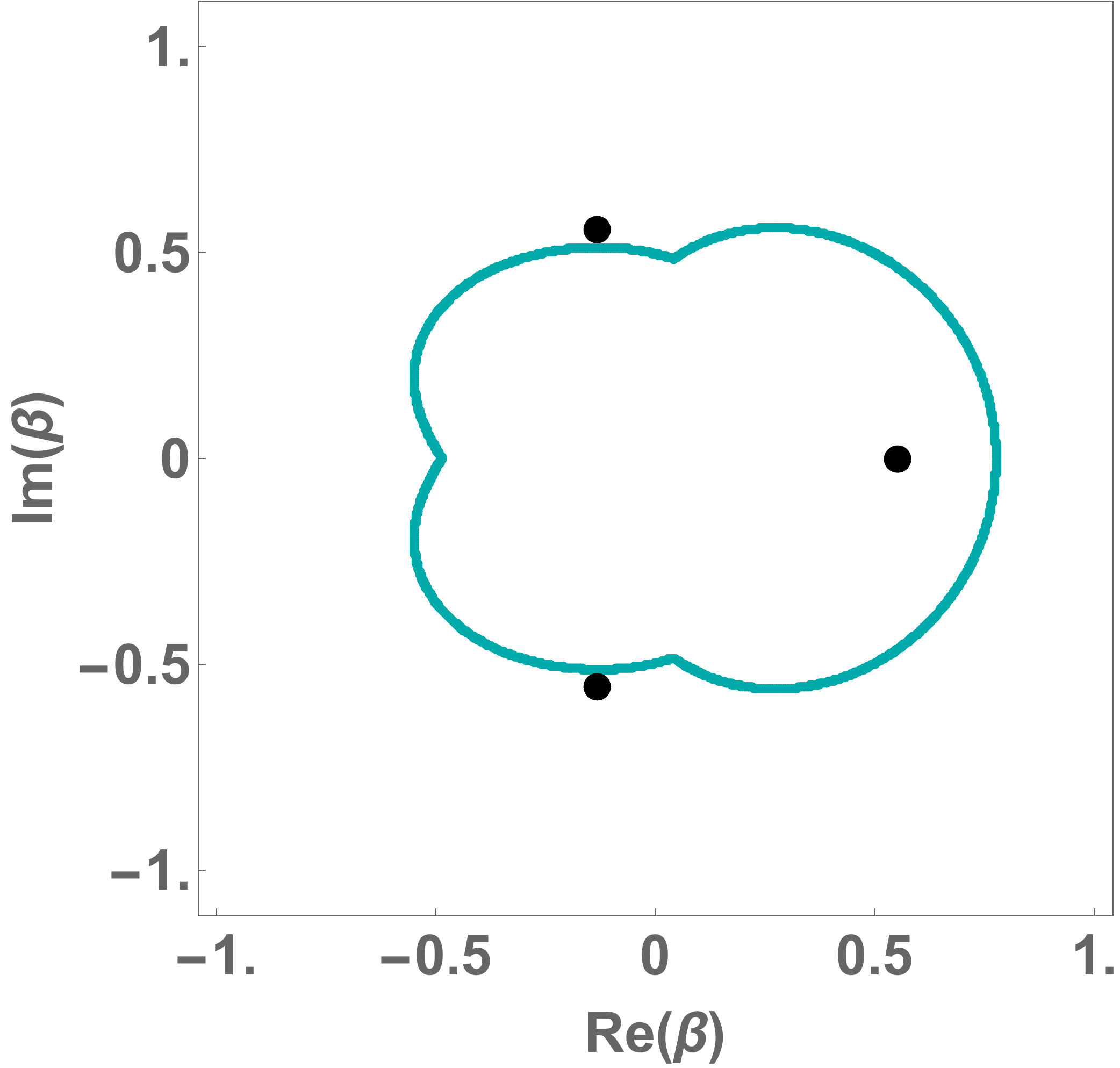}}
    \caption{The sub-GBZs $GBZ_{\pm}$~(cyan and orange loops, respectively) of the two-band model Eq.~(\ref{fwmodel}) and zeros~(black dots) of $ch[\beta,E]$ for selected (a) $E=2.02\in\mathbb{C}_{+}$ and (b) $E=-1.42\in\mathbb{C}_{-}$, respectively. The fourth root is not shown for being outside this region. The parameters are the same as Fig.~\ref{suppfigs1}.}
    \label{suppfigs2}
\end{figure}

\section{Exact relation between biorthogonal EBB-eigenstates and non-Hermitian skin effect}
\label{appendixB}
As a posterior, there exist $\mathcal{M}$ sub-GBZ spectra~(EBBs) and isolated edge modes with the arbitrary possible number $n_{e}$. We denote the eigenenergies and right~(left) eigenstates of the isolated edge modes as $E_{e}$ and $\ket{\Psi_{e}}_{R}$~($_{L}\bra{\Psi_{e}}$), $e=1,2,\ldots,n_{e}$, respectively. Consider an arbitrary eigenenergy of the $\mu$th EBB $E_{\mu}(\beta_{0})$ with $\beta_{0}\in GBZ_{\mu}$; the corresponding characteristic equation $\det{\left[E_{\mu}(\beta_{0})-H(\beta)\right]}=0$ produces $p+q$ solutions of $\beta$ ordered as $|\beta_{1}|\leq\ldots\leq|\beta_{p}|=|\beta_{p+1}|\leq\ldots\leq|\beta_{p+q}|$, where we assume the absence of zero and multiple solutions of $\beta$ and $\beta_{0}=\beta_{p}$ without loss of generality~\cite{fu2022}. The right eigenstate with respect to the bulk band energy $E_{\mu}(\beta_{0})$ formally reads
\begin{align}
    \label{rightstate}
    \ket{\Psi_{\mu}(\beta_{0})}_{R}=\sum_{j=1}^{p+q}\alpha_{j}^{\mu}\ket{\psi_{j}^{\mu}}_{R},
\end{align}
where $\ket{\psi_{j}^{\mu}}_{R}=\sum_{x=1}^{L}\beta_{j}^{x}\ket{u_{j}^{\mu}}_{R}\ket{x}$ and $H(\beta_{j})\ket{u_{j}^{\mu}}_{R}=E_{\mu}(\beta_{0})\ket{u_{j}^{\mu}}_{R}$. The coefficients $\alpha_{j}^{\mu}$ must satisfy boundary equations induced by OBC, i.e.,
\begin{align}
    \label{boundaryeq}
    \left(\begin{matrix}
    f_{1}^{\mu}\left(\beta_{1},E_{\mu}(\beta_{0})\right)&\ldots&f_{1}^{\mu}\left(\beta_{p},E_{\mu}(\beta_{0})\right)&f_{1}^{\mu}\left(\beta_{p+1},E_{\mu}(\beta_{0})\right)&\ldots&f_{1}^{\mu}\left(\beta_{p+q},E_{\mu}(\beta_{0})\right)\\
    \vdots&\vdots&\vdots&\vdots&\vdots&\vdots\\
    f_{p}^{\mu}\left(\beta_{1},E_{\mu}(\beta_{0})\right)&\ldots&f_{p}^{\mu}\left(\beta_{p},E_{\mu}(\beta_{0})\right)&f_{p}^{\mu}\left(\beta_{p+1},E_{\mu}(\beta_{0})\right)&\ldots&f_{p}^{\mu}\left(\beta_{p+q},E_{\mu}(\beta_{0})\right)\\
    g_{1}^{\mu}\left(\beta_{1},E_{\mu}(\beta_{0})\right)\beta_{1}^{L}&\ldots&g_{1}^{\mu}\left(\beta_{p},E_{\mu}(\beta_{0})\right)\beta_{p}^{L}&g_{1}^{\mu}\left(\beta_{p+1},E_{\mu}(\beta_{0})\right)\beta_{p+1}^{L}&\ldots&g_{1}^{\mu}\left(\beta_{p+q},E_{\mu}(\beta_{0})\right)\beta_{p+q}^{L}\\
    \vdots&\vdots&\vdots&\vdots&\vdots&\vdots\\
    g_{q}^{\mu}\left(\beta_{1},E_{\mu}(\beta_{0})\right)\beta_{1}^{L}&\ldots&g_{q}^{\mu}\left(\beta_{p},E_{\mu}(\beta_{0})\right)\beta_{p}^{L}&g_{q}^{\mu}\left(\beta_{p+1},E_{\mu}(\beta_{0})\right)\beta_{p+1}^{L}&\ldots&g_{q}^{\mu}\left(\beta_{p+q},E_{\mu}(\beta_{0})\right)\beta_{p+q}^{L}
    \end{matrix}\right)
    \left(\begin{matrix}
        \alpha_{1}^{\mu}\\
        \vdots\\
        \alpha_{p}^{\mu}\\
        \alpha_{p+1}^{\mu}\\
        \vdots\\
        \alpha_{p+q}^{\mu}
    \end{matrix}\right)=0,  
\end{align}
where $f_{n}^{\mu}\left(\beta_{j},E_{\mu}(\beta_{0})\right)$ and $g_{n}^{\mu}\left(\beta_{j},E_{\mu}(\beta_{0})\right)$ are polynomials of $\beta_{j}$ with finite values deduced by the boundary conditions~\cite{yokomizo2019,fu2022}. We observe that the first $p$ equations and the last $q$ equations of Eq.~(\ref{boundaryeq}) take the forms
\begin{align}
    \label{coefficientseq}
    f_{n}^{\mu}\left(\beta_{1}\right)\alpha_{1}^{\mu}+\ldots+f_{n}^{\mu}\left(\beta_{p}\right)\alpha_{p}^{\mu}+f_{n}^{\mu}\left(\beta_{p+1}\right)\alpha_{p+1}^{\mu}+\ldots+f_{n}^{\mu}\left(\beta_{p+q}\right)\alpha_{p+q}^{\mu}&=0\nonumber\\
    g_{n}^{\mu}\left(\beta_{1}\right)\beta_{1}^{L} \alpha_{1}^{\mu}+\ldots+g_{n}^{\mu}\left(\beta_{p}\right)\beta_{p}^{L} \alpha_{p}^{\mu}+g_{n}^{\mu}\left(\beta_{p+1}\right)\beta_{p+1}^{L} \alpha_{p+1}^{\mu}+\ldots+g_{n}^{\mu}\left(\beta_{p+q}\right)\beta_{p+q}^{L} \alpha_{p+q}^{\mu}&=0,
\end{align}
respectively, where we have omitted the variable $E_{\mu}(\beta_{0})$ of functions $f_{n}^{\mu}$ and $g_{n}^{\mu}$ for simplicity. Due to $|\beta_{p}|=|\beta_{p+1}|$ generating $GBZ_{\mu}$, we compare the asymptotic behavior of the terms concerning coefficients $\alpha_{j}^{\mu},j\neq p,p+1$ with that concerning coefficients $\alpha_{p}^{\mu},\alpha_{p+1}^{\mu}$ in Eq.~(\ref{rightstate}) in the deep bulk. Here, we specify that $x$ is in the deep bulk if $\mathcal{P}, \mathcal{Q}\ll x\ll L-\mathcal{Q}+1$ in the thermodynamics limit. We assume that $\alpha_{p}^{\mu},\alpha_{p+1}^{\mu}$ are not toward infinite without loss of generality, that is, nonzero finite values. Due to the different orders of $\beta_{j}^{L}$, the terms concerning coefficients $\alpha_{j}^{\mu},j\neq p,p+1$ in the second set of equations in Eq.~(\ref{coefficientseq}) are either asymptotic toward to zero or the same order with the terms concerning coefficients $\alpha_{p}^{\mu},\alpha_{p+1}^{\mu}$ such that these equations hold. We discuss according to the following two cases. First, we consider the case $|\beta_{p}|=|\beta_{p+1}|=|\beta_{0}|<1$. If $|\beta_{j}|<|\beta_{0}|<1$, that is, $j=1,2,\ldots,p-1$, the corresponding coefficients $\alpha_{j}^{\mu}$ must be finite~($|\alpha_{j}^{\mu}\beta_{j}^{L}|\rightarrow 0$) or $|\alpha_{1}^{\mu}|\gg|\alpha_{2}^{\mu}|\gg\ldots\gg|\alpha_{p-1}^{\mu}|\gg 1$ to make the second set of equations in Eq.~(\ref{coefficientseq}) hold. However, the first set of equations in Eq.~(\ref{coefficientseq}) prevents the result of $|\alpha_{j}^{\mu}|\gg 1$; thus we obtain $|\alpha_{j}^{\mu}\beta_{j}^{x}|\ll|\alpha_{p}^{\mu}\beta_{p}^{x}|$ in the deep bulk. If $|\beta_{0}|<|\beta_{j}|<1$, that is, $j>p+1$, $O(\alpha_{j}^{\mu}\beta_{j}^{L})\sim O(\alpha_{p}^{\mu}\beta_{p}^{L})$ makes the second set of equations in Eq.~(\ref{coefficientseq}) hold, leading to $|\alpha_{j}^{\mu}|\sim O(\beta_{p}^{L}/\beta_{j}^{L})\rightarrow 0$ and $|\alpha_{j}^{\mu}\beta_{j}^{x}|\ll|\alpha_{p}^{\mu}\beta_{p}^{x}|$ in the deep bulk. Note that $\alpha_{j}^{\mu}$ being a nonzero value, the exceptional situation, does not influence the left-localized asymptotic behavior of the corresponding bulk right eigenstate. If $|\beta_{p}|<1<|\beta_{j}|$, the second set of equations in Eq.~(\ref{coefficientseq}) requires $\alpha_{j}^{\mu}\leq O(\beta_{p}^{L}/\beta_{j}^{ L})$, leading to $|\alpha_{j}^{\mu}\beta_{j}^{x}|\ll|\alpha_{p}^{\mu}\beta_{p}^{x}|$ in the deep bulk. Second, we consider the case $|\beta_{0}|>1$. If $|\beta_{j}|<1<|\beta_{0}|$ or $1<|\beta_{j}|<|\beta_{0}|$, the second set of equations in Eq.~(\ref{coefficientseq}) requires $\alpha_{j}^{\mu}$ must be finite, leading to $|\alpha_{j}^{\mu}\beta_{j}^{x}|\ll|\alpha_{p}^{\mu}\beta_{p}^{x}|$ in the deep bulk. If $1<|\beta_{p}|<|\beta_{j}|$, the second set of equations in Eq.~(\ref{coefficientseq}) requires $\alpha_{j}^{\mu}\leq O(\beta_{p}^{L}/\beta_{j}^{ L})$, leading to $|\alpha_{j}^{\mu}\beta_{j}^{x}|\ll|\alpha_{p}^{\mu}\beta_{p}^{x}|$ in the deep bulk. In conclusion, the right EBB eigenstates Eq.~(\ref{rightstate}) are approximately 
\begin{align}
    \label{bulkrightstate}
    \braket{x|\Psi_{\mu}(\beta_{0})}_{R}&\sim \alpha_{p}^{\mu}\beta_{p}^{x}\ket{u_{p}^{\mu}}_{R}+\alpha_{p+1}^{\mu}\beta_{p+1}^{x}\ket{u_{p+1}^{\mu}}_{R}\nonumber\\
    &= \beta_{p}^{x}\left(\alpha_{p}^{\mu}\ket{u_{p}^{\mu}}_{R}+\alpha_{p+1}^{\mu}e^{i\theta x}\ket{u_{p+1}^{\mu}}_{R}\right)\nonumber\\
    &\equiv \beta_{0}^{x}\ket{\mathcal{N}_{\mu}(\beta_{0})}_{R},
\end{align}
where $x$ is in the deep bulk and $\beta_{p+1}=\beta_{p}e^{i\theta}$.  Noteworthily, the asymptotic behaviors in the deep bulk of the right EBB eigenstates Eq.~(\ref{rightstate}) are dominated by the points on $GBZ_{\mu}$, which is the precise significance of the non-Hermitian skin effect~(NHSE). As we approach the boundaries, the contributions in the right EBB eigenstates induced by the terms corresponding to the solutions $\beta$ of the characteristic equation away from GBZ emerge.

We turn to observe the left EBB eigenstates for completing the biorthogonal EBB eigenstates. The left eigenstate corresponding to the bulk band energy $E_{\mu}(\beta_{0})$ of $GBZ_{\mu}$ formally reads
\begin{align}
    \label{leftstate}
    \ket{\Psi_{\mu}(\beta_{0})}_{L}=\sum_{j=1}^{p+q}\alpha_{j}^{\prime\mu}\ket{\psi_{j}^{\mu}}_{L},
\end{align}
satisfying the eigenequation
\begin{align}
    \label{lefteigeneq}
    \hat{H}^{\dagger}\ket{\Psi_{\mu}(\beta_{0})}_{L}=E_{\mu}^{*}(\beta_{0})\ket{\Psi_{\mu}(\beta_{0})}_{L},
\end{align}
where $\ket{\psi_{j}^{\mu}}_{L}=\sum_{x=1}^{L}\left(\beta_{j}^{*}\right)^{-x}\ket{u_{j}^{\mu}}_{L}\ket{x}$, $H^{\dagger}(\beta_{j})\ket{u_{j}^{\mu}}_{L}=E_{\mu}^{*}(\beta_{0})\ket{u_{j}^{\mu}}_{L}$, and $H^{\dagger}(\beta_{j})=\sum_{n=-\mathcal{P}}^{\mathcal{Q}}t_{n}^{\dagger}(\beta_{j}^{*})^{n}$. Following the discussion for right EBB eigenstates, we immediately conclude that the left EBB eigenstates Eq.~(\ref{leftstate}) are approximately 
\begin{align}
    \label{bulkleftstate}
    \braket{x|\Psi_{\mu}(\beta_{0})}_{L}&\sim \alpha_{p}^{\prime\mu}(\beta_{p}^{*})^{-x}\ket{u_{p}^{\mu}}_{L}+\alpha_{p+1}^{\prime\mu}(\beta_{p+1}^{*})^{-x}\ket{u_{p+1}^{\mu}}_{L}\nonumber\\
    &= (\beta_{p}^{*})^{-x}\left(\alpha_{p}^{\prime\mu}\ket{u_{p}^{\mu}}_{L}+\alpha_{p+1}^{\prime\mu}e^{i\theta x}\ket{u_{p+1}^{\mu}}_{L}\right)\nonumber\\
    &\equiv (\beta_{0}^{*})^{-x}\ket{\mathcal{N}_{\mu}(\beta_{0})}_{L},
\end{align}
where $x$ is in the deep bulk in the thermodynamics limit and $\beta_{p+1}^{-1}=\beta_{p}^{-1}e^{-i\theta}$. Similarly, the asymptotic behaviors in the deep bulk of the left EBB eigenstates Eq.~(\ref{leftstate}) are dominated by the inverse of the points on $GBZ_{\mu}$, and the contributions induced by the terms corresponding to the solutions $\beta$ of the characteristic equation away from $GBZ_{\mu}$ emerge as we approach the boundaries. Note that the conjugation of Eq.~({\ref{bulkleftstate}}) should read
\begin{align}
    \label{phybiorthogonalbaisi}
    _{L}\braket{\Psi_{\mu}(\beta_{0})|x}&\sim _{L}\bra{u_{p}^{\mu}}\beta_{p}^{-x}\alpha_{p}^{\prime\mu*}+_{L}\bra{u_{p+1}^{\mu}}\beta_{p+1}^{-x}\alpha_{p+1}^{\prime\mu*}\nonumber\\
    &=\left(_{L}\bra{u_{p}^{\mu}}\alpha_{p}^{\prime\mu*}+_{L}\bra{u_{p+1}^{\mu}}e^{-i\theta x}\alpha_{p+1}^{\prime\mu*}\right)\beta_{p}^{-x}\nonumber\\
    &\equiv \left(_{L}\bra{u_{p}^{\mu}}\tilde{\alpha}_{p}^{\mu}+_{L}\bra{u_{p+1}^{\mu}}e^{-i\theta x}\tilde{\alpha}_{p+1}^{\mu}\right)\beta_{p}^{-x}\nonumber\\
    &\equiv_{L}\bra{\mathcal{N}_{\mu}(\beta_{0})}\beta_{0}^{-x}.
\end{align}
Combining the biorthogonal edge and EBB eigenstates, the unit in Hilbert space is given by~(assume the Hamiltonian is nondefective) 
\begin{align}
    \label{suppeq31}
    \sum_{\mu=1}^{\mathcal{M}}\sum_{\beta\in \left[GBZ_{\mu}\right]}\ket{\Psi_{\mu}(\beta)}_{RL}\bra{\Psi_{\mu}(\beta)}+\sum_{e=1}^{n_{e}}\ket{\Psi_{e}}_{RL}\bra{\Psi_{e}}=\mathbf{1}.
\end{align}
We call $\left[GBZ_{\mu}\right]$ the reduced sub-GBZ of energy band $E_{\mu}(\beta)$, which contains only one typically selected point $\beta_{p}$ for each so-called $n$-bifuration state $\ket{\Psi_{\mu}(\beta_{p})}$~\cite{wu2022}, with the solutions $\beta$ of characteristic equation concerning eigenenergy $E_{\mu}(\beta_{p})$ satisfying $|\beta_{1}|\leq\ldots=|\beta_{p}|=|\beta_{p+1}|=\ldots\leq|\beta_{p+q}|$~($n$ equal norms). Note that it is a discrete summation on $\left[GBZ_{\mu}\right]$; thus we cannot transform it to integral in the thermodynamics limit. Consequently, the non-defective Hamiltonian can be exactly diagonalized as
\begin{align}
    \label{hamiltoniandig}
    \hat{H}=\sum_{\mu=1}^{\mathcal{M}}\sum_{\beta\in \left[GBZ_{\mu}\right]}E_{\mu}(\beta)\ket{\Psi_{\mu}(\beta)}_{RL}\bra{\Psi_{\mu}(\beta)}+\sum_{e=1}^{n_{e}}E_{e}\ket{\Psi_{e}}_{RL}\bra{\Psi_{e}},
\end{align}
with $\hat{H}\ket{\Psi_{\mu}(\beta)}_{R}=E_{\mu}(\beta)\ket{\Psi_{\mu}(\beta)}_{R}$.
As usual, the edge states only contribute to the boundaries, which we will ignore in the deep bulk with $L\rightarrow+\infty$. 

\section{The completely biorthogonal basis of non-Hermitian systems}
\label{appendixC}
The single-particle retarded Green's function of 1D noninteracting non-Hermitian systems at zero temperature is defined as
\begin{eqnarray}
    \label{suppeq2}
    G(x,y;t)=-i\bra{0}\hat{c}_{x}(t)\hat{c}_{y}^{\dagger}(0)\ket{0}=-i\bra{x}e^{-i\hat{H}t}\ket{y},
\end{eqnarray}
where $\hat{c}_{x}(t)=e^{i\hat{H}t}c_{x}e^{-i\hat{H}t}$~($t>0$) is the annihilated operator in the Heisenberg picture, and $\ket{y}=c_{y}^{\dagger}\ket{0}$. As usual, we utilize the Bloch representation to handle the Green's function under the PBC,
\begin{align}
    \label{suppeq3}
    G(x,y;t)&=-i\int_{0}^{2\pi} dk \int_{0}^{2\pi}dk'\braket{x|k}\bra{k}e^{-i\hat{H}t}\ket{k'}\braket{k'|y}\nonumber\\
    &=-i\int_{0}^{2\pi}\frac{dk}{2\pi}e^{-iH(k)t}e^{ik(x-y)},
\end{align}
where $\ket{k}=\frac{1}{\sqrt{L}}\sum_{x=1}^{L}e^{ikx}\ket{x}$ is the Bloch wave function, expanding the single-particle Hilbert space and $\int_{0}^{2\pi}dk\ket{k}\bra{k}=\mathbf{1}$. 
Noteworthily, the integral of the wave vector $k$ is carried out through the continuous 1D first Brillouin zone~(BZ), which is valid in the thermodynamics limit $L\rightarrow\infty$, and reduces to the summation $\frac{1}{L}\sum_{k\in BZ}$ with $k=\frac{2\pi}{L}\alpha, \alpha=0,1,2,\ldots,L-1$ when $L$ is finite. Actually, $\ket{x}=(\ket{x,1},\ldots,\ket{x,\mathcal{M}})$ is a row vector of the Wannier functions at each internal degree of freedom, which reads $\braket{r|x,\mu}=w_{\mu}(x-r)=\frac{1}{\sqrt{2\pi}}\int_{0}^{2\pi}dk \phi_{\mu}(k,r)e^{-ikx},\mu=1,2,\ldots,\mathcal{M}$ representing in the coordinate space $\left\{r\in \mathbb{R}\right\}$. Here, the Bloch wave function represented in $\mathbb{R}$, $\braket{r|k,\mu}=\phi_{\mu}(k,r)$, is expressed independently for each internal degree of freedom, which means we just choose a simple ``plane-wave'' basis mathematically. We need to
diagonalize the Bloch Hamiltonian $H(k)$ to obtain the Bloch eigen-wave-functions. However, the ``plane-wave'' basis is convenient to calculate periodic-boundary Green's functions all the time.

When we refer to the OBC in non-Hermitian systems, the conventional BZ fails to produce Green's functions, since the open-boundary spectra correspond to the generalized Brillouin zone~(GBZ). The ``plane-wave" basis is not always valid in non-Hermitian systems with OBC, and we need to find a new basis to express Green's functions in general. Motivated by the biorthogonality of the eigenvectors of non-Hermitian matrices, we construct a minimally biorthogonal basis~(MBB) for the current non-Hermitian system~[Eq.~(\ref{generaltba})],
\begin{align}
    \label{suppeq4}
    \ket{\beta}_{R}&=\frac{1}{\sqrt{L}}\sum_{x=1}^{L}\beta^{x}\ket{x},\nonumber\\
    _{L}\bra{\beta}&=\frac{1}{\sqrt{L}}\sum_{x=1}^{L}\beta^{-x}\bra{x},
\end{align}
where $\beta=\mathscr{R}e^{i\theta}, \theta=\frac{2\pi}{L}m,m=0,1,2,\ldots,L-1$, with $\mathscr{R}$ being an arbitrary positive real number.
The biorthogonality of the MBB is given by
\begin{align}
    \label{suppeq5}
   _{L}\braket{\beta,\mu|\beta',\nu}_{R}=\delta_{\beta\beta'}\delta_{\mu\nu},
\end{align}
where
\begin{align}
    \label{suppeq6}
    \ket{\beta,\mu}_{R}&=\frac{1}{\sqrt{L}}\sum_{x=1}^{L}\beta^{x}\ket{x,\mu},\nonumber\\
    _{L}\bra{\beta,\mu}&=\frac{1}{\sqrt{L}}\sum_{x=1}^{L}\beta^{-x}\bra{x,\mu}.
\end{align}
More explicitly,
\begin{align}
   _{L}\braket{\beta,\mu|\beta',\nu}_{R}=\frac{1}{L}\sum_{x=1}^{L}\sum_{x'=1}^{L}\beta^{-x}\beta'^{x'}\braket{x,\mu|x',\nu}=\frac{1}{L}\sum_{x=1}^{L}\beta^{-x}\beta'^{x}\delta_{\mu\nu}
   =\frac{1}{L}\sum_{x=1}^{L}e^{i\frac{2\pi}{L}(-m+m')x}\delta_{\mu\nu}.\nonumber
\end{align}
If $m\neq m'$ and $-m+m'\equiv a\in\left\{1,2,\ldots,L-1\right\}$, 
\begin{align}
\sum_{x=1}^{L}e^{i\frac{2\pi}{L}(-m+m')x}=e^{i\frac{2\pi}{L}a}+e^{i2\frac{2\pi}{L}a}+\ldots+e^{iL\frac{2\pi}{L}a}=\frac{e^{i\frac{2\pi}{L}a}-e^{i(L+1)\frac{2\pi}{L}a}}{1-e^{i\frac{2\pi}{L}a}}=0;\nonumber
\end{align}
if $m=m'$, 
$\sum_{x=1}^{L}e^{i\frac{2\pi}{L}(-m+m')x}=L$,
thus $\frac{1}{L}\sum_{x=1}^{L}e^{i\frac{2\pi}{L}(-m+m')x}=\delta_{mm'}=\delta_{\beta\beta'}$. The completeness of MBB is given by
\begin{align}
    \label{suppeq7}
    \sum_{\beta}\ket{\beta}_{RL}\bra{\beta}=\mathbf{1}.
\end{align}
More explicitly,
\begin{align}
    \sum_{\beta}\ket{\beta}_{RL}\bra{\beta}
    =\frac{1}{L}\sum_{m=0}^{L-1}\sum_{x=1}^{L}\sum_{x'=1}^{L}\mathscr{R}^{x-x'}e^{i\frac{2\pi}{L}m(x-x')}\ket{x}\bra{x'}=\sum_{x=1}^{L}\ket{x}\bra{x}=\mathbf{1},\nonumber
\end{align}
where we have used $\frac{1}{L}\sum_{m=0}^{L-1}e^{i\frac{2\pi}{L}m(x-x')}=\delta_{xx'}$. In the thermodynamics limit, the completeness Eq.~(\ref{suppeq7}) becomes
\begin{align}
    \label{suppeq8}
    \frac{L}{2\pi}\int_{0}^{2\pi}d\theta\ket{\beta}_{RL}\bra{\beta}=\mathbf{1},
\end{align}
where $\left\{\beta=\mathscr{R}e^{i\theta},\theta\in\left[0,2\pi\right)\right\}$ forms a circle of radius $\mathscr{R}$.

\section{General form of open-boundary Green's functions}
\label{appendixD}
To obtain the general form of open-boundary Green's functions, we insert Eq.~(\ref{suppeq8}) into Eq.~(\ref{suppeq2}),
\begin{align}
    \label{suppeq9}
    G(x,y;t)&=-i\left(\frac{L}{2\pi}\right)\int_{0}^{2\pi}d\theta\braket{x|e^{-i\hat{H}t}|\beta}_{RL}\braket{\beta|y},
\end{align}
where $\beta=\mathscr{R}e^{i\theta}$. Next, we concentrate on the term $\braket{x|e^{-i\hat{H}t}|\beta}_{R}$, which is expanded as $\sum_{\alpha=0}^{+\infty}\frac{(-it)^{\alpha}}{\alpha!}\braket{x|\hat{H}^{\alpha}|\beta}_{R}$. First, we derive that 
\begin{align}
    \hat{H}\ket{\beta}_{R}&=\frac{1}{\sqrt{L}}\sum_{x'}\sum_{n\in \mathscr{D}}c_{x'}^{\dagger}t_{n}c_{x'+n}\sum_{x=1}^{L}\beta^{x}\ket{x}\nonumber\\
    &=\frac{1}{\sqrt{L}}\Bigg[\beta\sum_{n=0}^{\mathcal{Q}}\ket{1}t_{n}\beta^{n}
    +\ldots+\beta^{\mathcal{P}}\sum_{n=-\mathcal{P}+1}^{\mathcal{Q}}\ket{\mathcal{P}}t_{n}\beta^{n}+\sum_{x=\mathcal{P}+1}^{L-\mathcal{Q}}\beta^{x}\sum_{n=-\mathcal{P}}^{\mathcal{Q}}\ket{x}t_{n}\beta^{n}\nonumber\\
    &\qquad+\beta^{L-\mathcal{Q}+1}\sum_{n=-\mathcal{P}}^{\mathcal{Q}-1}\ket{L-\mathcal{Q}+1}t_{n}\beta^{n}+\ldots+\beta^{L}\sum_{n=-\mathcal{P}}^{0}\ket{L}t_{n}\beta^{n}\Bigg]\nonumber\\
    &=\frac{1}{\sqrt{L}}\Bigg[\ket{1}\left(H(\beta)\beta-\sum_{n=-1}^{-\mathcal{P}}t_{n}\beta^{n+1}\right)
    +\ldots+\ket{\mathcal{P}}\left(H(\beta)\beta^{\mathcal{P}}-t_{-\mathcal{P}}\right)+\sum_{x=\mathcal{P}+1}^{L-\mathcal{Q}}\ket{x}H(\beta)\beta^{x}\nonumber\\
    &\qquad+\ket{L-\mathcal{Q}+1}\left(H(\beta)\beta^{L-\mathcal{Q}+1}-t_{\mathcal{Q}}\beta^{L+1}\right)+\ldots+\ket{L}\left(H(\beta)\beta^{L}-\sum_{n=1}^{\mathcal{Q}}t_{n}\beta^{L+n}\right)\Bigg]\nonumber\\
    &=\frac{1}{\sqrt{L}}\sum_{x=1}^{L}\beta^{x}\ket{x}H(\beta)-\frac{1}{\sqrt{L}}\left(\sum_{n=-1}^{-\mathcal{P}}\ket{1}t_{n}\beta^{n+1}+\ldots+\ket{\mathcal{P}}t_{-\mathcal{P}}+\ket{L-\mathcal{Q}+1}t_{\mathcal{Q}}\beta^{L+1}+\ldots+\sum_{n=1}^{\mathcal{Q}}\ket{L}t_{n}\beta^{L+n}\right)\nonumber\\
    &=\ket{\beta}_{R}H(\beta)-\Big(\ket{1}\mathcal{B}_{1}(\beta)+\ldots+\ket{\mathcal{P}}\mathcal{B}_{\mathcal{P}}(\beta)+\ket{L-\mathcal{Q}+1}\mathcal{B}_{L-\mathcal{Q}+1}(\beta)+\ldots+\ket{L}\mathcal{B}_{L}(\beta)\Big)\nonumber,
\end{align} 
where we denote $\mathcal{B}_{1}(\beta)=\frac{1}{\sqrt{L}}\sum_{n=-1}^{-\mathcal{P}}t_{n}\beta^{n+1},\ldots,\mathcal{B}_{\mathcal{P}}(\beta)=\frac{1}{\sqrt{L}}t_{-\mathcal{P}},\mathcal{B}_{L-\mathcal{Q}+1}(\beta)=\frac{1}{\sqrt{L}}t_{\mathcal{Q}}\beta^{L+1},\ldots,\mathcal{B}_{L}(\beta)=\frac{1}{\sqrt{L}}\sum_{n=1}^{\mathcal{Q}}t_{n}\beta^{L+n}$, and $H(\beta)=\sum_{n=-\mathcal{P}}^{\mathcal{Q}}t_{n}\beta^{n}$ is called non-Bloch Hamiltonian. 

Second, by induction, we obtain 
\begin{align}
    \hat{H}^{2}\ket{\beta}_{R}&=\hat{H}\Big[\ket{\beta}_{R}H(\beta)-\Big(\ket{1}\mathcal{B}_{1}(\beta)+\ldots+\ket{\mathcal{P}}\mathcal{B}_{\mathcal{P}}(\beta)+\ket{L-\mathcal{Q}+1}\mathcal{B}_{L-\mathcal{Q}+1}(\beta)+\ldots+\ket{L}\mathcal{B}_{L}(\beta)\Big)\Big]\nonumber\\
    &=\ket{\beta}_{R}H^{2}(\beta)-\Big(\ket{1}\mathcal{B}_{1}(\beta)+\ldots+\ket{\mathcal{P}}\mathcal{B}_{\mathcal{P}}(\beta)+\ket{L-\mathcal{Q}+1}\mathcal{B}_{L-\mathcal{Q}+1}(\beta)+\ldots+\ket{L}\mathcal{B}_{L}(\beta)\Big)H(\beta)\nonumber\\
    &\quad-\hat{H}\Big(\ket{1}\mathcal{B}_{1}(\beta)+\ldots+\ket{\mathcal{P}}\mathcal{B}_{\mathcal{P}}(\beta)+\ket{L-\mathcal{Q}+1}\mathcal{B}_{L-\mathcal{Q}+1}(\beta)+\ldots+\ket{L}\mathcal{B}_{L}(\beta)\Big)\nonumber\\
    &=\ket{\beta}_{R}H^{2}(\beta)-\sum_{n\in\mathcal{D}}\Big(\ket{n}\mathcal{B}_{n}(\beta)H(\beta)+\hat{H}\ket{n}\mathcal{B}_{n}(\beta)\Big)\nonumber\\
    \hat{H}^{3}\ket{\beta}_{R}&=\hat{H}\Big[\ket{\beta}_{R}H^{2}(\beta)-\sum_{n\in\mathcal{D}}\Big(\ket{n}\mathcal{B}_{n}(\beta)H(\beta)+\hat{H}\ket{n}\mathcal{B}_{n}(\beta)\Big)\Big]\nonumber\\
    &=\ket{\beta}_{R}H^{3}(\beta)-\sum_{n\in\mathcal{D}}\ket{n}\mathcal{B}_{n}(\beta)H^{2}(\beta)-\sum_{n\in\mathcal{D}}\hat{H}\Big(\ket{n}\mathcal{B}_{n}(\beta)H(\beta)+\hat{H}\ket{n}\mathcal{B}_{n}(\beta)\Big)\nonumber\\
    &=\ket{\beta}_{R}H^{3}(\beta)-\sum_{n\in\mathcal{D}}\Big(\ket{n}\mathcal{B}_{n}(\beta)H^{2}(\beta)+\hat{H}\ket{n}\mathcal{B}_{n}(\beta)H(\beta)+\hat{H}^{2}\ket{n}\mathcal{B}_{n}(\beta)\Big)\nonumber\\
    \ldots\ldots\nonumber\\
    \hat{H}^{\alpha}\ket{\beta}_{R}&=\ket{\beta}_{R}H^{\alpha}(\beta)-\sum_{n\in\mathcal{D}}\sum_{\delta=0}^{\alpha-1}\hat{H}^{\alpha-1-\delta}\ket{n}\mathcal{B}_{n}(\beta)H^{\delta}(\beta),\nonumber
\end{align}
where $\mathcal{D}=\left\{1,\ldots,\mathcal{P},L-\mathcal{Q}+1,\ldots,L\right\}$ is the set of sites on the boundaries.
More explicitly, if 
\begin{align}
    \hat{H}^{\alpha-1}\ket{\beta}_{R}=\ket{\beta}_{R}H^{\alpha-1}(\beta)-\sum_{n\in\mathcal{D}}\sum_{\delta=1}^{\alpha-2}\hat{H}^{\alpha-2-\delta}\ket{n}\mathcal{B}_{n}(\beta)H^{\delta}(\beta),\nonumber
\end{align}
then 
\begin{align}
    \hat{H}^{\alpha}\ket{\beta}_{R}&=\hat{H}\Big[\ket{\beta}_{R}H^{\alpha-1}(\beta)-\sum_{n\in\mathcal{D}}\sum_{\delta=1}^{\alpha-2}\hat{H}^{\alpha-2-\delta}\ket{n}\mathcal{B}_{n}(\beta)H^{\delta}(\beta)\Big]\nonumber\\
    &=\Big(\ket{\beta}_{R}H(\beta)-\sum_{n\in\mathcal{D}}\ket{n}\mathcal{B}_{n}(\beta)\Big)H^{\alpha-1}(\beta)-\sum_{n\in\mathcal{D}}\sum_{\delta=1}^{\alpha-2}\hat{H}^{\alpha-1-\delta}\ket{n}\mathcal{B}_{n}(\beta)H^{\delta}(\beta)\nonumber\\
    &=\ket{\beta}_{R}H^{\alpha}(\beta)-\sum_{n\in\mathcal{D}}\ket{n}\mathcal{B}_{n}(\beta)H^{\alpha-1}(\beta)-\sum_{n\in\mathcal{D}}\sum_{\delta=1}^{\alpha-2}\hat{H}^{\alpha-1-\delta}\ket{n}\mathcal{B}_{n}(\beta)H^{\delta}(\beta)\nonumber\\
    &=\ket{\beta}_{R}H^{\alpha}(\beta)-\sum_{n\in\mathcal{D}}\sum_{\delta=0}^{\alpha-1}\hat{H}^{\alpha-1-\delta}\ket{n}\mathcal{B}_{n}(\beta)H^{\delta}(\beta).\nonumber
\end{align}
Consequently,
\begin{align}
    \label{suppeq10}
    \braket{x|e^{-i\hat{H}t}|\beta}_{R}&=\sum_{\alpha=0}^{+\infty}\frac{(-it)^{\alpha}}{\alpha!}\bra{x}\Big[\ket{\beta}_{R}H^{\alpha}(\beta)-\sum_{n\in\mathcal{D}}\sum_{\delta=0}^{\alpha-1}\hat{H}^{\alpha-1-\delta}\ket{n}\mathcal{B}_{n}(\beta)H^{\delta}(\beta)\Big]\nonumber\\
    &=\sum_{\alpha=0}^{+\infty}\frac{(-it)^{\alpha}}{\alpha!}\Big[\braket{x|\beta}_{R}H^{\alpha}(\beta)-\sum_{n\in\mathcal{D}}\sum_{\delta=0}^{\alpha-1}\braket{x|\hat{H}^{\alpha-1-\delta}|n}\mathcal{B}_{n}(\beta)H^{\delta}(\beta)\Big]\nonumber\\
    &=\sum_{\alpha=0}^{+\infty}\frac{(-it)^{\alpha}}{\alpha!}\Big[\frac{1}{\sqrt{L}}\beta^{x}H^{\alpha}(\beta)-\sum_{n\in\mathcal{D}}\sum_{\delta=0}^{\alpha-1}\braket{x|\hat{H}^{\alpha-1-\delta}|n}\mathcal{B}_{n}(\beta)H^{\delta}(\beta)\Big].
\end{align}
Finally, we obtain the general form of Green's function
\begin{align}
    \label{suppeq11}
    G(x,y;t)&=-i\left(\frac{L}{2\pi}\right)\int_{0}^{2\pi}d\theta\sum_{\alpha=0}^{+\infty}\frac{(-it)^{\alpha}}{\alpha!}\Big[\frac{1}{\sqrt{L}}\beta^{x}H^{\alpha}(\beta)-\sum_{n\in\mathcal{D}}\sum_{\delta=0}^{\alpha-1}\braket{x|\hat{H}^{\alpha-1-\delta}|n}\mathcal{B}_{n}(\beta)H^{\delta}(\beta)\Big]\,_{L}\braket{\beta|y}\nonumber\\
    &=-i\left(\frac{L}{2\pi}\right)\int_{0}^{2\pi}d\theta\sum_{\alpha=0}^{+\infty}\frac{(-it)^{\alpha}}{\alpha!}\Big[\frac{1}{\sqrt{L}}\beta^{x}H^{\alpha}(\beta)-\sum_{n\in\mathcal{D}}\sum_{\delta=0}^{\alpha-1}\braket{x|\hat{H}^{\alpha-1-\delta}|n}\mathcal{B}_{n}(\beta)H^{\delta}(\beta)\Big]\frac{1}{\sqrt{L}}\beta^{-y}\nonumber\\
    &=-i\left(\frac{L}{2\pi}\right)\int_{0}^{2\pi}d\theta\sum_{\alpha=0}^{+\infty}\frac{(-it)^{\alpha}}{\alpha!}\Big[\frac{1}{L}\beta^{x-y}H^{\alpha}(\beta)-\sum_{n\in\mathcal{D}}\sum_{\delta=0}^{\alpha-1}\frac{1}{\sqrt{L}}\beta^{-y}\braket{x|\hat{H}^{\alpha-1-\delta}|n}\mathcal{B}_{n}(\beta)H^{\delta}(\beta)\Big]\nonumber\\
    &=-i\left(\frac{1}{2\pi}\right)\int_{0}^{2\pi}d\theta\sum_{\alpha=0}^{+\infty}\frac{(-it)^{\alpha}}{\alpha!}\beta^{x-y}H^{\alpha}(\beta)+i\left(\frac{L}{2\pi}\right)\int_{0}^{2\pi}d\theta\sum_{\alpha=0}^{+\infty}\frac{(-it)^{\alpha}}{\alpha!}\sum_{n\in\mathcal{D}}\sum_{\delta=0}^{\alpha-1}\frac{1}{\sqrt{L}}\beta^{-y}\braket{x|\hat{H}^{\alpha-1-\delta}|n}\mathcal{B}_{n}(\beta)H^{\delta}(\beta)\nonumber\\
    &=-i\left(\frac{1}{2\pi}\right)\int_{0}^{2\pi}d\theta\beta^{x-y}e^{-iH(\beta)t}+i\left(\frac{L}{2\pi}\right)\int_{0}^{2\pi}d\theta\sum_{\alpha=0}^{+\infty}\sum_{n\in\mathcal{D}}\sum_{\delta=0}^{\alpha-1}\frac{1}{\sqrt{L}}\frac{(-it)^{\alpha}}{\alpha!}\beta^{-y}\braket{x|\hat{H}^{\alpha-1-\delta}|n}\mathcal{B}_{n}(\beta)H^{\delta}(\beta)\nonumber\\
    &=-i\oint_{|\beta|=\mathscr{R}}\frac{d\beta}{2\pi i\beta}\beta^{x-y}e^{-iH(\beta)t}+i\oint_{|\beta|=\mathscr{R}}\frac{d\beta}{2\pi i\beta}\sum_{\alpha=0}^{+\infty}\sum_{n\in\mathcal{D}}\sum_{\delta=0}^{\alpha-1}\frac{(-it)^{\alpha}}{\alpha!}\beta^{-y}\braket{x|\hat{H}^{\alpha-1-\delta}|n}\mathcal{B}_{n}(\beta)H^{\delta}(\beta),
\end{align}
where we have used $d\theta=\frac{d\beta}{i\beta}$ and rescaled $\mathcal{B}_{n}(\beta)$ as $\frac{1}{\sqrt{L}}\mathcal{B}_{n}(\beta)$ in the last equality. The first term of $G(x,y;t)$, denoted as $\mathcal{I}_{G}(x,y;t)\equiv-i\oint_{|\beta|=\mathscr{R}}\frac{d\beta}{2\pi i\beta}\beta^{x-y}e^{-iH(\beta)t}$, is the main part of open-boundary Green's functions, and the second term, denoted as $\mathcal{I}_{B}(x,y;t)\equiv i\oint_{|\beta|=\mathscr{R}}\frac{d\beta}{2\pi i\beta}\sum_{\alpha=0}^{+\infty}\sum_{n\in\mathcal{D}}\sum_{\delta=0}^{\alpha-1}\frac{(-it)^{\alpha}}{\alpha!}\beta^{-y}\braket{x|\hat{H}^{\alpha-1-\delta}|n}\mathcal{B}_{n}(\beta)H^{\delta}(\beta)$, is the contribution from the boundaries of the system, which tends toward vanished with $x,y$ being in the deep bulk as $L\rightarrow +\infty$. To see this clearly, we recall that $x,y$ are in the deep bulk if $\mathcal{P}, \mathcal{Q}\ll x,y\ll L-\mathcal{Q}+1$. We must always remember that the Green's function is the matrix value and the integral is performed over each matrix element independently. In addition, the integral of $\mathcal{I}_{B}$ is a polynomial of $\beta$ with only one pole $\beta=0$ for each matrix element; therefore the integral is irrelevant with positive real value $\mathscr{R}$. The nonzero contribution from $\mathcal{I}_{B}$ requires that there exist nonvanishing $\braket{x|\hat{H}^{\alpha-1-\delta}|n}$ and $\mathcal{I}_{c}\equiv i\oint_{|\beta|=\mathscr{R}}\frac{d\beta}{2\pi i\beta}\beta^{-y}\mathcal{B}_{n}(\beta)H^{\delta}(\beta)$ for some summation index $\left\{\alpha,n,\delta\right\}$. Since $\hat{H}$ shifts $\left\{\ket{n}, n\in\mathcal{D}\right\}$ at the left~(right) boundary to the right~(left) direction by a finite length, nonvanishing $\braket{x|\hat{H}^{\alpha-1-\delta}|n}$ requires $(\alpha-1-\delta)\rightarrow +\infty$~(i.e., large in the thermodynamics limit) with $\alpha\rightarrow +\infty$ if $x$ is in the deep bulk, which leads to that $\delta$ must be finite. According to the residue theorem, the integral $\mathcal{I}_{c}$ is nonzero only if the element of $\mathcal{B}_{n}(\beta)H^{\delta}(\beta)$ contains the $\beta^{y}$ term. Note that the order of $\beta$ of the element of $\mathcal{B}_{n}(\beta),n\in\left\{1,2,\ldots,\mathcal{P}\right\}$ is in the range $\left\{0,-1,\ldots,-\mathcal{P}+1\right\}$, while that of $\mathcal{B}_{n}(\beta),n\in\left\{L-\mathcal{Q}+1,\ldots,L-1,L\right\}$, is in the range $\left\{L+1,L+2,\ldots,L+\mathcal{Q}\right\}$, and the order of $\beta$ of the element of $H^{\delta}(\beta)$ is in the range $\left\{-\delta\mathcal{P},\ldots,\delta\mathcal{Q}\right\}$. We immediately find that the nonvanishing integral $\mathcal{I}_{c}$ requires $\delta\rightarrow +\infty$ when $y$ is also in the deep bulk, which leads to a contradiction. However, the finite $\delta$ is enough to make $\mathcal{I}_{c}$ nonvanishing when $y$ is located at the boundaries of the system. In turn, nonvanishing $\braket{x|\hat{H}^{\alpha-1-\delta}|n}$ requires finite $(\alpha-1-\delta)$ if $x$ is located at the boundaries, which results that there possibly exist nonzero contributions from $\mathcal{I}_{c}$ for both $y$ being in the bulk and the boundaries with $\alpha\in(0,+\infty)$, thus nonvanishing $\mathcal{I}_{B}$. Finally, we conclude that Green's functions contain a nonzero contribution from $\mathcal{I}_{B}$ when $x$ and~(or) $y$ are located at the boundaries, while $\mathcal{I}_{B}$ vanishes when both $x,y$ are in the deep bulk as $L\rightarrow +\infty$. In this paper, we concentrate on the main part $\mathcal{I}_{G}$ of open-boundary Green's functions, which implies the deep bulk information.

\section{Open-boundary Green's functions in frequency space}
\label{appendixE}
We transform the open-boundary Green's function in the deep bulk into frequency space,
\begin{align}
    \label{suppeq12}
    G(x,y;\omega)\equiv\int_{0}^{+\infty}dt\,\mathcal{I}_{G}(x,y;t)e^{i\omega t}.
\end{align}
Here, we should be careful to perform the matrix integral and treat it according to the matrix elements. Checking that for arbitrary constant matrix $K$
\begin{align}
    \label{suppeq13}
    \left[\int_{0}^{+\infty}dt\,e^{iKt}\right]_{\mu\nu}&=\left[\int_{0}^{+\infty}\sum_{n=0}^{+\infty}\frac{(it)^{n}}{n!}K^{n}d(Kt)K^{-1}\right]_{\mu\nu}\nonumber\\
    &=\int_{0}^{+\infty}\left(d(K_{\mu\rho}t)+\sum_{n=1}^{+\infty}\frac{(it)^{n}}{n!}K_{\mu\sigma_{1}}K_{\sigma_{1}\sigma_{2}}\ldots d(K_{\sigma_{n}\rho}t)\right)K^{-1}_{\rho\nu}\nonumber\\
    &=\left(K_{\mu\rho}t\Big|_{0}^{+\infty}+\sum_{n=1}^{+\infty}\frac{i^{n}}{n!}K_{\mu\sigma_{1}}K_{\sigma_{1}\sigma_{2}}\ldots K_{\sigma_{n}\rho}\frac{t^{n+1}}{n+1}\Big|_{0}^{+\infty}\right)K^{-1}_{\rho\nu}\nonumber\\
    &=\frac{1}{i}\left(i\delta_{\mu\rho}+\sum_{n=0}^{+\infty}\frac{(it)^{n+1}}{(n+1)!}K_{\mu\sigma_{1}}K_{\sigma_{1}\sigma_{2}}\ldots K_{\sigma_{n+1}\rho}\right)\Bigg|_{0}^{+\infty}K^{-1}_{\rho\nu}\nonumber\\
    &=\frac{1}{i}\left(\sum_{n=0}^{+\infty}\frac{(itK)^{n}_{\mu\rho}}{n!}\right)\Bigg|_{0}^{+\infty}K^{-1}_{\rho\nu}\nonumber\\
    &=\frac{1}{i}\left(\sum_{n=0}^{+\infty}\frac{\left(itK\right)^{n}_{\mu\rho}}{n!}\right)\Bigg|_{t\rightarrow +\infty}K^{-1}_{\rho\nu}-\frac{1}{i}\left(\sum_{n=0}^{+\infty}\frac{\left(itK\right)^{n}_{\mu\rho}}{n!}\right)\Bigg|_{t=0}K^{-1}_{\rho\nu}\nonumber\\
    &=i\left(K+i0^{+}\right)^{-1}_{\mu\nu},
\end{align}
we immediately obtain~(omit the factor $i0^{+}$)
\begin{align}
    \label{suppeq14}
    G(x,y;\omega)=\int_{0}^{+\infty}dt\left(-i\oint_{|\beta|=\mathscr{R}}\frac{d\beta}{2\pi i\beta}\beta^{x-y}e^{i\big(\omega-H(\beta)\big)t}\right)=\oint_{|\beta|=\mathscr{R}}\frac{d\beta}{2\pi i\beta}\frac{\beta^{x-y}}{\omega-H(\beta)}.
\end{align}
When we impose PBC to the system, we constrain $\beta^{L+1}=1$ and $\beta=e^{ik}$ in MBB, which are exactly the Bloch wave functions $\ket{k}$ in Eq.~(\ref{suppeq3}). Consequently, $\hat{H}\ket{k}=H(k)\ket{k}$ due to PBC, and $G(x,y;\omega)=\int_{0}^{2\pi}\frac{dk}{2\pi}\frac{{e^{ik(x-y)}}}{\omega-H(k)}$, in which the contributions from the boundary vanish naturally. However, the conventional topological invariants calculated under PBC cannot predict the topological edge modes in general, indicating the breakdown of the conventional bulk-boundary correspondence in non-Hermitian systems. Here, the topological edge modes in 1D non-Hermitian systems are the eigenenergies isolated from the continuous energy bands~(EBBs) under OBC~\cite{fu2022}. 

\section{Integral contours of open-boundary Green's functions}
\label{appendixF}
After obtaining the general form of open-boundary Green's functions, the urgent affair is choosing the integral contour for $G(x,y;\omega)$, since the contour with arbitrary $\mathscr{R}$ maybe not be physical. We can express Eq.~(\ref{suppeq14}) as
\begin{align}
    \label{suppeq15}
    G(x,y;\omega)=\oint_{|\beta|=\mathscr{R}}\frac{d\beta}{2\pi i\beta}\beta^{x-y}\frac{adj\left[\omega-H(\beta)\right]}{\det\left[\omega-H(\beta)\right]},
\end{align}
where $adj\left[\omega-H(\beta)\right]$ is the adjoint matrix of $\omega-H(\beta)$. Note that each element of $\beta^{x-y}adj\left[\omega-H(\beta)\right]$ is a polynomial of $\beta$, formally reading $\sum_{j}g_{j}\beta^{j}$. For each element of $\beta^{x-y}adj\left[\omega-H(\beta)\right]$, the integral surrounding the circle $|\beta|=\mathscr{R}$ in Eq.~(\ref{suppeq15}) results in the coefficients of the Laurent series $f(\beta,\omega)\equiv\det\left[\omega-H(\beta)\right]^{-1}$ expanded in a ring $\mathscr{R}_{1}(\omega)<\mathscr{R}< \mathscr{R}_{2}(\omega)$ multiplying corresponding $g_{j}$. Motivated by Ref.~\cite{xue2021simple}, consider the Toeplitz matrix $T(f)$ of Laurent series $f(\beta,\omega)$, whose elements are given by $T_{jk}=\oint_{|\beta|=\mathscr{R}}\frac{d\beta}{2\pi i\beta}\frac{\beta^{j-k}}{\det\left[\omega-H(\beta)\right]}$. We must always keep the formula $T(f)^{-1}=T(f^{-1})$ holding, thus leading to that $f(\beta,\omega)^{-1}=\det\left[\omega-H(\beta)\right]$ is smoothly interpolated to a unit $1$, which keeps this formula trivially. Since the winding number of $1$ surrounding the loop $|\beta|=\mathscr{R}$ vanishes, the winding number of $f(\beta,\omega)^{-1}$~[same for $f(\beta,\omega)$] enclosing this contour also vanishes. The continuous interpolation between $1$ and $f(\beta,\omega)^{-1}$ can be explicitly realized as follows. We define a continuous map $I:$ $[0,1]\rightarrow\mathbb{C}_{\mu}$ with $I(0)=0,I(1)=E_{0}$, and the interpolation is given by
$F(\beta,\omega_{\mu},\lambda)=E_{0}^{-\mathcal{M}}\det{\left[I(\lambda)(\omega-H(\beta))+(E_{0}-I(\lambda))\right]}$, which leads to $F(\beta,\omega,0)=1$ and $F(\beta,\omega,1)=f(\beta,\omega)^{-1}$. During $\lambda$ running in $[0,1]$, $F(\beta,\omega,\lambda)$ must be expressed in the contours keeping the vanishing winding number. The next task is to select the integral contours, namely find the physical expanded ring of Laurent series $f(\beta,\omega)$. 

The physical integral contour of $G(x,y;\omega)$ is the circle $|\beta|=\mathscr{R}$ on which the winding number of $ch(\beta,\omega)^{-1}\equiv f(\beta,\omega)$ vanishes. Note that the roles of zeros and poles of $ch(\beta,\omega)$ and $f(\beta,\omega)$ exchange, but it does not matter to the result of the winding number. Recalling the properties of sub-GBZs in Appendix~\ref{appendixA}, we find that all of the physical integral contours $|\beta|=\mathscr{R}$ are $|\beta_{p}(\omega)|<\mathscr{R}<|\beta_{p+1}(\omega)|$, which are equivalent to $GBZ_{\mu}$ with respect to $f(\beta,\omega)$ for any given $\omega\in\mathbb{C}_{\mu}$ and $\omega\not\in E_{\mu}[GBZ_{\mu}]$, denoted as $\omega_{\mu},\mu=1,2,\ldots,\mathcal{M}$ hereafter. In other words, $f(\beta,\omega_{\mu})$ is analytic in the ring $\mathscr{R}_{1}(\omega_{\mu})<\mathscr{R}<\mathscr{R}_{2}(\omega_{\mu})$ with $\mathscr{R}_{1}(\omega_{\mu})=|\beta_{p}(\omega_{\mu})|$, $\mathscr{R}_{2}(\omega_{\mu})=|\beta_{p+1}(\omega_{\mu})|$, and all of the closed integral contours in this ring are homotopic to each other. Finally, the open-boundary Green's function is given by 
\begin{align}
    \label{suppeq19}
    G(x,y;\omega_{\mu})=\oint_{GBZ_{\mu}}\frac{d\beta}{2\pi i\beta}\frac{\beta^{x-y}}{\omega_{\mu}-H(\beta)}.
\end{align}
Noteworthily, all of the sub-GBZs with respect to EBBs $E_{\mu}(\beta)$ are degenerate at BZ for Hermitian systems or non-Hermitian systems without the skin effect, and consequently, Eq.~(\ref{suppeq19}) reduces to 
\begin{align}
    \label{suppeq20}
    G(x,y;\omega)=\int_{0}^{2\pi}\frac{dk}{2\pi}\frac{e^{ik(x-y)}}{\omega-H(k)},
\end{align}
which is exactly the conventional form of Green's functions expressed in BZ. We emphasize that Eq.~(\ref{suppeq19}) is equivalent to the open-boundary Green's function only when the two correlated points $x,y$ are in the deep bulk in the thermodynamics limit, as well as Eq.~(\ref{suppeq20}) for the cases of Hermitian systems or non-Hermitian systems without the skin effect, while the Green's function under PBC is exactly Eq.~(\ref{suppeq20}) for any $x,y$ in both Hermitian and non-Hermitian systems. 

Mathematically, each monomial term of matrix element of $\beta^{x-y}adj\left[\omega-H(\beta)\right]$ within the integral Eq.~(\ref{suppeq15}) takes the form 
\begin{align}
    \label{suppeq21}
    G(x,y;\omega)=\oint_{|\beta|=\mathscr{R}}\frac{d\beta}{2\pi i\beta}\beta^{x-y}\frac{g_{j}\beta^{j}}{\det\left[\omega-H(\beta)\right]},
\end{align}
and we can regard $\frac{\beta^{j'}}{\det\left[\omega-H(\beta)\right]}$ with any integers $j'\neq0$ as the Laurent series to choose the integral contours. The sequent integral contours can be any closed loops in any rings, which are not the ring $|\beta_{p}(\omega_{\mu})|<\mathscr{R}<|\beta_{p+1}(\omega_{\mu})|$ anymore. These integral contours are not homotopic to $GBZ_{\mu}$, resulting in nonphysical Green's functions, since only $GBZ_{\mu}$ produces the physical continuous bulk spectrum~(EBB). Therefore, we obtain the physical integral contours for open-boundary Green's functions in the deep bulk only via regarding $f(\beta,\omega)$ as the allowed Laurent series. 
\end{widetext}

\bibliography{reference}

\end{document}